\documentclass[a4paper,11pt]{article}
\pdfoutput=1
\usepackage{jcappub}
\usepackage{setspace}
\usepackage[utf8]{inputenc} 
\usepackage{graphicx}
\usepackage{epstopdf}
\usepackage[body={17cm, 22.5cm},right=2.cm]{geometry}
\usepackage{amssymb,amsmath,amsfonts}
\usepackage{dcolumn}
\usepackage{mathrsfs}
\usepackage{array}
\usepackage{nonfloat}
\usepackage{soul}

\usepackage{multirow}
\usepackage{psfrag}
\pdfoutput=1

\usepackage{simpler-wick}

\definecolor{linkblue}{rgb}{0,0,0.8}
\definecolor{linkgreen}{rgb}{0,0.5,0}

\hypersetup{pdfpagemode=UseNone, pdfstartview=FitH, linkcolor=linkblue, %
            citecolor=linkgreen, urlcolor=linkblue}
            
\usepackage[numbers,sort&compress]{natbib}

\numberwithin{equation}{section}

\def\la{\langle}
\def\ra{\rangle}
\def\beq{\begin{equation}}
\def\eeq{\end{equation}}
\def\d{\partial}

\newcommand{\vd}{\boldsymbol{d}}
\newcommand{\vk}{\boldsymbol{k}}

\newcommand{\vq}{\boldsymbol{q}}
\newcommand{\vp}{\boldsymbol{p}}
\newcommand{\vl}{\boldsymbol{\ell}}

\newcommand{\vx}{\boldsymbol{x}}

\newcommand{\vu}{\boldsymbol{u}}

\newcommand{\vL}{\boldsymbol{L}}
\newcommand{\vtheta}{\boldsymbol{\theta}}

\def\calL{\mathcal{L}}

\newcommand{\knl}{k_{\rm NL}}

\newcommand{\kpar}{k_\parallel}
\newcommand{\kparone}{k_{\parallel1}}
\newcommand{\kpartwo}{k_{\parallel2}}
\newcommand{\Iobs}{I_{\rm len}}

\newcommand{\dirac}{\delta_{\rm D}}
\newcommand{\kro}{\delta^{\rm K}}
\newcommand{\invMpc}{h\, {\rm Mpc}^{-1}\,}

\def\Omm{\Omega_{\rm m}}

\newcommand{\lp}{\left(}
\newcommand{\rp}{\right)}
\newcommand{\lb}{\left[}
\newcommand{\rb}{\right]}

\definecolor{purple}{rgb}{0.78,0.18,0.77}

\newcommand{\kerfs}[1]{F_{#1}^{\rm (s)}}

\newcommand\numberthis{\addtocounter{equation}{1}\tag{\theequation}}

\def\sec#1{Sec.~\ref{#1}}

\makeatletter
\newlength{\apb@width}
\newcommand{\autoparbox}[2][c]{\settowidth{\apb@width}{#2}\parbox[#1]{\apb@width}{#2}}

\makeatother

\title{Lensing reconstruction from line intensity maps: the impact of gravitational nonlinearity}

\author[a]{Simon Foreman,}
\author[b,c,d,e]{P.~Daniel Meerburg,}
\author[a]{Alexander van Engelen,}
\author[a]{and Joel Meyers}

\affiliation[a]{Canadian Institute for Theoretical Astrophysics,\\University of Toronto, 60 St.~George Street, Toronto, ON, Canada, M5S 3H8}
\affiliation[b]{Kavli Institute for Cosmology, Madingley Road, Cambridge, UK, CB3 0HA}
\affiliation[c]{DAMTP, Centre for Mathematical Sciences, Wilberforce Road, Cambridge, UK, CB3 0WA}
\affiliation[d]{Kapteyn Astronomical Institute,\\ University of Groningen, P.O. Box 800, 9700 AV Groningen, The Netherlands}
\affiliation[e]{Van Swinderen Institute for Particle Physics and Gravity,\\ University of Groningen,
Nijenborgh 4, 9747 AG Groningen, The Netherlands}

\abstract{
We investigate the detection prospects for gravitational lensing of three-dimensional maps from upcoming line intensity surveys, focusing in particular on the impact of gravitational nonlinearities on standard quadratic lensing estimators. Using perturbation theory, we show that these nonlinearities can provide a significant contaminant to lensing reconstruction, even for observations at reionization-era redshifts. However, we show how this contamination can be mitigated with the use of a ``bias-hardened'' estimator. Along the way, we present an estimator for reconstructing long-wavelength density modes, in the spirit of the ``tidal reconstruction" technique that has been proposed elsewhere, and discuss the dominant biases on this estimator. After applying bias-hardening, we find that a detection of the lensing potential power spectrum will still be challenging for the first phase of SKA-Low, CHIME, and HIRAX, with gravitational nonlinearities decreasing the signal to noise by a factor of a few compared to forecasts that ignore these effects. On the other hand, cross-correlations between lensing and galaxy clustering or cosmic shear from a large photometric survey look promising, provided that systematics can be sufficiently controlled. We reach similar conclusions for a single-dish survey inspired by CII measurements planned for CCAT-prime, suggesting that lensing is an interesting science target not just for 21cm surveys, but also for intensity maps of other lines.
}

\begin{document}
\maketitle
\flushbottom

\section{Introduction}
\label{sec:introduction}

Gravitational lensing is the process by which photons are deflected by gradients of gravitational potentials as they traverse the universe.  If one can measure the impact of these deflections on observed images of the sky, it is possible to reconstruct a map of the potentials that caused the deflections. The prospect of such a map is incredibly useful for cosmology, as the underlying large-scale structure carries the imprints of everything from the Universe's initial conditions to the precise behavior of the dark energy which is accelerating the cosmic expansion at recent times. Sufficiently deep potential wells can have drastic effects on observed images at small scales, and while such ``strong lensing" events can sometimes be used for cosmology, we will focus on the opposite ``weak lensing" regime, where the strongest detections are typically accomplished through the analysis of the statistical imprint of much weaker lensing effects.

Such analyses have been successfully carried out in a variety of contexts. Lensing of cosmic microwave background (CMB) fluctuations was first detected about a decade ago in cross-correlation with the clustering of luminous objects~\cite{Smith:2007rg,Hirata:2008cb}. The current state of the art is a $\sim$40$\sigma$ detection of the projected auto spectrum of the relevant gravitational potentials by the Planck collaboration~\cite{Ade:2015zua}, and future measurements are predicted to reach $\sim$500$\sigma$~\cite{Abazajian:2016yjj}, enabling tight constraints on cosmological physics, particularly with respect to neutrino masses when combined with other probes~\cite{Kaplinghat:2003bh}. However, CMB lensing is limited in that we can only use a single two-dimensional ``screen" to measure the deflections, implying a limited number of available Fourier modes to use in the reconstruction process, and only grants access to a single 2d projection of the full 3d distribution of gravitational potentials.

Lensing can also be measured from correlations between the measured shapes of galaxies. The Dark Energy Survey has performed these measurements with a total detection significance of $\sim$26$\sigma$ in their Year 1 dataset~\cite{Troxel:2017xyo}, and more precise measurements are to come, both from that collaboration and future large galaxy surveys. This procedure also faces certain limitations. Large number densities of resolved galaxies are needed for a significant measurement of shape correlations, and this necessitates the use of photometric redshifts, which have sizable uncertainties that must be carefully accounted for, and which require complicated algorithms to reduce. Furthermore, accurate shape measurements are contingent on having strict control over systematics such as the telescope's point-spread function or biases in the method used to process raw images into catalogues of ellipticities. Finally, so-called ``intrinsic alignments" of galaxies with their surrounding environments can mimic the lensing signal and must be carefully modeled. (See Ref.~\cite{Mandelbaum:2017jpr} for a recent review of all of these issues.) 

In the presence of these difficulties, and given the promise of lensing to improve our knowledge about cosmology, one is led to ask whether we can measure lensing by other means. Since any measurements of a source field at cosmological distances will be subject to the effects of lensing, in principle, {\em any} source field with known statistics can be used to reconstruct the intervening gravitational potentials. For example, lensing of the Lyman-$\alpha$ forest~\cite{Croft:2017tur,Metcalf:2017qty} and the cosmic infrared background~\cite{Schaan:2018yeh} have recently been investigated. A particularly exciting example is neutral hydrogen, which is ubiquitously distributed throughout the Universe at all times after recombination, and can absorb or emit photons with a wavelength of 21cm as it undergoes a spin-flip transition of the proton-electron pair. Upon detection of these photons, the redshifts of the sources can then be precisely determined by a measurement of the photons' wavelengths. Consequently, the redshifted 21cm field traces the 3-dimensional distribution of neutral hydrogen. Measurements of this field have been identified as being able to supply information about the first collapsed objects and how they eventually reionized the Universe~\cite{Furlanetto:2006jb,2012RPPh...75h6901P}, and at lower redshifts, being able to constrain the Universe's recent expansion history via the baryon acoustic oscillation scale (e.g.~\cite{Chang:2007xk}).

Several works related to lensing of 21cm fluctuations exist in the literature. Early works suggested making use of pre-reionization 21cm absorption measurements using CMB lensing techniques~\cite{Cooray:2003ar}, or using either a position-space variance map or shape measurements of individually-resolved minihalos~\cite{Pen:2003yv}. Magnification of number-counts of 21cm-emitting galaxies (pursuant to the availability of sufficient angular resolution) was also identified as a possible option~\cite{Zhang:2005pu,Zhang:2005eb}. It was suggested in Ref.~\cite{Sigurdson:2005cp} that 21cm lensing measurements might be useful in de-lensing CMB B-modes that act as a contaminant to estimates of primordial tensor modes, but this was found to be a very futuristic prospect.

In Ref.~\cite{Zahn:2005ap}, the 2d Fourier-space quadratic estimator used in CMB lensing~\cite{Hu:2001tn,Hu:2001kj} was extended to observations in 3d, with an eye to future applications to reionization-era 21cm measurements. Other estimators have also been discussed, including configuration-space correlation-function-based estimators~\cite{Metcalf:2006ji,Metcalf:2008gq} similar to those for the CMB~\cite{Seljak:1998aq,Zaldarriaga:1998te}, mixed configuration-Fourier space estimators~\cite{Metcalf:2006ji}, and estimators for the lensing convergence and shear instead of the deflection angle~\cite{Lu:2007pk,Lu:2009je}. Fourier-space estimators have been further investigated, in the presence of a Poisson component of the source field power spectrum~\cite{Pourtsidou:2013hea,Pourtsidou:2014pra} and in simulations~\cite{Romeo:2017zwt}. An alternative method for calculating the effect of lensing on 21cm observations, involving a Wilsonian cutoff-based approach that leads to a system of differential equations, was investigated in Ref.~\cite{Mandel:2005xh}. In addition to the applications mentioned above, it has been proposed that 21cm lensing could be used to detect massive halos~\cite{Metcalf:2006ji,Hilbert:2007jda}, measure galaxy cluster masses~\cite{Kovetz:2012pt}, or test the origin of the CMB cold spot~\cite{Kovetz:2012jq}. Finally, detection of curl lensing modes induced by inflationary gravitational waves could help to verify the primordial origin of large-scale B-modes in the CMB~\cite{Sheere:2016yqu}, or even constrain the tensor-to-scalar ratio to high precision~\cite{Book:2011dz}.

In addition to 21cm, there are other emission lines that one could use to map out the Universe in certain regimes. Many of these lines are too faint to detect from all but the brightest emitting objects, but this hurdle can be overcome by aggregating all emission at a given wavelength into broad, arcminute-scale pixels, a technique known as ``line intensity mapping." A community effort is beginning to mobilize around this idea, exploring its applications to open questions in star formation, galaxy evolution, and cosmology; for a recent summary, see Ref.~\cite{Kovetz:2017agg}. Lensing reconstruction using these maps has not been well-explored outside of the 21cm case, aside from Ref.~\cite{Cooray:2003ar}, who mentioned the infrared background from the first stars as a possible source field. As with 21cm, the benefit of mapping an emission line is that redshift information can be used to construct a 3d map. Such a map will have limited angular resolution compared to galaxy surveys, but this is precisely the regime of CMB measurements, for which there exist well-developed and actively-used tools for lensing reconstruction. 

In contrast to the CMB, however, the lower-redshift structure traced by line intensity maps will be subject to significant gravitational evolution, which induces nonlinearities into an otherwise linear field with Gaussian statistics.  Even at high redshifts, at which the source field will be more linear {\em at a given comoving scale} than at lower redshifts, nonlinearities will become important at sufficiently small scales, which often overlap with the expected angular resolutions of intensity mapping surveys at those redshifts.  As we will demonstrate in this paper, these nonlinearities can act as an extra source of noise on lensing power spectra, and can also generate additive biases both at map level and power-spectrum level.  These effects are sometimes remarked upon, but their quantitative impact on forecasts or formalisms has often been omitted in previous work. Notable exceptions are Refs.~\cite{Lu:2007pk,Lu:2009je}, in which gravitational nonlinearities were incorporated into a lensing estimator by calibrating against $N$-body simulations, and Ref.~\cite{Schaan:2018yeh}, which has investigated related effects in lensing of the cosmic infrared background.

Our overall goal is to quantify the impact of these nonlinearities on the Fourier-space quadratic estimator from Ref.~\cite{Zahn:2005ap}, first in a general setting and then for specific intensity mapping surveys that are ongoing or planned.  We will do so using large-scale structure perturbation theory up to a certain order; this approach is only valid within a certain regime (which we quantify, and which overlaps well with the regime accessible to most observations we consider), but has the advantage of being describable analytically, enabling more control over predictions than in a simulation- or fitting-function-based approach. 
\goodbreak

Our paper is organized as follows:
\begin{itemize}
\item {\bf \sec{sec:lensing-estimator}}: We first review the derivation of the quadratic lensing estimator from Ref.~\cite{Zahn:2005ap}. We then set up and sketch our perturbation theory calculation of the leading effects of gravitational non-linearities on this estimator, presenting the final expressions but relegating the details to Apps.~\ref{app:kernels}-\ref{app:fkernels}.  We visualize the size of these effects as a function of redshift and angular resolution, and show that they significantly increase the lensing reconstruction noise (i.e.\ the noise per mode) over the standard assumptions of linearity and Gaussianity.
\item {\bf \sec{sec:bh}}: We show how the technique of ``bias-hardening"~\cite{Namikawa:2012pe} can be used to modify the lensing estimator to subtract off a sizable portion of the gravitational bias, and quantify the advantages and disadvantages of this technique in different circumstances. We also comment on the application of quadratic estimators in this paper to ``tidal reconstruction"~\cite{Pen:2012ft,Zhu:2015zlh,Zhu:2016esh}, which can reconstruct long-wavelength modes of the matter density at the source redshift.
\item {\bf \sec{sec:forecasts}}: We perform forecasts for the detectability of lensing, either in the auto spectrum or in cross-correlation with galaxy clustering or cosmic shear from an LSST-like survey, for phase~1 of SKA-Low, CHIME, HIRAX, and a single-dish CII intensity mapping survey.  The results of these forecasts are summarized in Table~\ref{table:sn1} and Fig.~\ref{fig:ccat-sn}. Overall, we find that a detection of the lensing auto spectrum in these surveys will be challenging, but cross-correlations with a large photometric survey may be detectable (pursuant to systematics being controlled at the appropriate level).
\item {\bf \sec{sec:design}}: Using HIRAX as our base configuration, we examine which improvements in experimental or survey design would yield the greatest benefits to a lensing analysis. Various configurations are summarized in Table~\ref{table:superhirax}; in brief, we find that increasing the number of dishes in an interferometer would confer the largest improvement from a lensing point of view, as this would add longer baselines (increasing the angular resolution on the sky) while decreasing the thermal noise contribution across all observed scales.
\item {\bf \sec{sec:conclusions}}: We conclude by discussing related topics that would be worth pursuing in future work.
\end{itemize}

In this work, we will use same the background cosmology as the Planck 2015 lensing analysis~\cite{Ade:2015zua}: a spatially-flat Lambda cold dark matter model with physical baryon density $\Omega_{\rm b}h^2 = 0.0222$, physical matter density $\Omm h^2 = 0.1245$, physical neutrino density $\Omega_{\nu} h^2 = 0.00064$, dimensionless Hubble parameter $h=0.6712$, amplitude and slope of primordial perturbations of $A_{\rm s}=2.09\times 10^{-9}$ and $n_{\rm s}=0.96$, both evaluated at a pivot scale of $k_{\rm pivot}=0.05{\rm Mpc}^{-1}$. We will use the following conventions for integrals in 2d and 3d Fourier space:
\beq
\int_{\vl} \equiv \int \frac{d^2\vl}{(2\pi)^2}\ , \quad \int_{\vk} \equiv \int \frac{d^3\vk}{(2\pi)^3}\ .
\eeq
We will use $\dirac$ and $\kro$ to denote Dirac delta functions and Kronecker deltas, respectively. These are not to be confused with the nonlinear matter overdensity $\delta \equiv (\rho-\bar{\rho})/\bar{\rho}$, or $\delta_n$, the contribution to the overdensity at $n$th order in perturbation theory (introduced in \sec{sec:pt}).

\section{Lensing estimator}
\label{sec:lensing-estimator}

\subsection{Observations in 3d}
\label{sec:observationsin3d}

Consider observations of an intensity field in 3d at time $\tau$, $I_{\rm 3d}(\vx;\tau)$. Following Ref.~\cite{Zahn:2005ap}, we use the two-dimensional angular wavevector  $\vl$ as the transverse Fourier coordinate instead of $\vk_\perp$, the comoving spatial wavevector perpendicular to the line of sight. We define
\beq
I(\vl,\kpar;\tau) \equiv \int \frac{dx_\parallel}{\calL} \, e^{-i\kpar x_\parallel}
	\int d^2\vtheta\, e^{-i\vl\cdot\vtheta} I_{\rm 3d}(\chi\vtheta,x_\parallel;\tau)
	= \frac{1}{\calL \chi^2} I_{\rm 3d}(\vl/\chi,\kpar;\tau) \ ,
	\label{eq:idef}
\eeq
working in the flat sky approximation.
We have defined $\calL$ as the comoving line-of-sight thickness within which we observe the intensity, and draw attention to the notational distinction between $I_{\rm 3d}$ and $I$ (the latter of which matches the definition of $\hat{I}$ from Ref.~\cite{Zahn:2005ap}).
Note that we have assumed that we can neglect time-evolution of $I_{\rm 3d}$ within the line-of-sight range of our observations, such that we do not need to Fourier transform along the light-cone in the line-of-sight direction; thus, we take $\tau$ to denote the time corresponding to the mean of the observed redshift interval, and take the comoving distance $\chi$ to be evaluated at $\tau$.\footnote{Note that this ``observed redshift interval" need not span an entire survey. In a typical case, the full observed range would be broken up into smaller redshift bins, with quantities evaluated at the mean redshift of each bin. We adopt this approach in the forecasts in \sec{sec:forecasts}.} Henceforth, we will drop the $\tau$ argument from all quantities.

We define the angular power spectrum of $I(\vl,\kpar)$ by
\beq
C_\ell(\kpar) \equiv \calL^{-1}\chi^{-2} P_I\lp\sqrt{ \frac{\ell^2}{\chi^2}+\kpar^2} \rp\ ,
	\label{eq:clidef}
\eeq
where $P_I$ is the 3d power spectrum of $I_{\rm 3d}$, so that
\beq
\left\la I(\vl,k_{\parallel}) I(\vl',-k_{\parallel}) \right\ra
	= (2\pi)^2\dirac(\vl+\vl') C_\ell(\kpar)\ .
\eeq
We similarly define $C_\ell^{\rm tot}(\kpar)$ as the angular power spectrum of $I$, including the experimental noise contribution $C_\ell^{\rm N}(\kpar)$:
\beq
C_\ell^{\rm tot}(\kpar) =  C_\ell(\kpar) + C_\ell^{\rm N}(\kpar)\ .
\label{eq:cltotdef}
\eeq
As indicated earlier, we will consider observations within a finite comoving thickness $\calL$, implying that the available $\kpar$ modes will be discrete. We index these modes by $j$, defining $k_{\parallel \alpha}=2\pi j_\alpha/\calL$. For ease of notation, we will sometimes use continuous notation for $\kpar$, with appropriate substitutions understood, e.g.\ $(2\pi) \dirac(\kparone-\kpartwo) \to \calL \kro_{j_1j_2}$.

\subsection{Review of quadratic estimator in 3d}
\label{sec:estimator-review}

The effect of gravitational lensing is typically described by the lensing potential $\phi(\vtheta;\tau)$, given by
\beq 
\phi(\vtheta;\chi_{\rm s}) = \frac{2}{c^2} \int_0^{\chi_{\rm s}} 
	d\chi \frac{\chi_{\rm s}-\chi}{\chi_{\rm s} \chi}
	\Phi(\chi\vtheta,\chi;\tau[\chi])\ ,
\eeq
in a flat universe, where $\chi_{\rm s}$ is the comoving distance to the sources that are being lensed and $\Phi$ is the gravitational potential, related to the matter overdensity $\delta$ by
\beq
\nabla^2\Phi(\vx,\tau) = \frac{3}{2} \Omm(\tau) a(\tau)^2 H(\tau)^2 \delta(\vx,\tau)\ .
\eeq
In the Limber approximation \cite{Limber:1953,Kaiser:1991qi}, the angular power spectrum of $\phi$ can be written as 
\beq \label{eq:clphiphilimber}
C_L^{\phi\phi} \approx \frac{9}{c^4} \int_0^{\chi_{\rm s}} d\chi \frac{\chi^2}{L^4} 
	\lp \frac{\chi_{\rm s}-\chi}{\chi_{\rm s}\,\chi} \rp^2 \Omm(\tau[\chi])^2 a(\tau)^4 H(\tau[\chi])^4
	P_{\delta\delta}(L/\chi;\tau[\chi])\ .
\eeq
In this paper, we will take $\chi_{\rm s}$ to be the distance to the mean redshift of the observed volume, and assume that the source field within that volume is all lensed by the same $\phi$.  In our forecasts for specific surveys below, we ensure to set $\calL$ such that $C_L^{\phi\phi}$ varies by less than $\sim 10\%$ over the thickness of each slab.

Observations of an intensity field will be affected by gravitational lensing via a re-mapping of the observed angular coordinates by a deflection field $\vd(\vtheta)$:
\beq
\Iobs(\vx_\perp,x_\parallel) = I_{\rm 3d}(\vx_\perp+\chi\vd(\vtheta),x_\parallel) \ .
\label{eq:iobs}
\eeq
In the limit of weak deflections, we can write $d_i(\vtheta) = (\d/\d\theta_i) \phi(\vtheta)$, and Taylor-expand Eq.~\eqref{eq:iobs} around $\phi=0$:
\begin{align*}
\Iobs(\vx_\perp,x_\parallel) &=
	I_{\rm 3d}(\vx_\perp,x_\parallel) + \chi \frac{\d}{\d\theta_a} \phi(\vtheta) 
	\cdot \frac{\d}{\d x_\perp^a} I_{\rm 3d}(\vx_\perp,x_\parallel) + \mathcal{O}(\phi^2)\ .
	\numberthis
\end{align*}
 In $(\vl,\kpar)$-space, this becomes
\beq
\Iobs(\vl,\kpar) = \calL^{-1} \chi^{-2} I_{\rm 3d}(\vl/\chi,\kpar)
	- \calL^{-1}\chi^{-2} \int_{\vl'} \vl'\cdot(\vl-\vl') \phi(\vl-\vl') I_{\rm 3d}(\vl'/\chi,\kpar) 
	+ \mathcal{O}(\phi^2)\ .
\eeq
If we consider an ensemble average over realizations of $I$ with $\phi$ held fixed, we obtain 
\begin{align*}
&\left\la \Iobs(\vl_1,k_{\parallel 1}) \Iobs(\vl_2,k_{\parallel 2}) \right\ra_{\phi\,{\rm fixed}} \\
&\qquad=  
(2\pi)^3\dirac(\vl_1+\vl_2) \dirac(k_{\parallel1}+k_{\parallel2}) \, 
	\calL^{-1 } C_{\ell_1}(\kparone) \\
&\qquad\quad + (2\pi) \dirac(k_{\parallel1}+k_{\parallel2}) \calL^{-1} 
	 \lp \vl_1+\vl_2 \rp \cdot
	\lb \vl_1 C_{\ell_1}(\kparone) +  \vl_2 C_{\ell_2}(\kpartwo) \rb \phi(\vl_1+\vl_2) \\
&\qquad\quad + \mathcal{O}(\phi^2)\ ;
	\numberthis
	\label{eq:iiavgphifixed}
\end{align*}
thus, lensing induces off-diagonal angular correlations in the observed intensity, and this fact can be used to construct an estimator for the associated mode of the lensing potential.

In detail, we can write a quadratic estimator for $\phi$ in the form
\beq
\hat{\phi}(\vL;\kpar) \equiv \int_{\vl} g_\phi(\vl,\vL-\vl,\kpar) 
	I_{\rm obs}(\vl,\kpar) I_{\rm obs}(\vL-\vl,-\kpar)\ ,
	\label{eq:phihatdef}
\eeq
where $I_{\rm obs}$ is equal to $\Iobs$ plus instrumental noise.  (Note that the $\kpar$ argument of the estimator indicates that modes of $I$ with parallel wavenumber $\kpar$ are being used to estimate $\phi(\vL)$, not that the desired mode of $\phi$ has parallel wavenumber $\kpar$.)  Since $\hat \phi(\vL;-\kpar) =\hat \phi(\vL;\kpar)$ we will restrict to positive values of $\kpar$ when we sum over radial modes below.

We can derive the choice of $g_\phi$ that minimizes the variance of $\hat{\phi}$ subject to the condition that the estimator is unbiased, $\langle\hat{\phi}(\vL;\kpar)\rangle=\phi(\vL)$, with the variance calculated assuming that the source intensity field is Gaussian. This yields
\beq
g_\phi(\vl,\vL-\vl,\kpar) \equiv N_{\phi\phi}^{\rm (G)}(L,\kpar) \frac{f_\phi(\vl,\vL-\vl,\kpar)}
	{C_{\ell}^{\rm tot}(\kpar) C_{|\vL-\vl|}^{\rm tot}(\kpar)}\ ,
	\label{eq:gphidef}
\eeq
where
\beq
f_\phi(\vl_1,\vl_2,\kpar) \equiv
	 (\vl_1+\vl_2)\cdot \lb \vl_1 C_{\ell_1}(\kpar)
	 + \vl_2 C_{\ell_2}(\kpar) \rb, 
\eeq
and
\beq
N_{\phi\phi}^{\rm (G)}(L,\kpar) \equiv  \lb \int_{\vl}
	\frac{f_{\phi}(\vl,\vL-\vl,\kpar)^2}
	{C_{\ell}^{\rm tot}(\kpar) C_{|\vL-\vl|}^{\rm tot}(\kpar)} \rb^{-1}\ .
	\label{eq:nphiphidef}
\eeq
With these choices, the variance of $\hat{\phi}(\vL;\kpar)$ is equal to $N_{\phi\phi}^{\rm (G)}(L,\kpar)$.  Note that for $\kpar = 0$, including for the case of a single source screen such as the CMB, there is an extra factor of 2 in the denominator of Eq.~\eqref{eq:gphidef} and the denominator of the integrand of \eqref{eq:nphiphidef}.  

In the $L\ll \ell$ limit and for a Gaussian source field, Eq.~\eqref{eq:phihatdef} optimally combines the information from the lensing convergence and shear. It is also possible to construct separate estimators for convergence and shear (e.g.~\cite{Bucher:2010iv,Lu:2007pk,Lu:2009je}), which can confer certain advantages in the presence of strong intrinsic non-Gaussianities. As we describe later, in this work we explicitly restrict ourselves to the regime of weak non-Gaussianity of the source field, where we expect the variance of the estimator in Eq.~\eqref{eq:phihatdef} to remain close to optimal. 

In the Gaussian approximation for $I$, the noise associated with the estimator in Eq.~\eqref{eq:phihatdef} for a given mode $\phi(\vL)$ is uncorrelated for different values of $\kpar$. Therefore, estimates from different $\kpar$ values can be combined with inverse-variance weighting to yield an optimal estimate of $\phi(\vL)$, with variance
\beq
N_{\phi\phi}^{\rm (G,combined)}(L) = \lb \sum_{j\geq j_{\rm min}} 
	N_{\phi\phi}^{\rm (G)}(L,2\pi j/\calL)^{-1} \rb^{-1}\ .
	\label{eq:nphiphigcomb}
\eeq
Note that similar expressions in the literature (e.g.~\cite{Zahn:2005ap,Romeo:2017zwt}) often have an extra factor of 2 in the denominator of the integrand of Eq.~\eqref{eq:nphiphidef}. In these works, the sum in Eq.~\eqref{eq:nphiphigcomb} runs over positive and negative values of $j$. Since $\hat \phi(\vL;-\kpar) =\hat \phi(\vL;\kpar)$, this sum is equivalent to twice the sum over only positive $j$ values, and is therefore equivalent to the expressions we present here, which are for $j > 0$.

\subsection{Gravitational contributions}
\label{sec:gravity-terms}

The estimator in \sec{sec:estimator-review} assumed that lensing is the only source of off-diagonal mode-couplings in the source intensity field. For the CMB, this is an excellent approximation, since other sources of off-diagonal correlations or  non-Gaussianity are subdominant and there exist well-developed techniques to model and remove these contaminants (e.g.~\cite{vanEngelen:2013rla,Osborne:2013nna}). On the other hand, a line intensity map will trace the underlying matter perturbations to a large extent, and these perturbations will be non-Gaussian due to gravitational evolution, with a magnitude that increases with time and wavenumber. Even at redshifts that might be considered ``high," if sufficiently small angular scales are used for lensing reconstruction, there is a danger that these scales could have a sizeable level of gravitational non-Gaussianity. If this non-Gaussianity is not properly accounted for, the induced mode-coupling in the source field will bias the estimate of the lensing field constructed from intensity maps. In this section, we set up a framework in which one can quantify these effects as a function of the properties of a given observation.

\subsubsection{Setup of perturbative calculation}
\label{sec:pt}

In the regime where nonlinearities from gravity are non-negligible but weak, they can be described using perturbation theory (e.g.~\cite{Bernardeau:2001qr}). In this approach, the effective fluid equations that describe the matter density and velocity at large scales are solved perturbatively around the solution to the linearized equations. If we denote\footnote{We will always take $\delta$ to be the standard definition of the matter overdensity, evaluated on 3d coordinates in either configuration or Fourier space, and therefore do not use the ``3d" subscript used on $I$ in Secs.~\ref{sec:observationsin3d} and~\ref{sec:estimator-review}.} the linear overdensity by $\delta_1(\vk;\tau)$, the corrections arising from nonlinear terms in the fluid equations can be written as an expansion in $\delta_1$, involving kernels $\kerfs{n}$ that describe nonlinearities via couplings between different linear modes:
\begin{align*}
\delta(\vk;\tau) &= \sum_{n=1}^\infty \delta_n(\vk;\tau) \\
&= \sum_{n=1}^\infty \int_{\vq_1}\cdots \int_{\vq_n} (2\pi)^3\dirac(\vk-\vq_1-\cdots-\vq_n)
	\kerfs{n}(\vq_1,\cdots,\vq_n) \delta_1(\vq_1;\tau)\cdots\delta_1(\vq_n;\tau)\ .
	\numberthis
	\label{eq:deltanexpansion}
\end{align*}
The form of the $\kerfs{n}$ kernels is given by a recurrence relation that can be derived from the fluid equations (e.g.~\cite{Bernardeau:2001qr}); in this work, we will only need $\kerfs{2}$ and $\kerfs{3}$, which are given in App.~\ref{app:kernels}. When correlation functions of $\delta$ are calculated using the expansion above, they will involve a series of terms composed of convolution integrals over factors of the kernels and the linear power spectrum. These terms can be described using the language of Feynman diagrams, with those containing no unevaluated integrals referred to as being ``tree-level," while those with $n$ integrals are called ``loop corrections," because the corresponding diagram contains $n$ closed loops around which the ``momenta" (wavenumbers) are unconstrained.

In the modern view, this perturbation theory should be understood as an effective field theory, valid only for wavenumbers smaller than some nonlinear scale~$\knl$ where fluctuations become too large to admit a perturbative treatment~\cite{Baumann:2010tm,Carrasco:2012cv}. The expansion in Eq.~\eqref{eq:deltanexpansion} must be supplemented by a parallel expansion in ``counterterms" with a similar form, which cure the sensitivity of loop integrals to high wavenumbers where the effective fluid description is no longer valid, and also parametrize further physical effects of couplings between long and short modes. These terms arise from an effective stress tensor in the Euler equation, itself written as an expansion in long-wavelength degrees of freedom and spatial derivatives, consistent with the symmetries expected to be obeyed at large distances. A growing body of work (e.g.~\cite{Pajer:2013jj,Carrasco:2013sva,Baldauf:2014qfa,Angulo:2014tfa,Baldauf:2015zga,McQuinn:2015tva,Foreman:2015lca,Bertolini:2015fya,Perko:2016puo}) has shown that the addition of these terms can improve the behavior of the perturbative expansion, both from the perspective of theoretical consistency and matching nonlinear measurements from $N$-body simulations.

In this work, we wish to quantify the contribution of gravitational nonlinearities to the variance of the quadratic lensing estimator, which ultimately arises from the trispectrum of the source intensity field. We restrict ourselves to the tree-level computation, which is $\mathcal{O}(\delta_1^6)$ in the trispectrum, requiring up to third order in the expansion in Eq.~\eqref{eq:deltanexpansion}. This has the advantage of computational simplicity, since there are no loop integrals to perform. Furthermore, it is not necessary to include any of the counterterms mentioned above, since they are only needed in loop-level calculations.

\begin{figure}[t]
\begin{centering}
\includegraphics[width=1\textwidth]{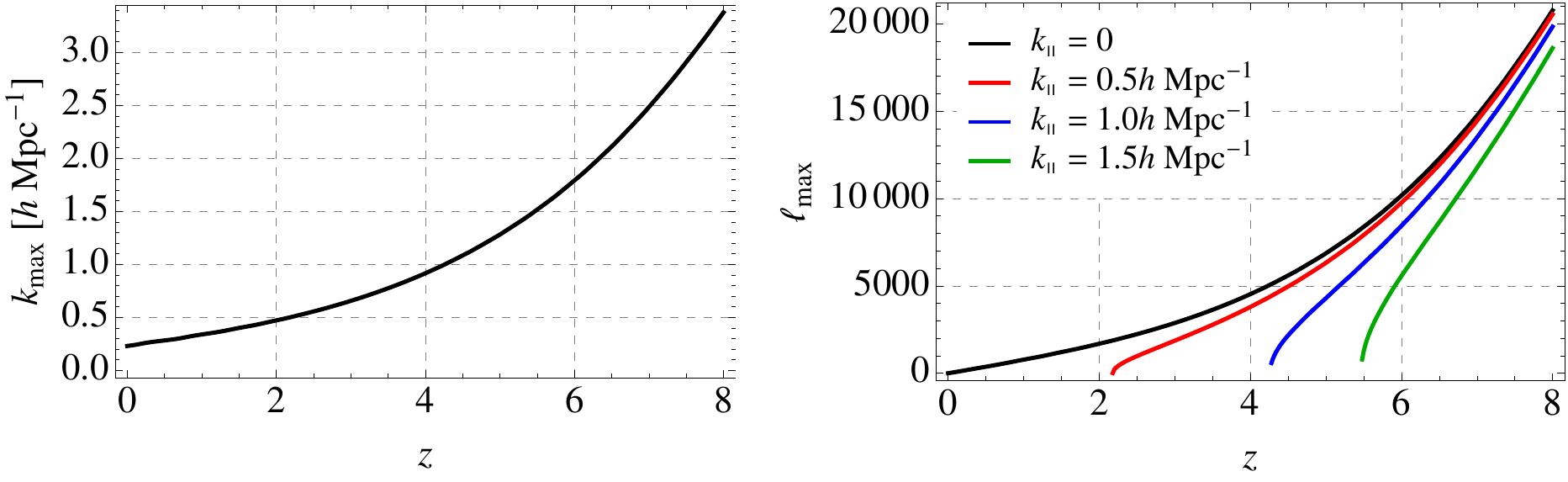}
\caption{\label{fig:treelevel-validity}
Estimate for the range of validity of our tree-level perturbation theory calculation of the matter trispectrum, as a function of redshift. Deviations from the tree-level approximations for all matter $n$-point functions are expected to be of the same order, so as a proxy for the maximum wavenumber ({\em left panel}) or angular multipole ({\em right panel}) where the tree-level trispectrum is valid, we compute where the nonlinear power spectrum as given by Halofit~\cite{Takahashi:2012em} deviates from the linear spectrum by 30\%.  In the right panel, we use $\ell_{\rm max}^2=\chi(z)^2(k_{\rm max}^2-\kpar^2)$, and plot different curves corresponding to the $\kpar$ values given in the legend.
}    
\end{centering}
\end{figure}

We can estimate where the tree-level computation is valid by examining where the nonlinear trispectrum begins to deviate from it by a significant margin. As a proxy for this, we can examine where the nonlinear matter power spectrum begins to deviate from the linear spectrum, since these deviations are expected to be of the same order for different $n$-point functions. We therefore compute, as a function of redshift, the wavenumber at which the Halofit~\cite{Takahashi:2012em} nonlinear power spectrum fitting function for our chosen cosmology differs from the linear spectrum by 30\%; this is shown in the left panel of Fig.~\ref{fig:treelevel-validity}.\footnote{Another way to estimate the validity of the tree-level computation would be to compare the one-loop power spectrum to the linear spectrum. We have checked that this gives comparable results to what we find from Halofit.   Also, the threshold of 30\% is roughly where biases from neglected higher-order terms are expected to become significant, but is not a unique choice. In an application to real data, one would need to investigate the biases arising from higher-order terms in more detail. For example, this could naturally be accomplished with simulations of the lensed intensity field.} In the right panel, we translate this wavenumber into an $\ell$ value using $\ell_{\rm max}^2 \sim \chi(z)^2 ( k_{\rm max}^2-\kpar^2 )$. We find that for $\kpar=0$, we are limited to $\ell\lesssim 2000$ ($\theta\gtrsim 5'$) at $z\sim 2$, while for $z\gtrsim 4$ our computation will be valid at least for $\ell \lesssim 5000$ ($\theta\gtrsim 2'$), with these numbers decreasing with increasing $\kpar$. These limits are generally within the ranges expected of current and future line intensity mapping experiments, except at very low redshifts. For observations with angular resolutions approaching these scales, extension of our results to one-loop order (or examination of appropriate simulations) would be necessary to precisely quantify where the tree-level picture begins to fail. 

To proceed, we will assume a linear relationship between the intensity $I_{\rm 3d}$ and the matter overdensity~$\delta$ (recall that temporal arguments are present for $I_{\rm 3d}$ and $\delta$, but we are omitting them from our notation):
\beq
I_{\rm 3d}(\vx) = b(z) \delta(\vx)\ .
\eeq
There are several real-world effects that could alter this relationship, such as nonlinear bias, shot noise in the discrete sources that are emitting the intensity (investigated in the context of lensing reconstruction in Refs.~\cite{Pourtsidou:2013hea,Pourtsidou:2014pra}), redshift space distortions beyond linear order, or the influence of fluctuations in other quantities (such as the ionization fraction during reionization, in the case of 21cm measurements at the appropriate redshifts). Since we are focused on the impact of purely gravitational nonlinearities in this work, we will neglect these effects, each of which must however be investigated in detail before the estimators can be applied to data.

\subsubsection{Gravitational mode-couplings}
\label{sec:gravmodecouplings}

We can now begin to extend the calculations in \sec{sec:estimator-review} to include mildly nonlinear gravitational effects. We expand the observed intensity in powers of both $\phi$ and $\delta_1$, bringing some prefactors to the left-hand side for brevity and omitting the redshift-dependence of $b$:
\begin{align*}
b^{-1} \calL \chi^{2} \Iobs(\vl,\kpar) &= \delta_1(\vl/\chi,\kpar) \\
&\quad -  \int_{\vl'} \vl'\cdot(\vl-\vl') \phi(\vl-\vl') \delta_1(\vl'/\chi,\kpar) \\
&\quad +  \left. \int_{\vq} \kerfs{2}(\vq,\vk-\vq) \delta_1(\vq) \delta_1(\vk-\vq)
	\right|_{\vk=(\vl/\chi,\kpar)} \\
&\quad +  \left. \int_{\vq} \int_{\vp}
	 \kerfs{3}(\vq,\vp,\vk-\vq-\vp) \delta_1(\vq) \delta_1(\vp) \delta_1(\vk-\vq-\vp)
	\right|_{\vk=(\vl/\chi,\kpar)} \\
&\quad -  \int_{\vl'} \vl'\cdot(\vl-\vl') \phi(\vl-\vl') 
	\left. \int_{\vq} \kerfs{2}(\vq,\vk'-\vq) \delta_1(\vq) \delta_1(\vk'-\vq)
	\right|_{\vk'=(\vl'/\chi,\kpar)} \\
&\quad + \mathcal{O}(\phi^2 \delta_1^1) + \mathcal{O}(\phi^1 \delta_1^3)
	 + \mathcal{O}(\phi^0 \delta_1^4)\ . \numberthis
	\label{eq:iexpansion}
\end{align*}
In addition to expanding up to $\mathcal{O}(\phi)$ and $\mathcal{O}(\delta_1^3)$, we have written the first mixed term, at $\mathcal{O}(\phi\delta_1)$, for illustrative purposes.

In the absence of nonlinearities from gravity ($\kerfs{2}=\kerfs{3}=0$), ensemble-averaging the two-point function of~$I$ with~$\phi$ held fixed would again result in Eq.~\eqref{eq:iiavgphifixed}. Now, let us consider the opposite situation, in which lensing is absent ($\phi=0$), and we take an ensemble average over all modes of $\delta_1$ except for those within a narrow wavenumber bin centered on $\vk=(\vL/\chi,\kpar)$. Assuming $\vk_i=(\vl_i/\chi,k_{\parallel i})$ for $i=1,2$ are not in this bin,
the analogous expression to Eq.~\eqref{eq:iiavgphifixed} is then
\begin{align*}
&(b^{-1} \calL\chi^2)^2 \left\la \Iobs(\vl_1,k_{\parallel 1}) \Iobs(\vl_2,k_{\parallel 2}) 
	\right\ra_{\phi=0;\, \delta_1(\vL/\chi,\kpar)\,{\rm fixed}} \\
&\qquad =  \left. \int_{\vq\,|\, \vq\,\not\in\,{\rm bin},\, \vk_2-\vq\,\not\in\,{\rm bin}}
	\kerfs{2}(\vq,\vk_2-\vq) \left\la \delta_1(\vk_1) \delta_1(\vq) \delta_1(\vk_2-\vq) \right\ra
	\right|_{\vk_i=(\vl_i,k_{\parallel i})}
	+ \lb 1\leftrightarrow 2 \rb \\
&\qquad\quad +  \left. 2\int_{\vq\,|\, \vq\,\in\,{\rm bin},\, \vk_2-\vq\,\not\in\,{\rm bin}}
	\kerfs{2}(\vq,\vk_2-\vq)\delta_1(\vq) 
	\left\la \delta_1(\vk_1)  \delta_1(\vk_2-\vq) \right\ra \right|_{\vk_i=(\vl_i,k_{\parallel i})}
	+ \lb 1\leftrightarrow 2 \rb \\
&\qquad = 2 \, \delta_1\!\lp\frac{\vl_1+\vl_2}{\chi},k_{\parallel1}+k_{\parallel2}\rp
	\lb \kerfs{2}(\vk_1+\vk_2,-\vk_1) 
	P_{\delta 1}\!\lp \sqrt{\frac{\ell_1^2}{\chi^2}+k_{\parallel1}^2 } \rp + \lb 1\leftrightarrow2 \rb  
	\rb_{\vk_{\perp i}=\vl_i/\chi} \\
&\qquad\quad\times
	\Theta\!\lb \lp\frac{\vl_1+\vl_2}{\chi},k_{\parallel1}+k_{\parallel2}\rp \in\, {\rm bin} \rb\ ,
	\numberthis
	\label{eq:iiavgdeltafixed}
\end{align*}
where $P_{\delta 1}$ is the linear matter power spectrum, and  we define $\Theta(\cdot)$ to equal one if its argument is true and zero otherwise. The second line above evaluates to zero because $\delta_1$ is assumed to be Gaussian. 
In brief, in Eq.~\eqref{eq:iiavgdeltafixed} we have found that
\beq
\left\la \Iobs(\vl_1,k_{\parallel 1}) \Iobs(\vl_2,k_{\parallel 2}) 
	\right\ra_{\delta_1(\vL/\chi,\kpar)\,{\rm fixed}}
	\propto \delta_1\!\lp\vL/\chi,\kpar\rp
	\text{ if }\, \vl_1+\vl_2 = \vL,\, \kparone+\kpartwo=\kpar\ .
	\label{eq:iisketchdelta}
\eeq
This is essentially the same statement as Eq.~\eqref{eq:iiavgphifixed}, which says that
\beq
\left\la \Iobs(\vl_1,k_{\parallel 1}) \Iobs(\vl_2,k_{\parallel 2}) 
	\right\ra_{\phi\,{\rm fixed}}
	\propto \phi(\vL)
	\text{ if }\, \vl_1+\vl_2 = \vL,\, \kparone+\kpartwo=0\ .
\eeq
Since $\phi$ and $I$ are uncorrelated for any value of $\vL$, we do not need to specify which modes of $\phi$ are held fixed in the ensemble average, but for the equivalent statement for $\delta_1$, more care is required, as seen in Eq.~\eqref{eq:iiavgdeltafixed}. Once this is done, we arrive at the following conclusion: just as we can use couplings between modes of the source field to estimate the (longer) lensing mode that is creating the coupling, we can also use mode-mode couplings to estimate the value of a longer mode of the overdensity that is creating the coupling via gravitational evolution. We have used a somewhat contrived ensemble averaging operation to derive this fact, but such an operation is effectively realized in analysis of data or simulations if the scales used for lensing reconstruction are disjoint from the scales at which the modes of $\phi$ or $\delta_1$ are being reconstructed.

We wish to highlight the relationship between non-stationary and non-Gaussian statistics manifested here. In general, the two concepts are separate: stationarity (also referred to as ``statistical homogeneity" in studies of large-scale structure) causes off-diagonal correlations to vanish, while non-Gaussianity is signalled by the presence of connected $n$-point functions that are not just products of the 2-point function. In an ensemble over density modes of all wavevectors, gravitational evolution induces non-Gaussianity but {\em not} non-stationarity. However, in an ensemble in which some (typically long-wavelength) density modes are held fixed and others (typically of shorter wavelength) fluctuate, gravitational evolution induces non-stationarity of the ensemble's statistics, as manifested in Eq.~\eqref{eq:iisketchdelta}. The fixed long modes create an inhomogeneous background for the short modes, breaking translation-invariance of the short modes' statistics. The leading-order cause for this breaking is the presence of tidal effects, as expected from the equivalence principle.

In the case where $\kparone+\kpartwo=0$, Eqs.~\eqref{eq:iiavgphifixed} and~\eqref{eq:iiavgdeltafixed} can be summarized by
\begin{align*}
&\left\la \Iobs(\vl,\kpar) \Iobs(\vL-\vl,-\kpar) 
	\right\ra_{\phi,\,\delta_1(\vL/\chi,0) \,{\rm fixed}} \\
&\qquad =  (2\pi)^2\dirac(\vL) C_{\ell}(\kpar) 
	+ f_\phi(\vl,\vL-\vl,\kpar,-\kpar) \phi(\vL) 
	+ f_\delta(\vl,\vL-\vl,\kpar,-\kpar) \delta_1(\vL/\chi,0) + \cdots\ ,
	\numberthis
	\label{eq:iiavgbothfixed}
\end{align*}
where
\begin{align} \label{eq:fphidef2}
f_\phi(\vl_1,\vl_2,\kparone,\kpartwo) &\equiv 
	 (\vl_1+\vl_2)\cdot \lb \vl_1 C_{\ell_1}(\kparone)
	 + \vl_2 C_{\ell_2}(\kpartwo) \rb \ , \\
f_{\delta}(\vl_1,\vl_2,\kparone,\kpartwo) &\equiv 2\calL^{-1} \chi^{-2} 
	\lb \kerfs{2}\!\lp-(\vl_1/\chi,\kparone),([\vl_1+\vl_2]/\chi,\kparone+\kpartwo)\rp
	 C_{\ell_1}(\kparone) + 
	\lb 1\leftrightarrow 2 \rb \rb\ .
	\label{eq:fdeltadef}
\end{align}
In Eq.~\eqref{eq:iiavgbothfixed}, we have absorbed the bias $b^2$ into the source field angular power spectrum $C_\ell(\kpar)$.

The $\kerfs{3}$ term from Eq.~\eqref{eq:iexpansion} will not contribute to Eq.~\eqref{eq:iiavgbothfixed}.  Following the same logic as Eq.~\eqref{eq:iiavgdeltafixed}, we find that if none of the four $\delta_1$ factors is evaluated at $\vk\sim (\vL/\chi,\kpar)$, then they will Wick-contract in pairs, and the resulting contribution will be an irrelevant diagonal correlation (proportional to $\dirac(\vL)$). We will assume that the  mode of $\delta$ held fixed in the average is much longer than either of the external modes in the two point function; thus, only at most one of the three $\delta_1$ factors under the integral will come out of the average, and the ensemble average of the remaining three $\delta_1$ factors will vanish. Intuitively, the $\kerfs{3}$ term does not induce non-stationarity on the statistics of $\delta$, but it does contribute to the non-Gaussianity of the $\delta$ 4-point function, which will affect the variance of the $\hat{\phi}$ estimator. We will see precisely how in the next subsection.

Some of the $\mathcal{O}(\delta_1^n)$ terms in Eq.~\eqref{eq:iexpansion} will cross-correlate with $\mathcal{O}(\phi\, \delta_1^{n'})$ terms to modify the coefficient of $\phi(\vL)$ in Eq.~\eqref{eq:iiavgbothfixed}. Such modifications will constitute a sequence of nonlinear corrections to the $\delta$ power spectrum. This can be accounted for by using the nonlinear matter power spectrum rather than the linear power spectrum in $f_\phi$. We will make the same replacement in $f_\delta$, which, while not strictly self-consistent within perturbation theory, will incorporate part of the correlations between $\mathcal{O}(\delta_1^{n>3})$ terms in Eq.~\eqref{eq:iexpansion}.

\subsubsection{Quadratic estimators revisited}
\label{sec:quadrevisited}

We can now better characterize the impact of gravitational nonlinearities on the quadratic lensing estimator discussed above, and also define an analogous estimator for long modes of the matter overdensity. We write a general quadratic estimator as
\beq
\hat{X}(\vL;\kpar) \equiv \int_{\vl} g_X(\vl,\vL-\vl,\kpar)
	I_{\rm obs}(\vl,\kpar) I_{\rm obs}(\vL-\vl,-\kpar)
	\label{eq:genestimator}
\eeq
where $X\in\{\phi,\delta\}$. We have already seen the $X=\phi$ case, while the $X=\delta$ case is an estimator for a purely transverse (to the line of sight) mode of the linear overdensity, $\delta_1(\vL/\chi,0)$. We will comment further on the use of this estimator in reconstructing the matter distribution in \sec{sec:tidal}.  Note that, since we have neglected redshift evolution of the source field over the observed volume, the $\delta$ estimator is an estimator for $\delta_1(\vL/\chi,0)$ evaluated at the mean redshift of the observed range. 

The covariance between two such estimators can be split into Gaussian and non-Gaussian\footnote{Some of the contributions that we refer to as non-Gaussian also cause the statistics of the source field to be non-stationary, and have the same form as non-stationarity-inducing terms from lensing. Thus, one could argue that these specific contributions are not ``non-Gaussian," but rather reflect a coordinate transformation that does not affect the Gaussianity of the source field's statistics. With this caveat, we refer to any contribution to the left-hand side of Eq.~\eqref{eq:covXYsplit} that is not Eq.~\eqref{eq:covXYG} as non-Gaussian.} parts:
\begin{align*}
&\left\la \hat{X}(\vL_1,\kparone) \hat{Y}^*(\vL_2,\kpartwo) \right\ra 
	- \left\la \hat{X}(\vL_1,\kparone)  \right\ra \left\la \hat{Y}^*(\vL_2,\kpartwo)  \right\ra \\
&\qquad = (2\pi)^2\dirac(\vL_1-\vL_2) 
	\lp {\rm Cov}_{\rm G}\!\lb \hat{X}(\vL_1,\kparone),\hat{Y}^*(\vL_1,\kparone) \rb 
	+ {\rm Cov}_{\rm nG}\!\lb \hat{X}(\vL_1,\kparone),\hat{Y}^*(\vL_1,\kpartwo) \rb \rp \ ,
	\label{eq:covXYsplit}
	\numberthis
\end{align*}
where
\beq
{\rm Cov}_{\rm G}\!\lb \hat{X}(\vL,\kparone),\hat{Y}^*(\vL,\kpartwo) \rb 
	= \kro_{\kparone,\kpartwo}  \int_{\vl} g_X(\vl,\vL-\vl,\kparone) g_Y(\vl,\vL-\vl,\kparone)
	C_\ell^{\rm tot}(\kpar) C_{|\vL-\vl|}^{\rm tot}(\kpar)
	\label{eq:covXYG}
\eeq
and
\begin{align*}
 &{\rm Cov}_{\rm nG}\!\lb \hat{X}(\vL,\kparone),\hat{Y}^*(\vL,\kpartwo) \rb \\
&\qquad= \int_{\vl_1} \int_{\vl_2} g_X(\vl_1,\vL-\vl_1,\kparone) g_Y(\vl_2,\vL-\vl_2,\kpartwo) \\
&\qquad\qquad\qquad \times
	\left\la I_{\rm obs}(\vl_1,\kparone) I_{\rm obs}(\vL-\vl_1,-\kparone) 
	I_{\rm obs}(-\vl_2,-\kpartwo) I_{\rm obs}(-\vL+\vl_2,\kpartwo)
	\right\ra_{\rm c}\ ,
	\numberthis	\label{eq:ngcov}
\end{align*}
and we have assumed that the $g$ functions are real and that $\kparone,\kpartwo>0$.  The ``c" subscript in the last line of Eq.~\eqref{eq:ngcov} refers to the connected part of the $I_{\rm obs}$ four-point function. Next, we define
\beq
N_{XY}^{\rm (G)}(L,\kpar) \equiv  \lb \int_{\vl}
	\frac{f_{X}(\vl,\vL-\vl,\kpar,-\kpar)f_{Y}(\vl,\vL-\vl,\kpar,-\kpar)}
	{C_{\ell}^{\rm tot}(\kpar) C_{|\vL-\vl|}^{\rm tot}(\kpar)} \rb^{-1}\ .
	\label{eq:nxydef}
\eeq
We choose the filter $g_X$ as in \sec{sec:estimator-review}: by minimizing the Gaussian contribution to the variance of $\hat{X}$, under the condition that $\hat{X}$ is unbiased {\em in the absence of any other sources of mode-coupling}. These requirements fix $g_X$ to be
\beq
g_X(\vl,\vL-\vl,\kpar) \equiv N_{XX}^{\rm (G)}(L,\kpar) \frac{f_X(\vl,\vL-\vl,\kpar,-\kpar)}
	{C_{\ell}^{\rm tot}(\kpar) C_{|\vL-\vl|}^{\rm tot}(\kpar)}\ ,
	\label{eq:gxdef}
\eeq
which implies that 
\beq
{\rm Cov}_{\rm G}\!\lb \hat{X}(\vL,\kpar),\hat{Y}^*(\vL,\kpar) \rb 
	= \frac{N_{XX}^{\rm (G)}(L,\kpar) N_{YY}^{\rm (G)}(L,\kpar)}{N_{XY}^{\rm (G)}(L,\kpar)}\ .
\eeq

The estimators above were derived under the assumption that only one of either lensing or gravitational nonlinearity is present. In the presence of both effects, each estimator will acquire a bias\footnote{See Ref.~\cite{Namikawa:2012pe} for a discussion of similar biases that arise in other contexts.}, if we again consider an ensemble average where we hold $\phi$ and a single long mode of $\delta$ fixed: 
\begin{align*}
\left\la \hat{\phi}(\vL;\kpar) \right\ra_{\phi,\,\delta(\vL/\chi,0)\, {\rm fixed}} &= \phi(\vL) 
	 +   \frac{N^{\rm (G)}_{\phi\phi}(L,\kpar)}{N^{\rm (G)}_{\phi\delta}(L,\kpar)} \delta_1(\vL/\chi,0) \ ,
	\numberthis 
	\label{eq:ephibias} \\
\left\la \hat{\delta}(\vL;\kpar) \right\ra_{\phi,\,\delta(\vL/\chi,0)\, {\rm fixed}}
	 &=  \delta_1(\vL/\chi,0)  
	 + \frac{N_{\delta\delta}^{\rm (G)}(L,\kpar)}{N^{\rm (G)}_{\phi\delta}(L,\kpar)} \phi(\vL)\ .
	\numberthis 
	\label{eq:edeltabias}
\end{align*}
Thus, if we take the covariance of Eq.~\eqref{eq:ephibias}, averaging over all fluctuations (including $\phi$ and all modes of $\delta_1$), we find that the power spectrum of $\hat{\phi}$ will contain the following terms:
\begin{align*}
\left\la  \hat{\phi}(\vL,\kpar) \hat{\phi}^*(\vL',\kpar)  \right\ra
	&\supset (2\pi)^2 \dirac(\vL-\vL')
	\lb   C_L^{\phi\phi}  + \calL\chi^{2}  
	\lp \frac{N_{\phi\phi}^{\rm (G)}(L,\kpar)}{N^{\rm (G)}_{\phi\delta}(L,\kpar)} \rp^2 P_{\delta 1}(L/\chi) \rb\ ,
	\numberthis \label{eq:didisignal}
\end{align*}
where $P_{\delta 1}$ is the linear matter power spectrum.
Both of these terms can also be derived directly from the  connected four-point function of $I_{\rm obs}$ that appears in Eq.~\eqref{eq:ngcov}. However, there are further non-Gaussian contributions that we have not encountered yet, but that should be considered in a complete calculation at the order we are working at.

To enumerate these contributions, let us examine the various contractions that occur within the two-point function of $\hat{\phi}$. Schematically, we can write
\beq
\hat{\phi} \sim \Iobs \Iobs\ ,
\qquad \Iobs \sim \delta_1 + \delta_1\phi + \delta_1\delta_1 
	+ \delta_1\delta_1\delta_1 + \cdots\ ,
\eeq
with the ellipsis denoting higher-order terms. Thus, $\la \hat{\phi} \hat{\phi} \ra$ will contain the following contractions:
\begin{enumerate}
\item $\la \wick{\c\delta_1\c\delta_1}\; \wick{\c\delta_1\c\delta_1}\ra$: 
This gives the Gaussian term $N_{\phi\phi}^{\rm (G)}$.
\item $\la \wick{\c\delta_1 \c\delta_1\c\phi\; \c\phi\,\c\delta_1\c\delta_1}\ra$: 
This gives a term $\propto C_L^{\phi\phi}$, which is the signal we aim to extract from the covariance. The $g_\phi$ filter in the quadratic estimator is chosen such that the prefactor of this term is unity.
(Here and below, each $\phi$ is taken to come from the same factor of $\Iobs$ as the adjacent $\delta_1$.)
\item $\la \wick{\c1\delta_1 \c2\delta_1\c3\phi\; \c3\phi\, \c2\delta_1\c1\delta_1}\ra$: 
This gives an additive correction to $N_{\phi\phi}^{\rm (G)}$ term, encapsulating the leading effects of lensing on power spectrum of $\Iobs$. As mentioned at the end of \sec{sec:gravmodecouplings}, this term (and similar terms appearing at higher order) can be incorporated into $N_{\phi\phi}^{\rm (G)}$ simply by using the lensed, nonlinear intensity power spectrum in the filters in the estimator.
\item $\la \wick{\c1\delta_1 \c2\delta_1\c3\phi\; \c3\phi\, \c1\delta_1\c2\delta_1}\ra$: 
This gives an integral that convolves $C_L^{\phi\phi}$ with factors of the source power spectrum. In CMB lensing, this term is known as a ``secondary contraction" that yields the so-called ``$N^{(1)}$ bias"~\cite{Kesden:2003cc}. Current CMB lensing measurements have reached sufficiently high precision that this term must be modeled~\cite{Hanson:2010rp}.  In App.~\ref{app:n1bias}, we show that this term is subdominant to the other terms we consider, and can safely be neglected for the purposes of this paper. 
\item $\la \wick{ \c\delta_1 \c\delta_1\c\delta_1\; \c\delta_1\c\delta_1\c\delta_1}\ra$: 
Analogous to the second term in this list, this contraction gives a term proportional to $P_{\delta 1}(L/\chi)$, the power spectrum of long modes of the density that are coupled to shorter modes by gravitational evolution. This term can be disentangled from the leading effects of lensing by defining ``bias-hardened" estimators, which we discuss in \sec{sec:bhlensing}.  For the same reasons, this term can also be targeted by procedures aimed at reconstructing the power spectrum of long-wavelength modes, as we will discuss in \sec{sec:tidal}.  In the language of perturbation theory, this term is contained in the $T_{2211}$ trispectrum diagram (e.g.~\cite{Bertolini:2015fya}).
\item $\la \wick{\c1\delta_1 \c2\delta_1\c3\delta_1\; \c3\delta_1\c2\delta_1\c1\delta_1}\ra$:
Analogous to the third term in this list, this contraction gives an additive correction to $N_{\phi\phi}^{\rm (G)}$, which can be absorbed into $N_{\phi\phi}^{\rm (G)}$ by modifying the power spectra used in the filters.
\item $\la \wick{\c1\delta_1 \c2\delta_1\c3\delta_1\; \c3\delta_1\c1\delta_1\c2\delta_1}\ra$:
Analogous to the fourth term in this list, this contraction gives an integral that convolves different factors of the matter power spectrum together with some wavenumber-dependent kernels. This term will be generically be of at least the same order as the fourth term, and must be included.  This term is also contained in the $T_{2211}$ trispectrum diagram.
\item $\la \wick{\c1\delta_1 \c2\delta_1\; \c2\delta_1\c1\delta_1\c\delta_1\c\delta_1}\ra$: 
This term will be of at least the same order as the seventh term, and must also be included. In perturbation theory, it is known as the $T_{3111}$  diagram.
\end{enumerate}
In App.~\ref{app:grav-derivations}, we present the full derivation of  the terms from the list above that will be relevant in the subsequent analysis: terms 2, 4, 5, 7, and 8. We summarize the result below, omitting term~4 for the reasons given earlier:
\begin{align*}
\left\la  \hat{X}(\vL_1,\kparone) \hat{Y}^*(\vL_2,\kpartwo)  \right\ra 
	&= (2\pi)^2 \dirac(\vL_1-\vL_2) \lb  N^{\rm (G)}_{XY}(L_1,\kparone) \, \kro_{\kparone,\kpartwo} 
	 + N_{XY}^{({\rm nG,}\phi)}(L_1,\kparone,\kpartwo) \right. \\
&\qquad\qquad\qquad\qquad\qquad\left. 
	+\, N_{XY}^{({\rm nG,}P)}(L_1,\kparone,\kpartwo)
	 + N_{XY}^{\rm (nG,c)}(L_1,\kparone,\kpartwo) \rb\ ,
	 \numberthis
	 \label{eq:covxyfull}
\end{align*}
where
\begin{align*}
N_{XY}^{({\rm nG,}\phi)}(L,\kparone,\kpartwo)
	&= \frac{N_{XX}^{\rm (G)}(L,\kparone) N_{YY}^{\rm (G)}(L,\kpartwo)}
	{N_{X\phi}^{\rm (G)}(L,\kparone)N_{Y\phi}^{\rm (G)}(L,\kpartwo)}  C_{L}^{\phi\phi} \ , \\
N_{XY}^{({\rm nG,}P)}(L,\kparone,\kpartwo)
	&= \frac{N_{XX}^{\rm (G)}(L,\kparone) N_{YY}^{\rm (G)}(L,\kpartwo)}
	{N_{X\delta}^{\rm (G)}(L,\kparone)N_{Y\delta}^{\rm (G)}(L,\kpartwo)} 
	\calL \chi^2  P_{\delta 1}(L/\chi) \ , \numberthis
\end{align*}
and 
\begin{align*}
&N_{XY}^{\rm (nG,c)}(L,\kparone,\kpartwo) \\
	&\quad=  \int_{\vl_1} \int_{\vl_2} 
	g_X(\vl_1,\vL-\vl_1,\kparone) g_Y(\vl_2,\vL-\vl_2,\kpartwo)  \\
&\qquad\quad
	 \times \lb \calL \chi^2\lp P[\vl_1-\vl_2,\kparone-\kpartwo] 
	f_\delta(\vl_1,-\vl_2,\kparone,-\kpartwo) f_\delta(\vL-\vl_1,-\vL+\vl_2,-\kparone,\kpartwo) 
	\right. \right. \\
&\qquad\qquad\qquad\quad\left. 
	+\,   \lb \kparone \leftrightarrow -\kparone \rb  \rp \\
&\qquad\qquad + 6\calL^{-1} \chi^{-2} 
	C_{\ell_1}(\kparone) C_{\ell_2}(\kpartwo) \\
&\qquad\qquad\quad \times \lp P[\vL-\vl_1,\kparone]
	\left\{ \kerfs{3}((\vl_1,\kparone),(\vL-\vl_1,-\kparone),(-\vl_2,-\kpartwo))
	+ \lb \kpartwo\leftrightarrow -\kpartwo \rb \right\} \right. \\
&\qquad\qquad\qquad \left.\left.
	+\, P[-\vL+\vl_2,\kpartwo] 
	\left\{ \kerfs{3}((\vl_1,\kparone),(-\vl_2,-\kpartwo),(-\vL+\vl_2,\kpartwo))
	+ \lb \kparone\leftrightarrow -\kparone \rb \right\} \rp \rb\ ,
	\numberthis
	\label{eq:nxyngc}
\end{align*}
making use of the following shorthands:
\beq
P[\vl_i,k] \equiv P_{\delta 1}\!\lp \sqrt{\frac{\ell_i^2}{\chi^2}+k^2} \rp ,
\quad (\vl_i,k) \equiv (\ell_{i1}/\chi,\ell_{i2}/\chi,k)\ .
\eeq
The ``$P$" and ``c" superscripts on the non-Gaussian terms represent those proportional to the long mode power spectrum and those from the remaining connected contribution to the $\Iobs$ four-point function, respectively.  Importantly, when a given $\phi$ mode is reconstructed using the quadratic estimator, estimates obtained using  different $\kpar$ values have correlated noise, so the minimum variance combination would not use the simple inverse-variance weighting from Eq.~\eqref{eq:nphiphigcomb}. Instead, the optimal combination of the different estimators is given by~\cite{Hu:2001kj}
\beq
\hat{\phi}_{\rm combined}(\vL)
	= \frac{\sum_{j_1,j_2=j_{\rm min}}^{j_{\rm max}} \lb\mathcal{N}^{-1}_{\phi\phi}(L)\rb_{j_1 j_2}
	\hat{\phi}(\vL,\kpartwo)}{\sum_{j_1,j_2=j_{\rm min}}^{j_{\rm max} }
	\lb\mathcal{N}^{-1}_{\phi\phi}(L)\rb_{j_1 j_2}}\ ,
	\label{eq:phihatcombined}
\eeq
and has variance given by
\beq
N_{\phi\phi}^{\rm (full,combined)}(L) = \lp \sum_{j_1,j_2=j_{\rm min}}^{j_{\rm max} }
	\lb\mathcal{N}^{-1}_{\phi\phi}(L)\rb_{j_1 j_2} \rp^{-1}\ .
	\label{eq:ncombng}
\eeq
In these expressions, $\mathcal{N}^{-1}_{\phi\phi}(L)$ is the inverse of the noise covariance matrix  $\mathcal{N}_{\phi\phi}(L)$, which has components $\mathcal{N}_{\phi\phi, j_1 j_2}(L)$ given by
\beq
\mathcal{N}_{\phi\phi, j_1 j_2}(L) 
	= N^{\rm (G)}_{\phi\phi}(L,2\pi j_1/\calL) \, \kro_{j_1,j_2} 
	+ N_{\phi\phi}^{({\rm nG,}P)}(L,2\pi j_1/\calL,2\pi j_2/\calL)
	 + N_{\phi\phi}^{\rm (nG,c)}(L,2\pi j_1/\calL,2\pi j_2/\calL)\ .
	 \label{eq:noisecovmat}
\eeq
Even if the Gaussian term dominates over the non-Gaussian terms for a given $j$ (which is usually the case), these correlations imply that as estimates from more and more $j$ values are combined, the correlated terms will act to slow the pace at which the combined noise decreases with $j_{\rm max}$.

In the next subsection, we will investigate the relative sizes of the four terms in Eq.~\eqref{eq:covxyfull} as a function of redshift and angular resolution (i.e.~$\ell_{\rm max}$). For that investigation, and all other numerical computations in this paper, we use CAMB~\cite{Lewis:1999bs} to compute linear and nonlinear matter power spectra (the latter with Halofit~\cite{Takahashi:2012em}). The various integrals are then computed using an extended version of Copter~\cite{Carlson:2009it} that uses the Monte Carlo integration routines from the CUBA library~\cite{Hahn:2004fe}. We have checked that a subset of these computations agree with a separate Mathematica-based implementation.

\subsection{Application to generic intensity maps}
\label{sec:ideal-im}
To gain intuition for the different contributions to the covariance of the lensing estimator $\hat{\phi}(\vL,\kpar)$ for different $\kpar$ values, we will consider intensity observations made in several redshift bands, each with width $\Delta z = 0.5$ and centered on redshifts from 2 to 8. For the instrumental noise, we will consider a toy model in which the noise is zero for $\ell$ below some $\ell_{\rm max}$ and infinite for  $\ell>\ell_{\rm max}$. As stated earlier, we will assume that the intensity is a deterministic linear tracer of the matter overdensity; this approximation is most suspect at $z\sim8$, when reionization is taking place, but our focus here is on how the gravitational and lensing effects scale with redshift, rather than other physical phenomena which might affect our ability to perform lensing reconstruction. 

\begin{figure}[t]
\includegraphics[width=\textwidth]{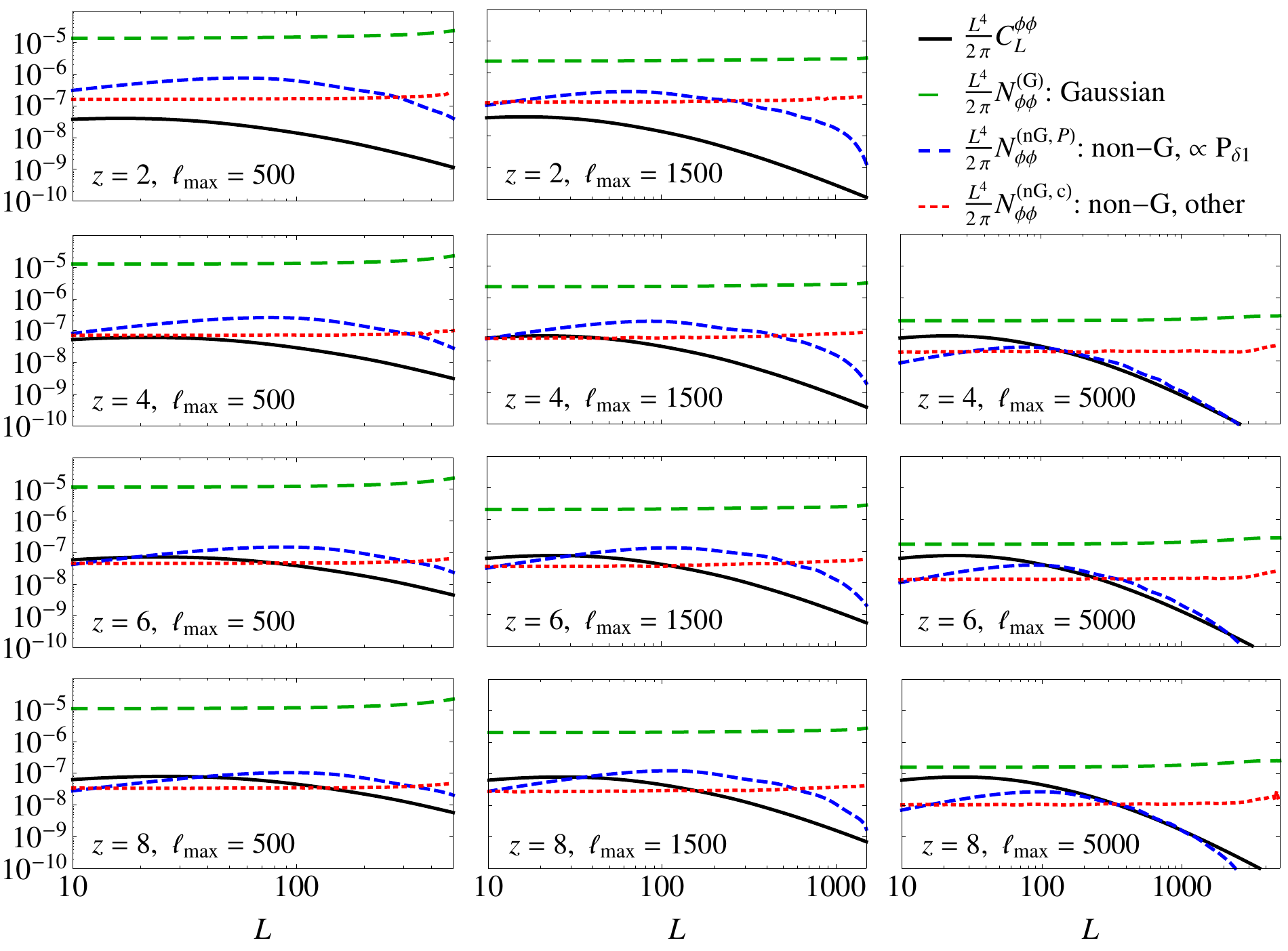}
\caption{\label{fig:singlej-noiseterms}
Different contributions to the power spectrum  of the quadratic lensing estimator $\hat{\phi}(\vL,\kpar)$ discussed in \sec{sec:quadrevisited}, assuming noise-free line intensity maps over $\Delta z=0.5$ and $\ell<\ell_{\rm max}$ (with infinite noise for $\ell>\ell_{\rm max}$). All plots are made with $\kpar\approx0.13\invMpc$, corresponding to $j=\{10,5,3,2\}$ for $z=\{2,4,6,8\}$. We consistently find large ranges of scales over which the non-Gaussian contributions from gravitational nonlinearity ({\em blue dashed and red dotted curves}) are comparable to the lensing potential power spectrum ({\em black solid curves}), although subdominant to the pure Gaussian contribution ({\em green dashed curves}). This conclusion is unchanged for observations with more realistic noise levels. Simple scaling arguments suffice to explain many of the trends seen in these panels, as discussed in the main text.
}    
\end{figure}

In Fig.~\ref{fig:singlej-noiseterms}, we show each of the four terms in Eq.~\eqref{eq:covxyfull} from observations made in four different redshift bands and at three different angular resolutions ($\ell_{\rm max}$ values). All curves are computed at a fixed value of $\kpar\approx0.13\invMpc$, corresponding to $j=\{ 10,5,3,2 \}$  for $z=\{ 2,4,6,8\}$.  The blue dashed and red dotted curves show the leading two effects of gravitational mode-coupling, one ($N_{\phi\phi}^{\rm (nG,P)}$) directly proportional to the power spectrum of long modes of the matter overdensity, the other  ($N_{\phi\phi}^{\rm (nG,c)}$) a convolution over various gravitational kernels and power spectra; both have been computed analytically for the first time in this work. We have not plotted any terms that are off-diagonal in $\kpar$ (i.e. $N_{\phi\phi}^{\rm (nG,c)}(L,\kparone,\kpartwo)$ or $N_{\phi\phi}^{({\rm nG,}P)}(L,\kparone,\kpartwo)$ with $\kparone\neq\kpartwo$), but they typically have similar amplitude and shape to the diagonal terms. We find that in many of the cases in Fig.~\ref{fig:singlej-noiseterms}, at least one of the gravitational terms dominates over the lensing power spectrum over a wide range of scales, and therefore these terms must be handled appropriately in order to accomplish a robust detection of lensing. Several characteristics of these curves can be straightforwardly understood:

\begin{itemize}
\item First, the Gaussian noise per mode, $N^{\rm (G)}_{\phi\phi}$, essentially counts the inverse of the number of angular modes up to $\ell_{\rm max}$ used in the lensing reconstruction. This number is constant in redshift but scales like $\ell_{\rm max}^{-2}$, implying the same scaling for $N^{\rm (G)}_{\phi\phi}$~\cite{Hanson:2010rp}. In the presence of more realistic noise on source field measurements, $N^{\rm (G)}_{\phi\phi}$ counts the number of modes that can be ``imaged" by virtue of being above the noise level.
\item Second, the shapes of the three gravitational curves can be understood by examining the $\ell\gg L$ limits of their associated expressions, since the largest contribution to each term will come from $\ell$-modes for which this is true. We give the $\ell\gg L$ limits of $f_\phi$ and $f_\delta$ in App.~\ref{app:flimits}, and by using these in the expressions in Secs.~\ref{sec:gravmodecouplings} and~\ref{sec:quadrevisited}, we find that in this limit, $N^{\rm (G)}_{\phi\phi}$ and $N_{\phi\phi}^{\rm (nG,c)}$ scale like $L^{-4}$, while $N_{\phi\phi}^{({\rm nG,}P)}$ scales like $L^{-4} P_{\delta 1}(L/\chi) $. Thus, when multiplied by $L^4$ as in Fig.~\ref{fig:singlej-noiseterms}, the first two terms take the form of white noise on $L^4 C_L^{\phi\phi}$, while the third term has roughly the shape of the linear power spectrum, with deviations occurring as $L$ approaches $\ell_{\rm max}$.
\item Third, at fixed $\ell_{\rm max}$, the magnitude of the non-Gaussian gravitational terms decreases with increasing redshift. This scaling can be accounted for by linear growth of the matter power spectrum (both non-Gaussian gravity terms scale like $P_{\delta 1}$), along with redshift-dependence of the prefactor $\calL^{-1} \chi^{-2}$.
\item However, at fixed redshift, the scaling of the gravitational terms with $\ell_{\rm max}$ is more complex, due to the possibility of cancellations occurring in the integrals as $\ell_{\rm max}$ is varied. The $N_{\phi\phi}^{\rm (nG,c)}$ term consistently decreases in amplitude as $\ell_{\rm max}$ increases, up to factor of a few from $\ell_{\rm max}=500$ to 5000 while $N_{\phi\phi}^{({\rm nG,}P)}$ can even scale non-monotonically with $\ell_{\rm max}$, due to chance cancellations inside the $N_{\phi\delta}^{\rm (G)}$ integral (which also depend on the value of $\kpar$). This is what causes the $N_{\phi\phi}^{({\rm nG,}P)}$ curves in the rightmost column of Fig.~\ref{fig:singlej-noiseterms} to be lower than in the other two columns.
\end{itemize}

\begin{figure}[t]
\includegraphics[width=1\textwidth]{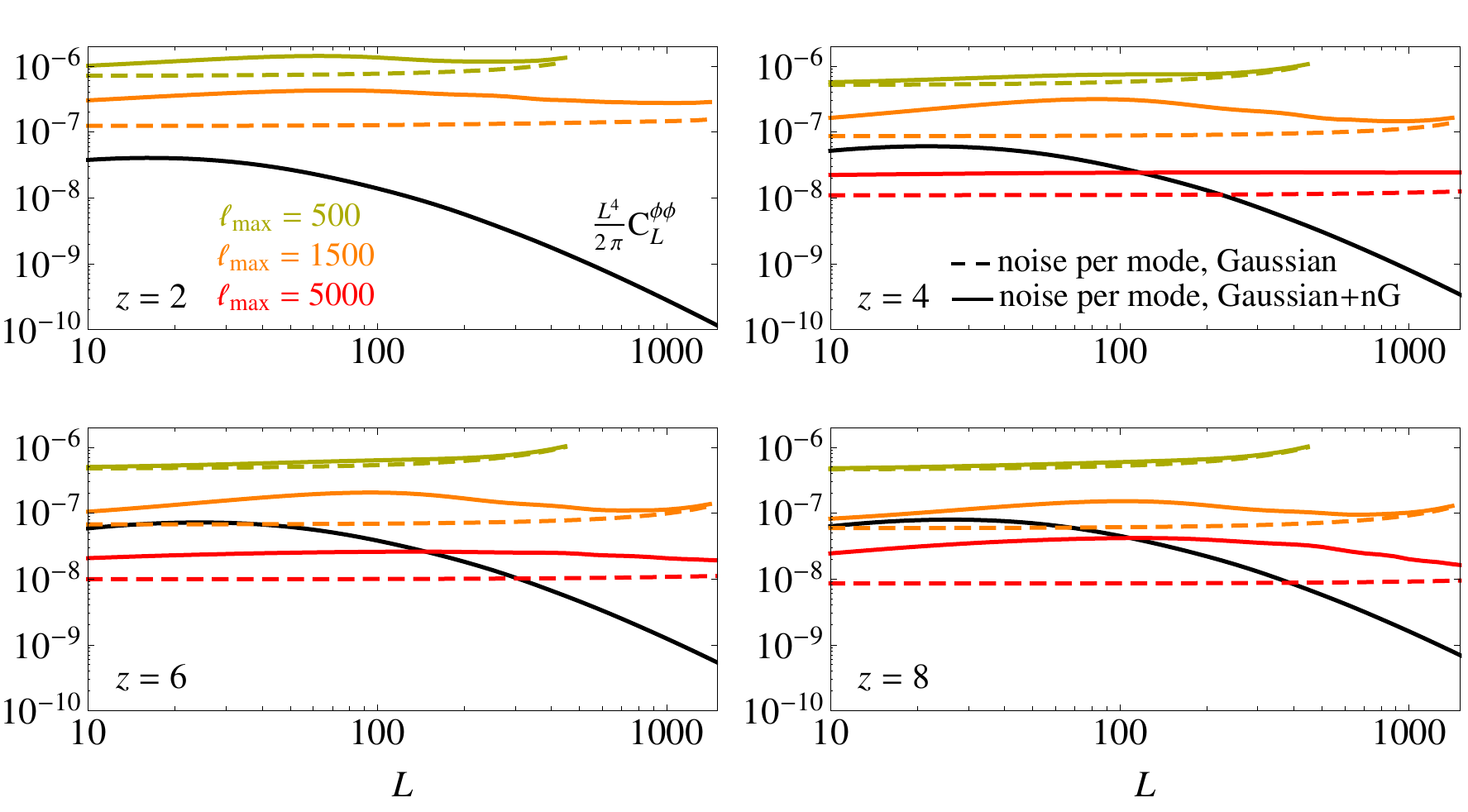}
\caption{\label{fig:combined_noise_gaus}
Contributions to the power spectrum of the combined lensing estimator when $3\leq j \leq 20$ are summed using Eq.~\eqref{eq:ncombng}, when the non-Gaussian gravitational contributions are either neglected {\em (dotted lines)} or included {\em (solid lines)}, for the same cases considered in Fig.~\ref{fig:singlej-noiseterms}. The importance of these contributions increases with the number of small-scale modes used in the lensing reconstruction, and can potentially increase the total contribution to the $\hat{\phi}$ power spectrum (which affects the noise per reconstructed $\phi$ mode, along with the additive bias and noise on $C_L^{\phi\phi}$) above the Gaussian contribution by a factor of several. We do not show the $\ell_{\rm max}=5000$ case for $z=2$, because this case is beyond the validity of our tree-level calculation (see Fig.~\ref{fig:treelevel-validity}). 
}    
\end{figure}

For a given $j$, the errorbar is typically dominated by Gaussian noise, which, being uncorrelated between different $j$ values, is reduced in the combined estimator. However, as more and more $j$ values are combined, eventually the Gaussian piece drops below the non-Gaussian piece, which, since it correlates different $j$ values, does not drop nearly as rapidly with $j_{\rm max}$. The result is that the combined noise per $\phi$ mode can generally be dominated by the non-Gaussian contribution. Fig.~\ref{fig:combined_noise_gaus} shows this combined noise, computed from Eq.~\eqref{eq:ncombng} for $3\leq j \leq 20$ when the non-Gaussian gravitational terms are either neglected or included. As expected, the importance of the non-Gaussian terms generally grows as smaller and smaller-scale modes are used in the reconstruction process, increasing the total noise per mode by as much as a factor of a few in the examples considered here.

\section{Bias-hardened estimators}
\label{sec:bh}

\subsection{Lensing}
\label{sec:bhlensing}

Figs.~\ref{fig:singlej-noiseterms} and~\ref{fig:combined_noise_gaus} make it clear that gravitational mode couplings will significantly affect lensing maps constructed from line intensity mapping observations using a CMB-lensing-type quadratic estimator, simply by virtue of these observations tracing structures that have begun to cluster nonlinearly on relevant scales. Mode by mode, these \emph{maps} will pick up a bias corresponding to modes of the matter overdensity [see Eq.~\eqref{eq:ephibias}].
An estimate of the lensing potential \emph{power spectrum} will  pick up both an additive bias and an extra source of statistical noise. Ref.~\cite{Schaan:2018yeh} has recently treated this by modifying the weights in the lensing estimator to downweight mode combinations that are coupled together through gravity or other nonlinear effects.  Here we explore a different approach.

In cases where multiple mode-couplings are present, each sourced by a different field (in this case, either $\phi$ or $\delta_1$), the corresponding quadratic estimators can be modified to subtract off unwanted mode-couplings~\cite{Namikawa:2012pe}. The idea is to form linear combinations of estimators for each effect that are free from bias from the other effect, at leading order. 
This technique has been applied to CMB lensing measurements to reduce unwanted mode couplings from the sky mask, anisotropic noise, and unresolved point sources~\cite{Ade:2015zua}. Since gravitational nonlinearities also induce mode couplings that could potentially mimic the effect of lensing, it is natural to apply this ``bias-hardening" procedure to the case at hand. 

For a fixed realization of $\phi(\vL)$ and $\delta_1(\vL/\chi, 0)$, the expectation values of our estimators for~$\phi$ and~$\delta_1$, Eqs.~\eqref{eq:ephibias} and~\eqref{eq:edeltabias}, can be written as (suppressing some arguments for brevity)
\beq
\lb \begin{array}{c} \left\la \hat{\phi}(\vL) \right\ra \\
	\left\la \hat{\delta}(\vL/\chi) \right\ra \\ \end{array} \rb
	=
	\lb \begin{array}{cc} 1 & N_{\phi\phi}^{\rm (G)}/N_{\phi\delta}^{\rm (G)} \\
	N_{\delta\delta}^{\rm (G)}/N_{\phi\delta}^{\rm (G)} & 1 \end{array} \rb
	\lb \begin{array}{c} \phi(\vL) \\ \delta_1(\vL/\chi,0)
	\end{array}\rb\ .
\eeq
Bias-hardened estimators $\hat{\phi}^{\rm H}$ and $\hat{\delta}^{\rm H}$ are formed simply by solving this system for the desired quantities. We obtain 
\begin{align*}
\hat{\phi}^{\rm H} &\equiv \frac{1}{1- 
	N_{\phi\phi}^{\rm (G)}N_{\delta\delta}^{\rm (G)} \!\lb N_{\phi\delta}^{\rm (G)}\rb^{-2}} 
	\lp \hat{\phi} - \lb N_{\phi\phi}^{\rm (G)} / N_{\phi\delta}^{\rm (G)} \rb \hat{\delta} \rp \ , \\
\hat{\delta}^{\rm H} &\equiv \frac{1}{1- 
	N_{\phi\phi}^{\rm (G)}N_{\delta\delta}^{\rm (G)} \!\lb N_{\phi\delta}^{\rm (G)}\rb^{-2}} 
	\lp \hat{\delta}- \lb N_{\delta\delta}^{\rm (G)} / N_{\phi\delta}^{\rm (G)} \rb \hat{\phi} \rp \ ,
	\numberthis
	\label{eq:ehatdefs}
\end{align*}
whose expectation values are then just $\phi(\vL)$  and $\delta_1(\vL/\chi,0)$, as desired. A short calculation yields the Gaussian part of the covariance of $\hat{\phi}^{\rm H}$:
\begin{align*}
{\rm Cov}_{\rm G}\!\lb \hat{\phi}^{\rm H}(\kparone),\hat{\phi}^{{\rm H}*}(\kpartwo) \rb
	&= \kro_{\kparone,\kpartwo} 
	\frac{N_{\phi\phi}^{\rm (G)}(\kparone)}
	{1- N_{\phi\phi}^{\rm (G)}(\kparone) N_{\delta\delta}^{\rm (G)}(\kparone) 
	\!\lb N_{\phi\delta}^{\rm (G)}(\kparone)\rb^{-2}}  \\
&\equiv \kro_{\kparone,\kpartwo}
	\frac{N_{\phi\phi}^{\rm (G)}(\kparone)}{1-\lb \rho(\hat{\phi},\hat{\delta})_{\kparone} \rb^2}\ ,
	\numberthis
	\label{eq:phihathgvar}
\end{align*}
where we have restored the $\kpar$ arguments but left $\vL$ implicit, assuming that all factors are evaluated at the same $\vL$ value.
In the above expression, $\rho(\hat{\phi},\hat{\delta})_{\kpar}$ is the correlation coefficient between the original~$\hat{\phi}$ and~$\hat{\delta}$ estimators, both evaluated at $\kpar$. This indicates that if the original estimators were highly correlated, the Gaussian part of the variance of the bias-hardened estimator $\hat{\phi}^{\rm H}$ will be boosted by a large amount, while if the estimators were relatively uncorrelated to start with, there will be little penalty in applying the bias-hardening procedure. The gravitational contribution to the covariance of $\hat{\phi}^{\rm H}$ is similarly given by
\begin{align*}
&\lp 1-\lb \rho(\hat{\phi},\hat{\delta})_{\kparone} \rb^2 \rp
	\lp 1-\lb \rho(\hat{\phi},\hat{\delta})_{\kpartwo} \rb^2 \rp
	{\rm Cov}_{\rm NG}\!\lb \hat{\phi}^{\rm H}(\kparone),\hat{\phi}^{{\rm H}*}(\kpartwo) \rb \\
&\qquad\qquad\qquad\qquad\qquad\qquad = N_{\phi\phi}^{\rm (nG,c)}(\kparone,\kpartwo)
	+
	\frac{N_{\phi\phi}^{\rm (G)}(\kparone) N_{\phi\phi}^{\rm (G)}(\kpartwo)}
	{N_{\phi\delta}^{\rm (G)}(\kparone) N_{\phi\delta}^{\rm (G)}(\kpartwo) }
	 N_{\delta\delta}^{\rm (nG,c)}(\kparone,\kpartwo) \\
&\qquad\qquad\qquad\qquad\qquad\qquad\quad	
	- \left\{ \frac{N_{\phi\phi}^{\rm (G)}(\kpartwo)}{N_{\phi\delta}^{\rm (G)}(\kpartwo)}
	 N_{\phi\delta}^{\rm (nG,c)}(\kparone,\kpartwo)
	+ \lb 1 \leftrightarrow 2 \rb \right\}\ .
	\numberthis
	\label{eq:phihatbhngcov}
\end{align*}

One can see by inspection of Eq.~\eqref{eq:phihatbhngcov} that the $N_{\phi\phi}^{({\rm nG,}P)}$ term from the variance of the original~$\hat{\phi}$ estimator cancels completely in the variance of~$\hat{\phi}^{\rm H}$. Thus, the bias-hardened estimator will be free of the negative consequences of this term: an additive mode-by-mode bias in lensing maps constructed by the original estimator [see Eq.~\eqref{eq:edeltabias}], and additive bias and noise on estimates of the lensing potential power spectrum. As mentioned earlier, the Gaussian contribution will be increased by a factor related to the correlation between the original~$\hat{\phi}$ and~$\hat{\delta}$ estimators, and we find that the remaining  non-Gaussian contribution, $N_{\phi\phi}^{\rm (nG,c)}$ (which cannot be removed with bias-hardening), is boosted by a similar amount [although it is less evident from the expression in Eq.~\eqref{eq:phihatbhngcov}]. However, if the original value of $N_{\phi\phi}^{({\rm nG,}P)}$ is greater than the boosted value of $N_{\phi\phi}^{\rm (nG,c)}$, the non-Gaussian additive bias on the lensing potential power spectrum will see a net reduction. The Gaussian additive bias on the power spectrum will be increased, but such a bias is much easier to subtract than non-Gaussian contributions, since it can measured from the observed realization of the sky and then subtracted~\cite{Dvorkin:2008tf, Namikawa:2012pe}.  Furthermore, once we combine lensing measurements from multiple $j$ values, the total noise per mode in the lensing maps can in some cases also see a net reduction, compared to the case of no bias hardening, if small enough angular scales are used.

The effects of lensing and gravitational nonlinearity can be distinguished because they induce distinct forms of anisotropy and scale-dependence in the correlations of small-scale modes; for a more detailed discussion, see App.~\ref{app:fconshear}. These distinctions disappear in the $\chi\kpar \gg \ell$ limit, preventing the bias-hardening procedure from separating the two sources of mode-coupling.
In particular, for $\chi\kpar \sim 2\ell_{\rm max}$, the correlation coefficient is within a few percent of unity, and increases further for higher $\kpar$. In this regime, the bias-hardened estimator will be too noisy to be useful. Thus, we can take $\kpar \sim 2\ell_{\rm max}/\chi$ as the maximum $\kpar$ value from which we can extract useful lensing information, even if there are shorter line-of-sight modes that are signal-dominated in a given observation. This can imply that the frequency resolution that is useful for lensing analysis is set by an instrument's effective angular resolution, rather than its frequency channel width or noise level.

\begin{figure}[t]
\includegraphics[width=1\textwidth]{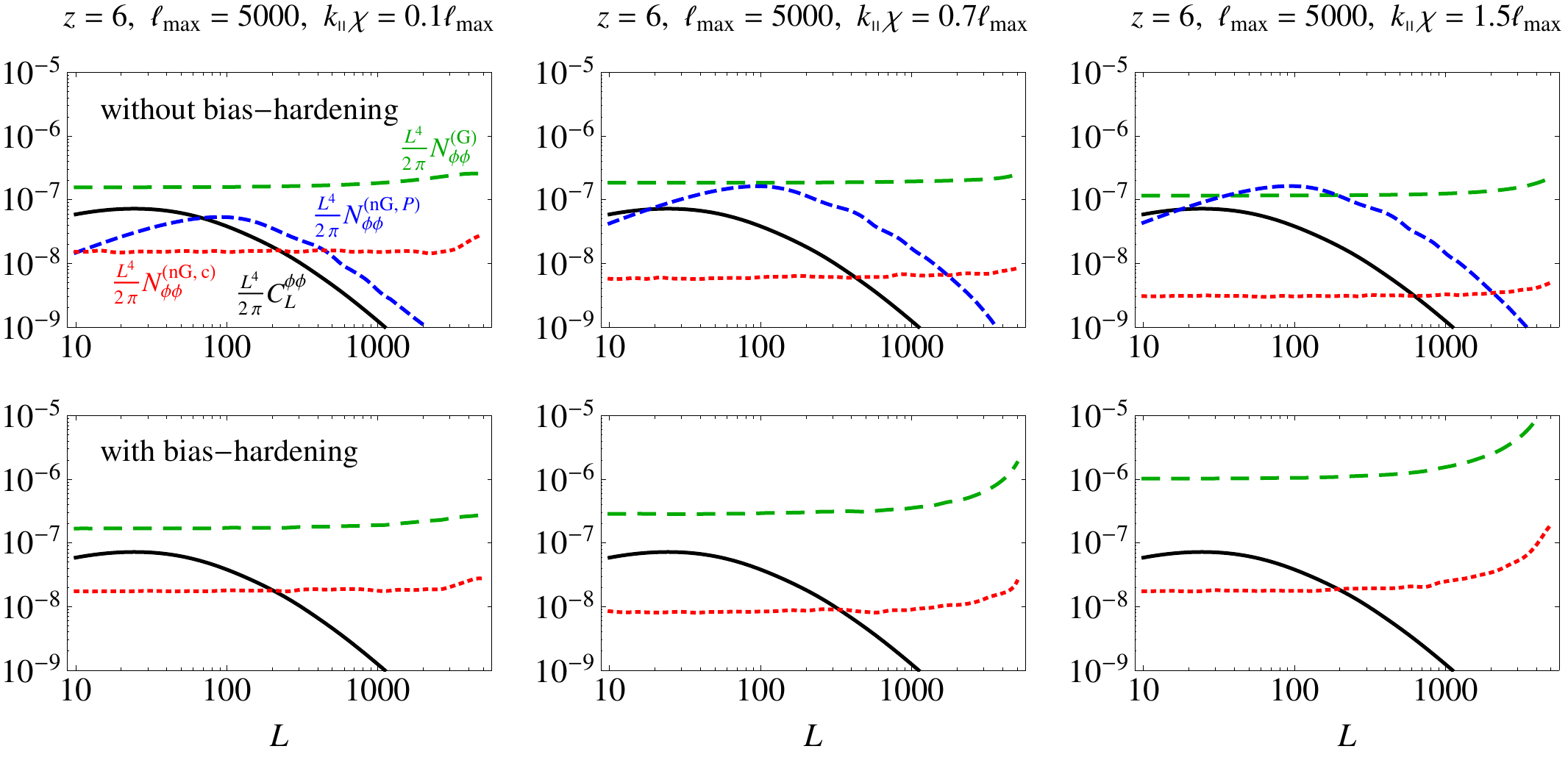}
\caption{\label{fig:singlej-bh}
Different contributions to the power spectrum of the quadratic lensing estimator $\hat{\phi}(\vL;\kpar)$ discussed in \sec{sec:quadrevisited}, with the same assumptions and color scheme as Fig.~\ref{fig:singlej-noiseterms}, but comparing the contributions to the non-bias-hardened ({\em upper panels}) and bias-hardened ({\em lower panels}) estimator's power spectrum for three choices of $\kpar$, at redshift $z=6$ and with $\ell_{\rm max}=5000$. By design, bias-hardening removes the $N_{\phi\phi}^{({\rm nG,}P)}$  contribution. For $\kpar\ll \ell_{\rm max}/\chi$, the other terms are not significantly affected, but once $\kpar \sim \ell_{\rm max}/\chi$, the other terms grow in amplitude, making the bias-hardened estimator noisier overall. However, as long as the $N_{\phi\phi}^{\rm (nG,c)}$ term does not grow beyond the original size of $N_{\phi\phi}^{({\rm nG,}P)}$, bias-hardening still removes the dominant source of non-Gaussian additive bias on an estimate of $C_L^{\phi\phi}$, which would otherwise be more difficult to model and remove than the Gaussian contribution.  }
\end{figure}

Fig.~\ref{fig:singlej-bh} shows an example of the individual contributions to the variance of~$\hat{\phi}$ (upper panels) and~$\hat{\phi}^{\rm H}$ (lower panels), for $\chi\kpar = 0.1$, 0.7, and 1.5 times $\ell_{\rm max}$. As~$\kpar$ increases towards $\ell_{\rm max}/\chi$, the Gaussian and connected non-Gaussian terms increase after bias-hardening, and for $\kpar>\ell_{\rm max}/\chi$, the increase is significant. However, in all cases there is a large reduction in the total non-Gaussian contribution to the noise per mode for $\ell\lesssim 200$, and similar conclusions are generic for other observations in ranges where $N_{\phi\phi}^{({\rm nG,}P)}\gg N_{\phi\phi}^{\rm (nG,c)}$ without bias hardening.

\begin{figure}
\includegraphics[width=0.9\textwidth, trim = 10 45 10 0 ]{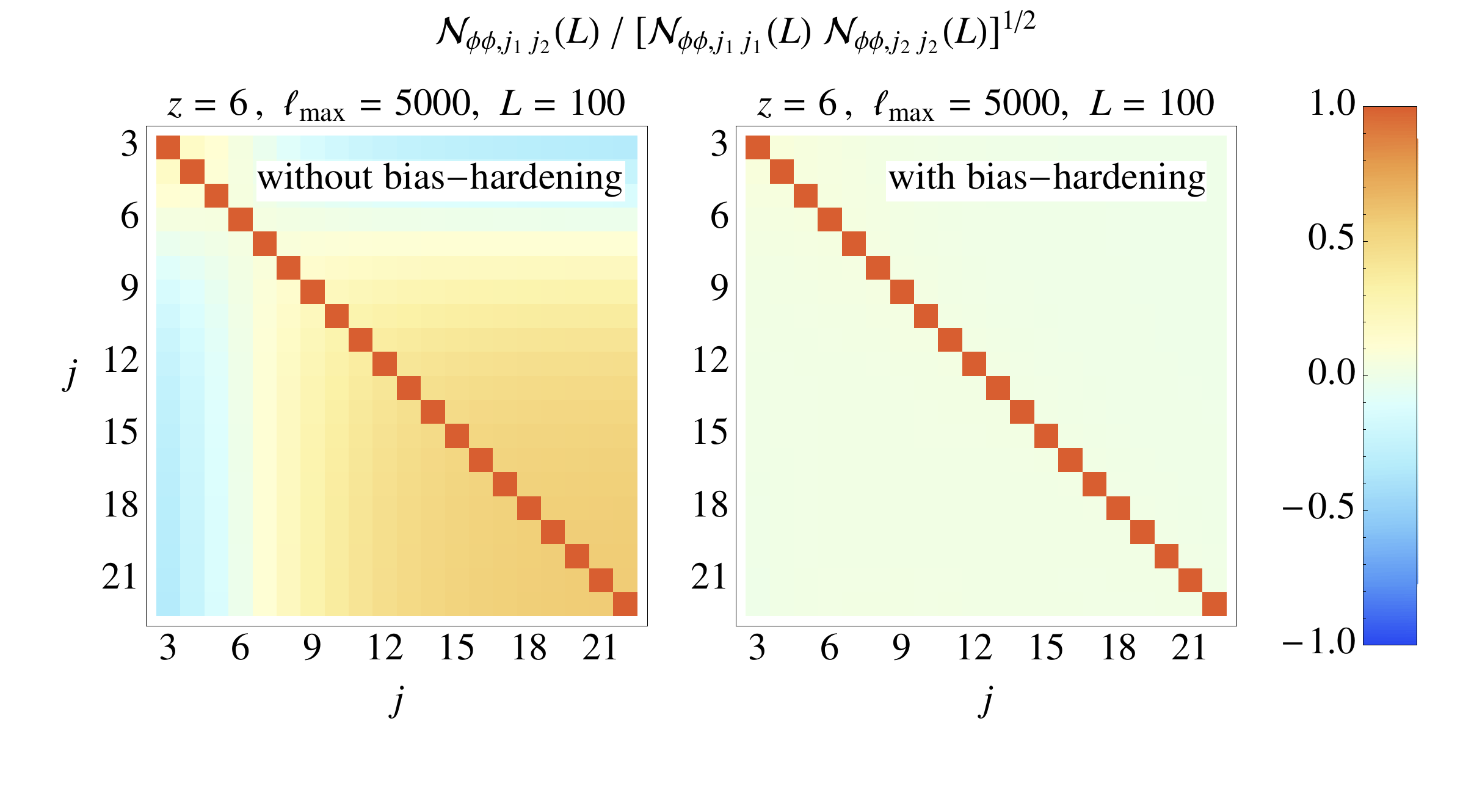}
\caption{\label{fig:noise-correlations-bh}
The noise correlation matrix, $\mathcal{N}_{\phi\phi, j_1 j_2}/[\mathcal{N}_{\phi\phi, j_1 j_1} \mathcal{N}_{\phi\phi, j_2 j_2}]^{1/2}$ with $\mathcal{N}_{\phi\phi, j_1 j_2}$ from Eq.~\eqref{eq:noisecovmat}, for the same redshift range and angular resolution as Fig.~\ref{fig:singlej-bh}, both before ({\em left panel}) and after ({\em right panel}) bias-hardening. The non-Gaussian contributions are significantly correlated between different values of $j$, but bias-hardening removes the dominant non-Gaussian contribution, and therefore  greatly reduces the correlation in the total noise. (Observations at lower angular resolution see a smaller reduction in their noise correlation matrix.)
}    
\end{figure}

As mentioned above, lensing reconstructions from different $j$ values are correlated by non-Gaussianities from gravity, and these correlations can also be greatly reduced using bias-hardening. Fig.~\ref{fig:noise-correlations-bh} shows the correlation matrix of the noise on $\hat{\phi}(\vL;\kpar)$, $\mathcal{N}_{\phi\phi, j_1 j_2}/[\mathcal{N}_{\phi\phi, j_1 j_1} \mathcal{N}_{\phi\phi, j_2 j_2}]^{1/2}$ with $\mathcal{N}_{\phi\phi, j_1 j_2}$ from Eq.~\eqref{eq:noisecovmat}, for the same redshift range and angular resolution as Fig.~\ref{fig:singlej-bh}, both before and after bias-hardening. Without bias-hardening, there are significant correlations between reconstructions from different $j$ values, which will limit the usefulness of combining them into a single lensing measurement, while after bias-hardening, these correlations have been substantially reduced, since the covariance matrix is then dominated by the (diagonal) Gaussian contribution. 

\begin{figure}[t]
\includegraphics[width=\textwidth, trim = 25 0 25 0 ]{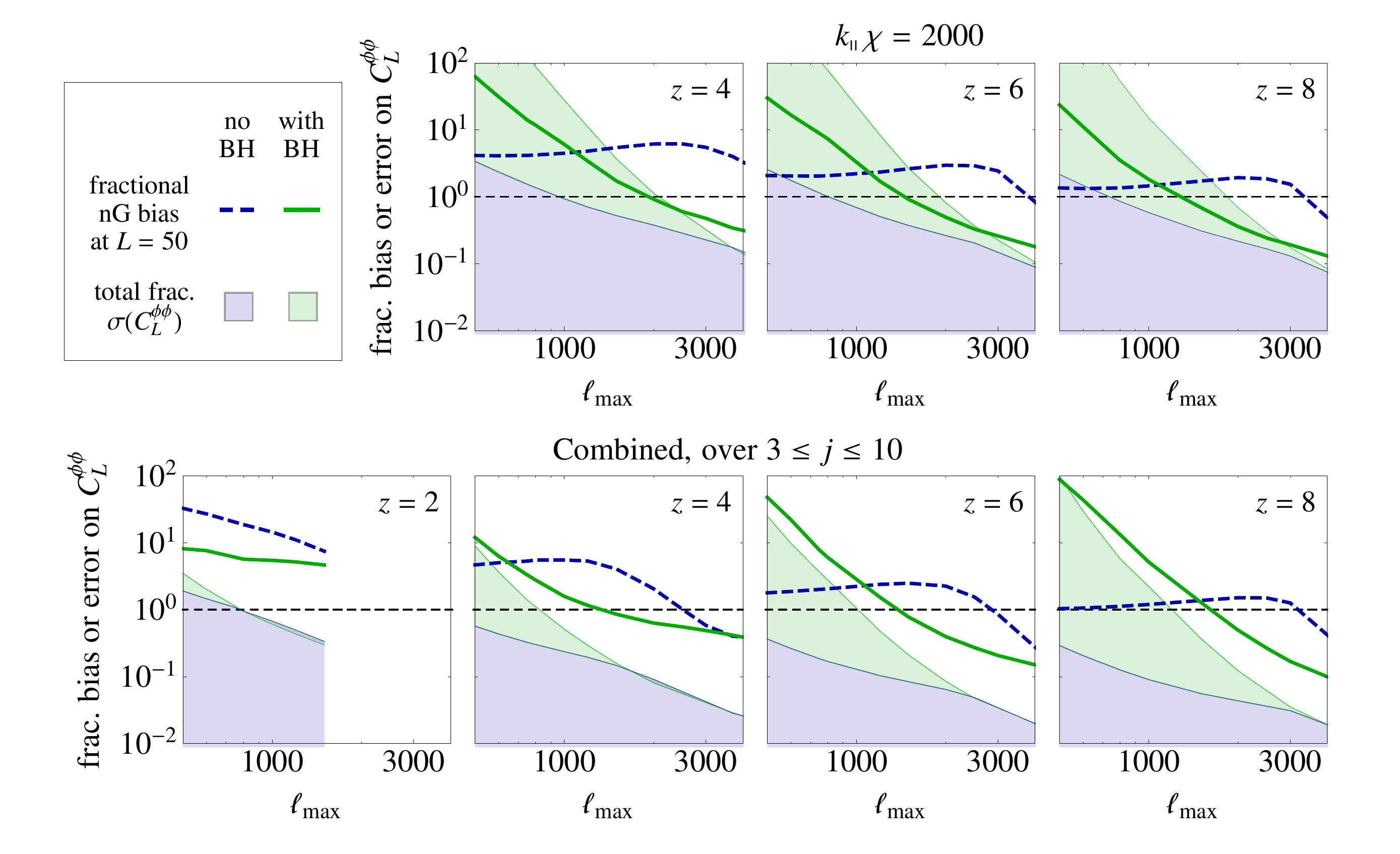}
\caption{\label{fig:ng_bias}
The fractional bias on an estimation of $C_L^{\phi\phi}$ at $L=50$ (representative of the bias on the overall amplitude of $C_L^{\phi\phi}$) arising from non-Gaussian terms in the lensing estimator covariance, for the same observations considered in Figs.~\ref{fig:singlej-noiseterms}-\ref{fig:noise-correlations-bh}. We show this bias both before ({\em blue dashed lines}) and after ({\em green solid lines}) applying bias-hardening. The shaded regions show the total errorbar on a measurement of the amplitude of $C_L^{\phi\phi}$.
The upper panels correspond to a fixed value of $\kpar\chi$, while the lower panels correspond to combining lensing estimates from $3\leq j\leq 10$.
For sufficiently high (low) $\ell_{\rm max}$, bias-hardening decreases (increases) the bias on a measurement of $C_L^{\phi\phi}$ (the reasons for the latter are explained in the main text). In a combined lensing estimate, the errorbar is reduced more than the bias, because the Gaussian contribution to the errorbar is reduced more rapidly (due to the lack of correlation between different $j$ values, in contrast to the non-Gaussian part). Thus, even with bias-hardening, the combined estimator exhibits a bias that exceeds the errorbar, and that will need to be modeled if the goal is to measure the lensing auto spectrum.
}
\end{figure}

As mentioned above, it is possible to subtract the realization of Gaussian noise from the power spectrum of a reconstructed lensing map~\cite{Dvorkin:2008tf, Namikawa:2012pe}, such that the dominant additive bias on $C_L^{\phi\phi}$ is given by the non-Gaussian contributions $N_{\phi\phi}^{({\rm nG,}P)}$ and $N_{\phi\phi}^{\rm (nG,c)}$. In particular, for maps constructed  using different $\kpar$ values in the estimator $\hat{\phi}(\vL,\kpar)$, the Gaussian contribution to each map can be subtracted, and the maps can then be combined according to Eq.~\eqref{eq:phihatcombined}. The non-Gaussian bias on the power spectrum of this combined map will then be given by
\begin{align*}
N_{\phi\phi}^{\rm (nG,combined)}(L) 
	&= N_{\phi\phi}^{\rm (full,combined)}(L)^2  \\
&\quad \times \sum_{j_1,j_2,j_3,j_4=j_{\rm min}}^{j_{\rm max}} 
	\lb\mathcal{N}^{-1}_{\phi\phi}(L)\rb_{j_1 j_2}
	\lb \mathcal{N}^{\rm (nG)}_{\phi\phi}(L) \rb_{j_2 j_3}
	\lb\mathcal{N}^{-1}_{\phi\phi}(L)\rb_{j_3 j_4}\ ,
	\numberthis
    \label{eq:ngbias}
\end{align*}
where $\mathcal{N}^{\rm (nG)}_{\phi\phi}(L)$ is given by Eq.~\eqref{eq:noisecovmat} with the Gaussian term subtracted off.

In Fig.~\ref{fig:ng_bias}, we plot this bias evaluated at $L=50$, which is roughly where the signal to noise on $C_L^{\phi\phi}$ peaks; therefore, the result is representative of the bias on a measurement of the overall amplitude of $C_L^{\phi\phi}$. We also plot the expected errorbar on this amplitude. The upper panels correspond to a fixed value of $\kpar\chi=2000$, while the lower panels correspond to combining estimators for $3\leq j \leq 10$. As we might expect from the discussion of Fig.~\ref{fig:singlej-bh}, the upper panels show that, for $\kpar\chi \lesssim \ell_{\rm max}$, bias-hardening indeed reduces this bias, because the (dominant) $N_{\phi\phi}^{({\rm nG,}P)}$ contribution is completely removed while the (subdominant) $N_{\phi\phi}^{\rm (nG,c)}$ contribution is basically unaltered. For $\kpar\chi \gtrsim \ell_{\rm max}$, however, bias-hardening generically {\em increases} the non-Gaussian bias beyond its unhardened value, again because the two mode-couplings that bias-hardening is designed to separate become less distinguishable in the high-$\kpar$ limit. The upper row of panels in Fig.~\ref{fig:ng_bias} illustrates this transition for $\kpar\chi=2000$. (They also show that the transition does not occur precisely at $\kpar\chi\sim\ell_{\rm max}$: when $\kpar\chi = 2000$, we find that the two bias curves cross closer to $\ell_{\rm max}\sim 1000$ than 2000.)

In the lower panels of Fig.~\ref{fig:ng_bias}, we again find that the non-Gaussian bias is decreased by bias-hardening for higher values of $\ell_{\rm max}$, but increases for lower $\ell_{\rm max}$ values. We also find that when estimates from several $j$ values are combined, the total lensing errorbar  is reduced to a much greater extent than the non-Gaussian bias. This is because, as explained in \sec{sec:ideal-im}, the Gaussian contribution to the lensing errorbar is reduced in the sum over $j$, while the non-Gaussian contribution, being correlated between different $j$ values, is not reduced by the same amount. A comparison between the upper and lower panels shows that the errorbar decreases more rapidly with $j_{\rm max}$ when bias-hardening is applied, as expected from the discussion of Fig.~\ref{fig:noise-correlations-bh}. 

For sufficiently high $\ell_{\rm max}$ values, bias-hardening is successful in reducing the total non-Gaussian bias. However, even after bias-hardening, this bias is still much larger than the statistical errorbar, indicating that some modeling efforts will likely be required to access the lensing power spectrum from measurements such as these.
Using different weights in the lensing estimator, as in Ref.~\cite{Schaan:2018yeh}, may also reduce the additive bias, but we leave a detailed investigation to future work.
Finally, we note that when $j_{\rm max}$ is such that $k_{\parallel{\rm max}}\gtrsim \ell_{\rm max}/\chi$, the non-Gaussian bias after bias-hardening tends to {\em increase} as $j_{\rm max}$ is increased, while the total errorbar saturates around that point, implying that choosing $j_{\rm max}$ such that $k_{\parallel{\rm max}}\sim \ell_{\rm max}/\chi$ will optimally minimize both the errorbar and bias on $C_L^{\phi\phi}$.
Meanwhile, lowering $j_{\rm min}$ (assuming that foregrounds do not prevent us from doing so) can significantly reduce the errorbar for lower $\ell_{\rm max}$, but does not affect the bias. 

Before moving on, we note that, for numerical evaluation, it is better to rewrite the denominator of Eq.~\eqref{eq:phihathgvar} like so:
\beq
1- N_{\phi\phi}^{\rm (G)}N_{\delta\delta}^{\rm (G)} \!\lb N_{\phi\delta}^{\rm (G)}\rb^{-2}
	= N_{\phi\phi}^{\rm (G)} N_{\delta\delta}^{\rm (G)}  
	\lp \lb N_{\phi\phi}^{\rm (G)} \rb^{-1} \lb N_{\delta\delta}^{\rm (G)} \rb^{-1}
	- \lb N_{\phi\delta}^{\rm (G)}  \rb^{-2} \rp \ ,
\eeq
and further rewrite the factor in parentheses as (dropping all $\kpar$ arguments for brevity)
\begin{align*}
&\lb N_{\phi\phi}^{\rm (G)}(L) \rb^{-1} \lb N_{\delta\delta}^{\rm (G)}(L) \rb^{-1}
	- \lb N_{\phi\delta}^{\rm (G)}(L)  \rb^{-2} \\
&\qquad =  \int_{\vl_1} \int_{\vl_2} 
	\frac{f_\phi(\vl_1,\vL-\vl_1) f_\delta(\vl_2,\vL-\vl_2)}{\lb C_{\ell}^{\rm tot} C_{|\vL-\vl|}^{\rm tot} \rb^2}
	\lp
	f_\phi(\vl_1,\vL-\vl_1) f_\delta(\vl_2,\vL-\vl_2)
	-  \lb \phi\leftrightarrow\delta \rb 
	 \rp  \ .
	\label{eq:bhgaus-combinedintegrand}
	\numberthis
\end{align*}
A similar procedure is advisable to combine the integrals on the right-hand side of Eq.~\eqref{eq:phihatbhngcov}. Evaluating these combined integrands reduces the requirements on numerical precision in cases where the results are close to zero. Similar manipulations are known to be useful for evaluating loop integrals in large-scale structure perturbation theory in certain cases~\cite{Carrasco:2013sva}.

\subsection{Tidal reconstruction}
\label{sec:tidal}

So far in this work, we have seen several situations in which the $N_{\phi\phi}^{({\rm nG,}P)}$ term (which is proportional to the power spectrum of long density modes) is comparable to, or even larger than, the lensing potential power spectrum which we have focused on as our signal of interest. (For example, compare the solid black and short-dashed blue curves in several panels of Fig.~\ref{fig:singlej-noiseterms}.) In these situations, it is natural to ask whether we can use quadratic estimators to reconstruct the long density modes themselves, treating lensing as a contaminant instead of the end goal. Indeed, this topic has already been explored several times in the literature: isotropic distortions of the matter power spectrum can be used to learn about so-called ``super-sample modes" that affect covariances between quantities of cosmological interest (e.g.~\cite{Takada:2013bfn,Li:2014jra}), while these modes can also be accessed through their effect on the gravitational tidal tensor on small scales, in a procedure often referred to as ``tidal reconstruction"~\cite{Pen:2012ft,Zhu:2015zlh,Zhu:2016esh}. Such a procedure could in principle yield cosmic-variance-free measurements of the logarithmic growth factor through a joint analysis of anisotropic galaxy power spectra and reconstructed long modes. The reconstructed modes could further be cross-correlated with CMB lensing or temperature, the latter yielding information about the integrated Sachs-Wolfe effect (e.g.~\cite{Zhu:2016esh,Moodley:inprep}).

We have presented an estimator for long modes of the overdensity in \sec{sec:quadrevisited} [specifically, Eq.~\eqref{eq:genestimator} for $\hat{\delta}$, using the filter $g$ given in Eq.~\eqref{eq:gxdef} with the mode coupling kernel $f_\delta$ given by Eq.~\eqref{eq:fdeltadef}]. Just as the quadratic lensing estimator in that section will be biased by gravitational nonlinearities, the corresponding $\delta$ estimator will be biased by lensing, but this can be overcome with the bias-hardening procedure from \sec{sec:bhlensing}. As with the lensing estimators, we have only presented calculations at tree-level in perturbation theory, and this sets the range of validity of our results. By contrast, Refs.~\cite{Pen:2012ft,Zhu:2015zlh,Zhu:2016esh} use configuration-space quadratic estimators in their tidal reconstruction, similar to those presented for lensing in Refs.~\cite{Lu:2007pk,Lu:2009je} (see also~\cite{Bucher:2010iv,Prince:2017sms}). When testing this approach on simulations, the authors first apply a Gaussianizing transform to the density field, allowing them to utilize smaller scales in the reconstruction procedure, at the possible expense of theoretical control over the results. Their estimators will also be contaminated by lensing, although this can likely be mitigated by making use of the full three-dimensional tidal tensor. A detailed comparison between their procedure and our estimators would be instructive, but we leave this for future work. As a separate topic, it would also be interesting to investigate whether the ``response'' formalism described in Refs.~\cite{Barreira:2017sqa,Barreira:2017kxd} could allow for filters that are valid at smaller scales  than the tree-level calculation we have carried out.

\begin{figure}[t]
\includegraphics[width=1\textwidth]{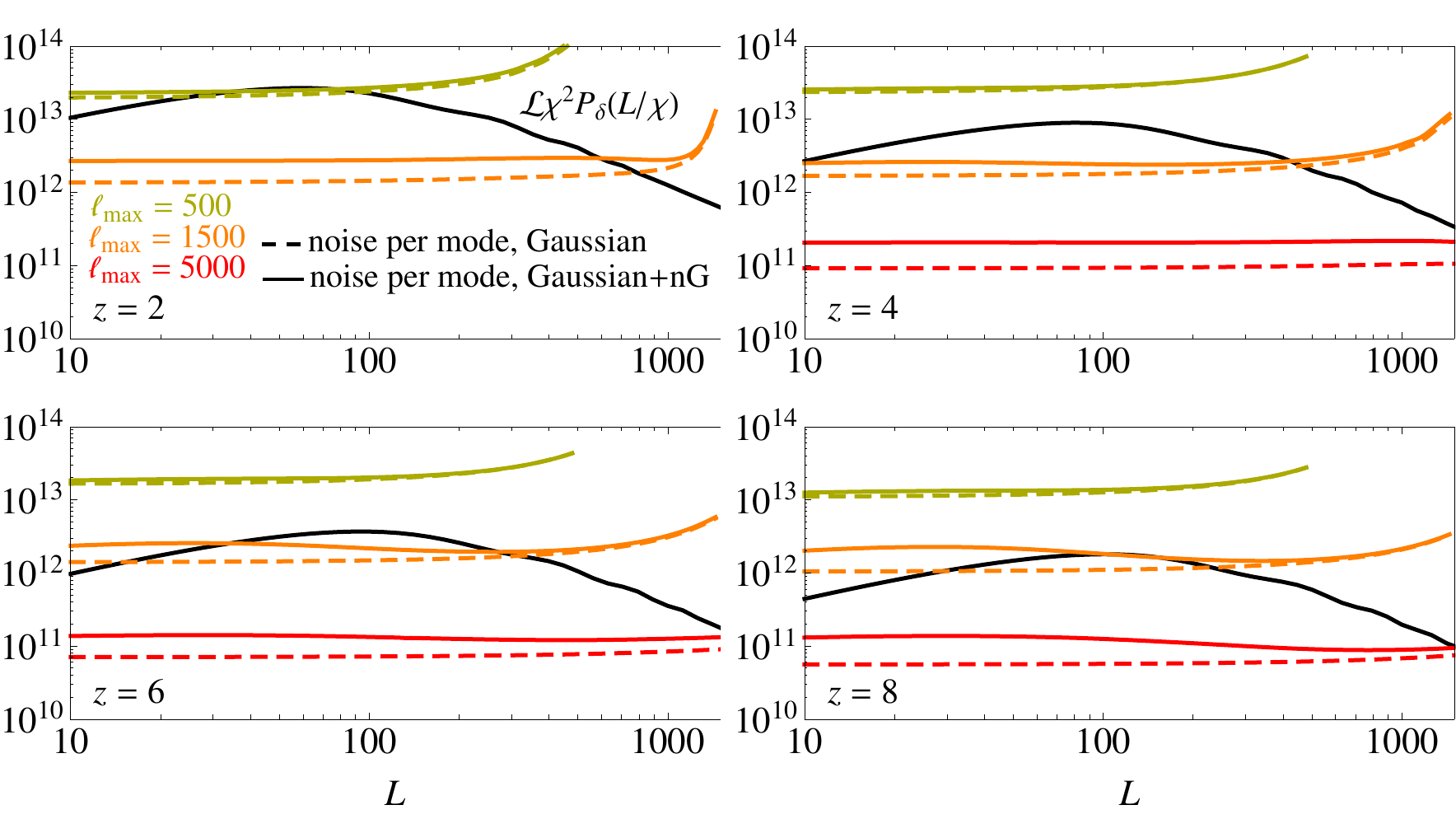}
\caption{\label{fig:combined_dd_gng_noise_bh}
As Fig.~\ref{fig:combined_noise_gaus}, but showing signal and noise curves corresponding to the density power spectrum, which can be reconstructed using the formalism we have presented. These ``noise curves" represent the noise per reconstructed density mode, and also additive bias and noise on the reconstructed density power spectrum. As in Fig.~\ref{fig:combined_noise_gaus}, we omit the $\ell_{\rm max}=5000$ case for $z=2$, because our tree-level calculation is not valid in that case. In general, these noises and bias have lower amplitude in this case than for lensing reconstruction. Thus, there is promise for a successful application of this technique to low-redshift intensity maps. 
}    
\end{figure}

In Fig.~\ref{fig:combined_dd_gng_noise_bh}, we repeat the computations of Fig.~\ref{fig:combined_noise_gaus}, but instead computing the contributions to a reconstruction of the power spectrum of long density modes, either neglecting or including the leading non-Gaussian terms (which now include a lensing contribution). The noise per reconstructed density mode is much lower in this case than for lensing reconstruction, regardless of the angular resolution assumed for the observations. Furthermore, the non-Gaussian contributions generally impart less bias to an estimate of the long-wavelength matter power spectrum, particularly at lower redshifts.

\begin{figure}[t]
\includegraphics[width=\textwidth, trim = 25 0 25 0]{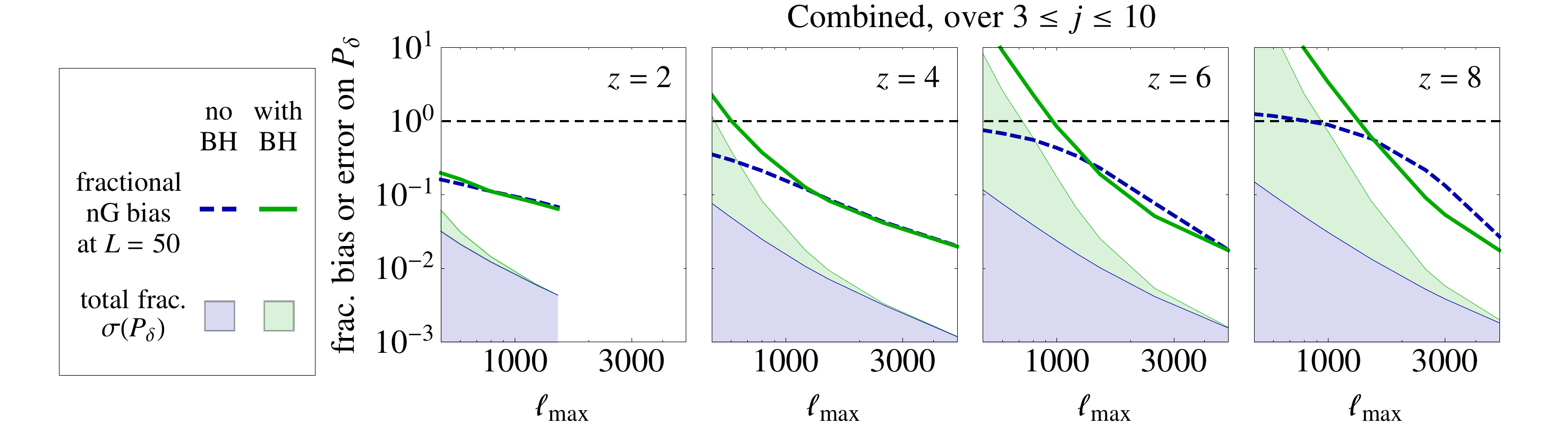}
\caption{\label{fig:ng_bias_dd}
As in the bottom row of Fig.~\ref{fig:ng_bias}, we show the fractional bias on an estimation of $P_\delta(L/\chi)$ at $L=50$ arising from non-Gaussian terms, which now include a lensing term and a gravitational term. The shaded regions show the total errorbar on a measurement of the amplitude of $P_\delta(L/\chi)$. As in lensing reconstruction, the combined estimator exhibits a bias that exceeds the errorbar. However, after bias-hardening, this bias only depends on an integral over several factors of $P_\delta$, and therefore it could possibly be used to enhance a measurement of $P_\delta$ using this technique.
}    
\end{figure}

In Fig.~\ref{fig:ng_bias_dd}, we repeat the computations of the bottom panels of Fig.~\ref{fig:ng_bias}, but compute the total errorbar on the measured amplitude of $P_\delta$, along with the bias from non-Gaussian terms. In this case, the bias arises from a lensing term and a tree-level gravitational term ($N_{\delta\delta}^{\rm (nG,c)}$, in our notation). Note that, at a given value of $\kpar$, the lensing contribution dominates the total bias when $\ell_{\rm max}\gtrsim \kpar\chi$, while the gravitational term dominates when $\ell_{\rm max}\lesssim \kpar\chi$. At high enough $\ell_{\rm max}$, bias-hardening reduces both the bias and total errorbar, but this does not occur within the range we plot in Fig.~\ref{fig:ng_bias_dd}. As in the case of lensing reconstruction, we find that the non-Gaussian bias far exceeds the errorbar in a combined tidal reconstruction estimator. After bias-hardening, the bias is completely determined by $N_{\delta\delta}^{\rm (nG,c)}$, which also depends on $P_\delta$ through an integral against filters and other factors. Thus, the presence of this bias could possibly enhance a measurement of $P_\delta$. Note that for cross-correlations of the reconstructed density with other tracers, this bias will not be present, but will contribute to the noise in the cross power spectrum. 

Overall, Fig.~\ref{fig:combined_dd_gng_noise_bh} indicates that reconstruction of the long-mode density power spectrum may be within reach of low-redshift intensity mapping surveys, such as those performed by CHIME, HIRAX, or GBT. This possibility would be well worth exploring in future work.

\section{Forecasts for specific observations}
\label{sec:forecasts}

In this section, we consider a few specific line intensity mapping efforts, and examine their expected performance in measuring lensing using the estimators we have discussed. Note that these forecasts should all be taken as best-case scenarios, due to our idealistic assumptions about both theoretical systematics (such as nonlinear bias in the source field) and instrumental issues (such as calibration).  In particular, uncertain knowledge of the statistics or linear bias of the source field will lead to a multiplicative offset on the lensing and tidal reconstructions \cite{Hu:2001kj}.  Where possible, we have included a rough accounting for uncertainties arising from foreground subtraction, but we have generally opted for simplicity over realism in this respect.

We consider SKA1-Low and CHIME/HIRAX observations of 21cm radiation in Secs.~\ref{sec:ska} and~\ref{sec:chime}, respectively. In \sec{sec:ccat}, we consider a single-dish intensity mapping survey motivated by the CCAT-prime telescope's planned CII observations. In \sec{sec:summary-and-xcorrs}, we summarize the predicted signal to noise on detections of either the lensing auto spectrum, or cross-correlations between lensing and galaxy clustering or cosmic shear from LSST.

\subsection{SKA1-Low}
\label{sec:ska}

As an example of line intensity maps observed at reionization-era redshifts, we will consider the low-frequency component of the Square Kilometre Array project, which will observe 21cm radiation from the spin-flip transition of neutral hydrogen between $\sim$50 and $\sim$350 MHz, corresponding to 
$3\lesssim z \lesssim 27$ (e.g.~\cite{Pritchard:2015fia}). Following previous work on 21cm lensing~\cite{Zahn:2005ap,Romeo:2017zwt}, we will consider a survey aimed at observing the epoch of reionization, covering 27deg$^2$ and $6<z<14$ (with the minimum redshift set roughly by where reionization is expected to have completed, and the maximum set roughly by where measurements of 21cm fluctuations become noise-dominated). We divide this range into bands that span 5MHz in the redshifted 21cm frequency.

In this era, in which the 21cm spin temperature far exceeds the mean CMB temperature, the 21cm brightness temperature is~\cite{Zahn:2005ap}
\beq
T(\vx;z) \approx 26 \lb 1+\delta_{\rm b}(\vx;z) \rb x_{\rm H}(\vx;z) 
	\lp \frac{\Omega_{\rm b}h^2}{0.022} \rp
	\lb \lp \frac{0.15}{\Omm h^2} \rp \lp \frac{1+z}{10} \rp \rb^{1/2} {\rm mK}\ ,
\eeq
where $\delta_{\rm b}$ is the baryon density contrast, $x_{\rm H}$ is the neutral fraction, and $\Omega_{\rm b}h^2$ and $\Omm h^2$ are the physical baryon and matter density parameters. We will consider a simplified situation in which the HI (and therefore baryon) distribution directly traces the underlying matter distribution, so that $\delta_{\rm b}=\delta$, and also neglect fluctuations in the neutral fraction (effectively assuming instantaneous reionization at $z=6$), setting $x_{\rm H}=1$. (We have also neglected the impact of fluctuations in the gas temperature.) In reality, each of these effects will need to be accounted for in a lensing analysis of 21cm observations around reionization, but in this work we are mainly focused on the interplay between nonlinear evolution of $\delta$ and lensing reconstruction, so these simplifications will suffice for our purposes.  The mean brightness temperature then becomes
\beq
\bar{T}(z) \approx 26  
	\lp \frac{\Omega_{\rm b}h^2}{0.022} \rp
	\lb \lp \frac{0.15}{\Omm h^2} \rp \lp \frac{1+z}{10} \rp \rb^{1/2} {\rm mK}\ ,
\eeq
and the corresponding 3d power spectrum (assuming no time evolution with the observed redshift band) is
\beq
P_T(\vk;z) = \bar{T}^2(z) \lp 1+f\mu^2 \rp^2 D_{\rm FoG}(\kpar,z) P_\delta(\vk;z)\ ,
\label{eq:pt21}
\eeq
with the squared factor accounting for the leading effect of redshift space distortions  on large scales, where $f\equiv \d\log D(a)/\d\log a$ is the logarithmic derivative of the linear growth factor $D(a)$ and $\mu^2\equiv \kpar^2/k^2$. We use the best-fit Lorentzian model\footnote{
For reference, this model is given by
\beq
D_{\rm FoG}(\kpar,z) = \lb 1+\frac{1}{2}\kpar^2\sigma_{\rm p}(z)^2 \rb^{-1}\ ,
\quad
\sigma_{\rm p}(z) = \lb 9.12 \,{\rm Mpc} \rb \lp 1+z \rp^{-1.15} \exp\lb -\lp\frac{z}{12.0}\rp^2 \rb\ ,
\eeq
where the numerical values were fitted to simulations of HI clustering at $1\lesssim z\lesssim 6$.
The authors of Ref.~\cite{Sarkar:2018gcb} note that this model may overestimate the HI velocity dispersion on small scales, and therefore the small-scale damping of the HI power spectrum caused by ``fingers of God," making this model a rather conservative choice to use in forecasting.
} from Ref.~\cite{Sarkar:2018gcb} for redshift-space ``finger of God" damping at small scales, $D_{\rm FoG}(\kpar,z)$.
 To connect with the formalism in the previous sections of this paper, the brightness temperature can be rewritten in terms of an intensity via the Rayleigh-Jeans law, but it is not necessary to explicitly perform this translation because the resulting prefactor will cancel out of any lensing-related expressions.

For the observational noise on the 21cm angular power spectrum, we take the interferometer thermal noise expression from Ref.~\cite{Romeo:2017zwt}, evaluated at the frequency corresponding to the mean of the observed frequency band:
\beq
C_\ell^{\rm N}(\kpar) = \lb \frac{\lambda^2}{A_{\rm e}} F\!\lp \frac{\nu}{\nu_{\rm c}} \rp \rb^2 
	\frac{T_{\rm sys}^2(\nu)}{N_{\rm pol} B t_0} \frac{1}{n(\vl/2\pi,\nu)}\ ,
	\quad
	F(x) \equiv \left\{
	\begin{array}{ll} 
	1\ , & x\leq1 \\
	x^2\ , & x > 1
	\end{array}
	\right. \ .
	\label{eq:ska-noise}
\eeq
We use the values for phase 1 of SKA-Low, also from Ref.~\cite{Romeo:2017zwt}: $A_{\rm e}=925$ m$^2$ is the effective receiving area of a single station, $\nu_{\rm c}=110$ MHz is the ``critical frequency" above which the effective receiving area receives a multiplicative correction of $(\nu_{\rm c}/\nu)^2$, $N_{\rm pol}=2$ is the number of polarizations per receiver, $B=5$ MHz is the observing bandwidth, and $t_0=2000$ hrs is the total observing time. The system temperature $T_{\rm sys}$ is the fundamental source of thermal noise in an antenna, and has contributions from the instrumental receiver temperature, set to 40 K, and galactic synchrotron radiation, which can be approximated as
\beq
T_{\rm sys}(\nu) = 40\,{\rm K} + 66\lp \frac{\nu}{300\,{\rm MHz}} \rp^{-2.55} \,{\rm K}\ .
\eeq
Finally, $n(\vu,\nu)$ is the time-averaged number density of baselines in the $uv$ plane, evaluated on $\vu=\vl/2\pi$ in Eq.~\eqref{eq:ska-noise}. For a circularly-symmetric array of receivers, this can be approximated by~\cite{Zahn:2005ap}
\beq
n(\vu,\nu) \approx \int d^2\vx\, \mathcal{P}(\vx + \lambda\vu) \mathcal{P}(\vx)\ ,
\eeq
where $\mathcal{P}(\vx)$ is the radial profile of the antenna distribution on the ground, with dimensions of length$^{-1}$. (Ref.~\cite{Zahn:2005ap} uses a different convention for $\mathcal{P}$: its version of this expression has a prefactor of $\lambda^2$, which requires that $\mathcal{P}$ have dimensions of length$^{-2}$.) This function must be normalized to equal the number of receiver pairs when integrated over the upper-half-$uv$-plane:
\beq
\int_{\rm UHP} d^2\vu \, n(\vu,\nu) = \frac{N_{\rm rec}(N_{\rm rec}-1)}{2}\ .
\eeq
We find that if $\mathcal{P}(\vx)$ is taken to be a Gaussian,
\beq
\mathcal{P}(\vx) = \frac{1}{\sqrt{2\pi}\sigma} e^{-x^2/2\sigma^2}\ ,
\eeq
with $\sigma=550$ m, the noise power spectrum resulting from Eq.~\eqref{eq:ska-noise} is a good match to that calculated in Ref.~\cite{Romeo:2017zwt} from a more realistic computation of $n(\vu,\nu)$ (shown in their Fig.~1), so we will use Eq.~\eqref{eq:ska-noise}  for our forecasts.

\begin{figure}
\includegraphics[width=\textwidth, trim = 25 20 25 0 ]{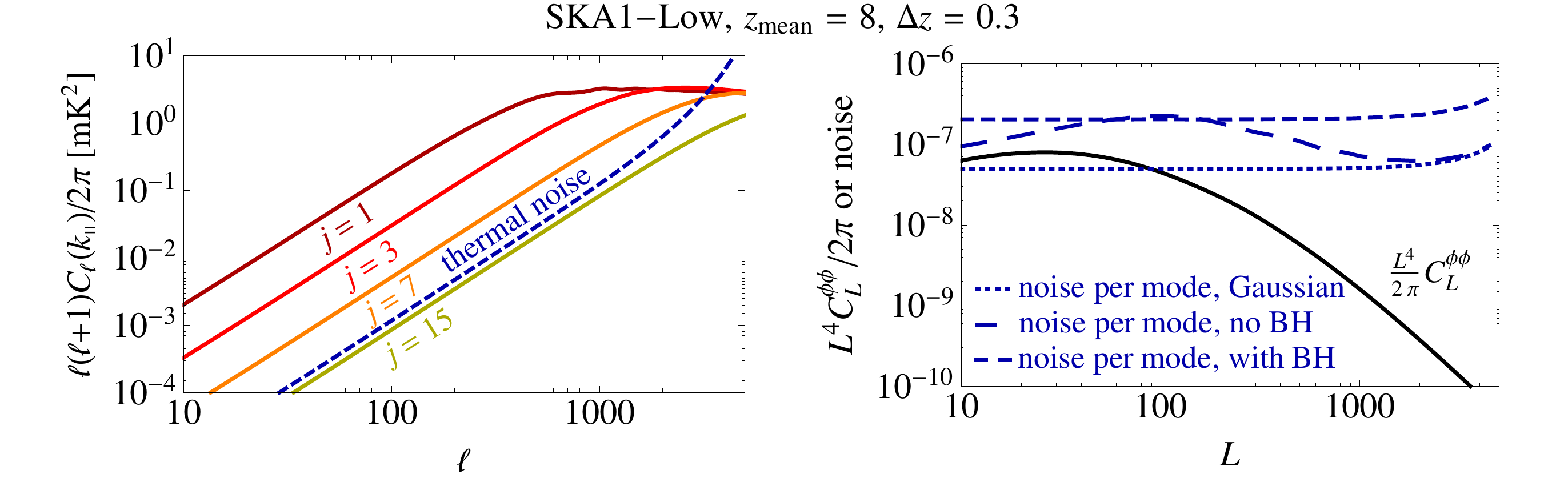}
\caption{\label{fig:ska}
{\em Left panel}: 21cm angular power spectra [related to the 3d intensity power spectrum by Eq.~\eqref{eq:clidef}] corresponding to different $j$ ($\kpar$) values (solid lines), along with the thermal noise angular power spectrum from Eq.~\eqref{eq:ska-noise} (dashed line), for a 5MHz band centered at $z=8$. At low $\ell$, the amplitude of the signal decreases with increasing $j$, until it falls below the thermal noise around $j\sim 13$.
{\em Right panel:} Lensing potential power spectrum (solid black line) and combined lensing reconstruction noise for $3\leq j \leq 20$, in non-bias-hardened (long-dashed), bias-hardened (short-dashed), and Gaussian-noise-only (dotted) cases. The nonlinearity of gravitational evolution increases the reconstruction noise substantially, and bias-hardening further increases the noise, but is necessary to remove the large gravitational bias that would otherwise contaminate the lensing estimator.
}    
\end{figure}

In the left panel of Fig.~\ref{fig:ska}, we show 21cm angular power spectra corresponding to different $j$ ($\kpar$) values, along with the thermal noise power spectrum from Eq.~\eqref{eq:ska-noise}, evaluated for a 5MHz band centered at $z=8$. The overall amplitude of the signal at low $\ell$ decreases with increasing $j$, until dropping below the noise around $j\sim 13$. (This number is about $j\sim17$ at $z=6$, and decreases monotonically with redshift.) The signal curves are flat in $C_\ell$ for $\kpar\gtrsim \ell/\chi$, and take the shape of the matter power spectrum for $\kpar\lesssim \ell/\chi$. The high-$\ell$ upturn in the noise power spectrum is due to the increasing sparsity of longer baselines in the interferometer, implying less sensitivity to small angular scales on the sky.

In the right panel, we show the noise per $\phi$ mode corresponding to the combined estimator for $3\leq j \leq 20$.  Using $j_\mathrm{min}=3$ (which corresponds to $k_{\parallel{\rm min}}\approx 0.4\invMpc$ and $0.2\invMpc$ for 5MHz bands at $z\approx 6$ and 20 respectively) is a rough way to incorporate some of the effects of foregrounds, which will prevent measurements of low-$\kpar$ modes due to the expected smoothness of the dominant foregrounds' spectral behavior.  While one could consider also excluding modes within a wedge-shaped region in Fourier space, to account for beam chromaticity that mixes measurements line-of-sight and transverse modes~\cite{Pober:2014lva}, recent work suggests that these modes might be recoverable  more effectively than previously thought \cite{Koopmans2017}. Taking $j_{\rm max}=20$ incorporates the contribution from the first few noise-dominated radial modes.

The non-bias-hardened curve takes the shape of the matter power spectrum at low $\ell$, due to the dominance of the $N_{\phi\phi}^{({\rm nG,}P)}$ term, and is generally much larger than the contribution of Gaussian noise alone. Meanwhile, once bias-hardening is applied, the $N_{\phi\phi}^{({\rm nG,}P)}$ contribution is removed and the remaining variance scales like white noise on $L^4 C_L^{\phi\phi}$. At most scales, bias-hardening increases the overall noise per mode, but will be necessary for many applications in order to prevent large biases in the reconstructed lensing maps and associated power spectra; indeed, in these forecasts, the non-Gaussian bias on the power spectrum, defined in Eq.~\eqref{eq:ngbias}, decreases by a factor of $\sim$10 after bias-hardening is applied. Note that, as discussed in \sec{sec:bh}, modes with $\kpar \gtrsim \ell_{\rm max}/\chi$ generally have a minimal impact on the noise of the bias-hardened estimator. For the redshift band centered on $z=8$, modes with $j\gtrsim 10$ contribute negligibly to the total lensing detection. 

A more optimistic stance on foreground cleaning could be incorporated by changing~$j_{\rm min}$ from 3 to 1, and this decreases the noise per $\phi$ mode by about a factor of 1.5. If one is instead more pessimistic, increasing $j_{\rm min}$ to 6 (corresponding to $\sim 0.8\invMpc$ and $0.4\invMpc$ at $z=6$ and 20 respectively), the lensing noise increases by roughly a factor of~2.5. Thus, the success of foreground subtraction will play a large role in determining the final precision of lensing measurements.

A direct comparison of these results with the simulation-based approach of Ref.~\cite{Romeo:2017zwt} is difficult due to different implementations of the lensing estimator and thermal noise computations; in particular, slight differences in the assumed $n(\vu,\nu)$ at high $\ell$ can make order-one differences in the results, due to the importance of long baselines for lensing reconstruction.   Nevertheless, the relative amplitudes of the different curves indicate that gravitational non-Gaussianities can have a large effect on the lensing estimator, even at the high redshifts probed by SKA-Low.

\subsection{CHIME \& HIRAX}
\label{sec:chime}

As an example of line intensity maps at lower redshifts, we will consider CHIME and HIRAX, two experiments which will both observe 21cm radiation between 400 and 800 MHz, corresponding to $0.8\lesssim z \lesssim 2.5$. CHIME~\cite{2014SPIE.9145E..22B}, located at a radio-quiet site in British Columbia, Canada, is an array of four 20m $\times$ 100m cylindrical dishes, each outfitted with 256 feeds positioned along the axis of each cylinder. HIRAX~\cite{Newburgh:2016mwi}, located in South Africa, will be a compact array of 32 $\times$ 32 6m dishes. Both experiments aim to measure baryon acoustic oscillations, and also to perform wide searches for radio transients.

At the relevant redshifts, the mean 21cm brightness temperature can be written as (e.g.~\cite{Shaw:2013wza})
\beq
\bar{T}(z) \approx 0.3  
	\lp \frac{\Omega_{\rm HI}}{10^{-3}} \rp
	\lp \frac{\Omm+(1+z)^{-3}\Omega_\Lambda}{0.29} \rp^{-1/2} \lp \frac{1+z}{2.5} \rp^{1/2} {\rm mK}\ ,
\eeq
with $\Omega_{\rm HI} \approx 5\times 10^{-4}$~\cite{Masui:2012zc}. 
The temperature power spectrum $P_T$ is then given by the following modification of Eq.~\eqref{eq:pt21}:
\beq
P_T(\vk;z) = \bar{T}^2(z) \lp b_{\rm HI}(z)+f\mu^2 \rp^2 D_{\rm FoG}(\kpar,z) P_\delta(\vk;z)\ ,
\eeq
where we use the model from Ref.~\cite{Castorina:2016bfm} for the linear HI bias $b_{\rm HI}(z)$.
We have neglected shot noise, because the average of the shot noise models in Ref.~\cite{Castorina:2016bfm} is no higher than the thermal noise for all redshifts we consider for these surveys, and comfortably below the thermal noise at the redshifts where the signal to noise peaks.

In principle, one could adapt the thermal noise computation from \sec{sec:ska} for these instruments, by computing $n(\vu,\nu)$ and also making appropriate modifications to account for a drift-scan observation strategy. However, for a close-packed distribution of feeds or dishes, one may use a simpler formalism in which the instrument is treated as having a single large collecting area observing the sky with multiple simultaneous beams~\cite{Seo:2009fq} (this approach, realizable in practice with FFT beamforming~\cite{2017arXiv171008591M}, contains the same information as stacking measurements from redundant baselines). In this case, the noise angular power spectrum is given by~\cite{Seo:2009fq,Shaw-comm}
\beq
C_\ell^{\rm N}(\kpar) 
	= \frac{T_{\rm sys}^2(\nu)}{t_{\rm pix} B}  \frac{A_{\rm pix}}{W(\ell)}\ ,
\eeq
where $A_{\rm pix}$ and $t_{\rm pix}$ are the angular area and observing time per angular ``pixel" observed on the sky, given respectively by
\beq
A_{\rm pix} = \frac{\lambda^2}{n_{\rm x} n_{\rm y} A_{\rm e}}\ , \quad
	t_{\rm pix} = \frac{A_{\rm pix}}{4\pi f_{\rm sky}} n_{\rm beams} t_0\ .
	\label{eq:apixtpix}
\eeq
In $A_{\rm pix}$, $n_{\rm x}$ and $n_{\rm y}$ are the number of feeds or dishes along two orthogonal axes on the ground, and, as before, $A_{\rm e}$ is the effective collecting area of a single element. For CHIME, $n_{\rm x}=4$, $n_{\rm y}=256$, and $A_{\rm e}=(20\times 80/256)\,{\rm m}^2 = 6.25$m$^2$ (since only 80m of each cylinder are instrumented with feeds), while for HIRAX, $n_{\rm x}=n_{\rm y}=32$ and $A_{\rm e}= (\pi 3^2/4)\,{\rm m}^2 \approx 28$m$^2$. In FFT beamforming, the maximum number of beams with non-redundant information is $n_{\rm beams}=(2n_{\rm x}-1)(2n_{\rm y}-1)$~\cite{2017arXiv171008591M}, so we use that for each experiment. The window function $W(\ell)$ encodes the effective scale-dependence of the instrument response. A reasonable choice for this is \cite{Shaw-comm}
\beq
W(\ell) = \Lambda\!\lp \frac{\ell \lambda/2\pi}{L_{\rm cyl}} \rp
	\Lambda\!\lp \frac{\ell \lambda/2\pi}{N_{\rm cyl} W_{\rm cyl}} \rp\ 
\eeq
for CHIME, and
\beq
W(\ell) = \Lambda\!\lp \frac{\ell \lambda/2\pi}{n_x D_{\rm dish}} \rp
	\Lambda\!\lp \frac{\ell \lambda/2\pi}{n_y D_{\rm dish}} \rp\ 
\eeq
for HIRAX, using the triangular response function 
\beq
    \Lambda(x)= 
\begin{cases}
    1-|x| , & |x|\leq1\\
    0,       & |x|\geq1
\end{cases}
\eeq
with arguments that are simply ratios of angular scales projected on the ground to the total dimensions of the interferometer. Finally, in both cases we assume $f_{\rm sky}=0.5$ and $t_0 = 2.5$ years of observing time (5~calendar years, assuming that roughly half of that time is usable for 21cm observations~\cite{Shaw-comm}).

\begin{figure}[t]
\includegraphics[width=\textwidth, trim = 25 20 25 0 ]{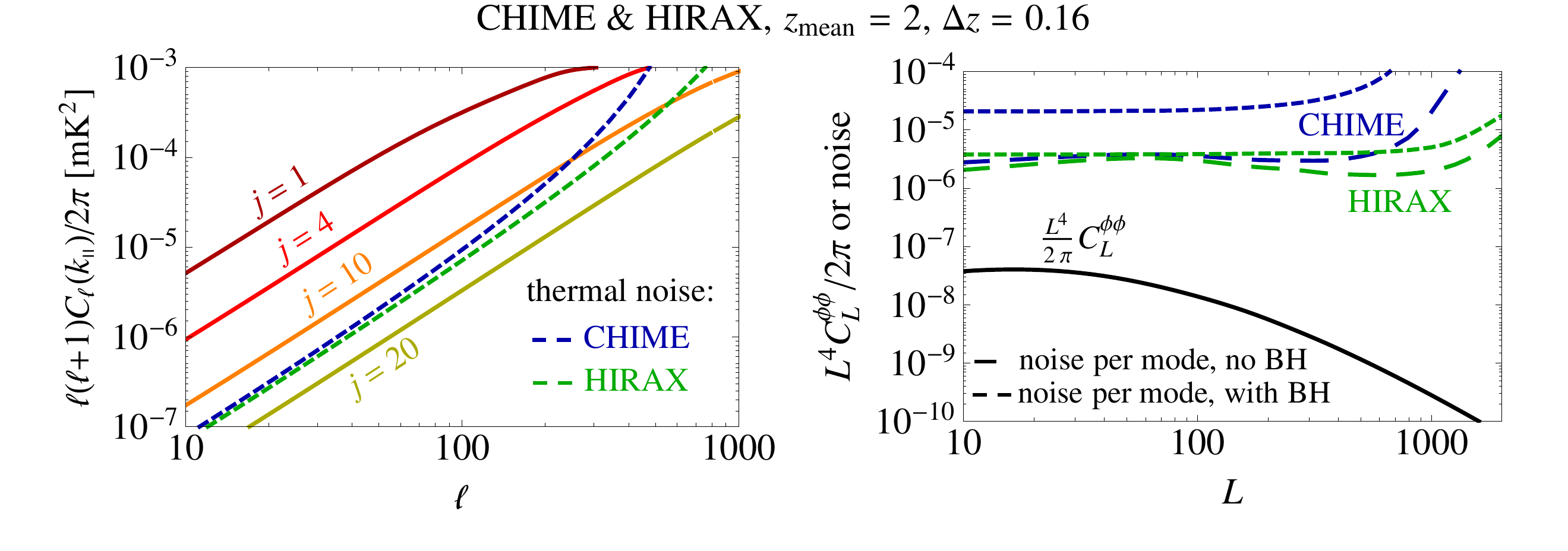}
\caption{\label{fig:chime}
{\em Left panel:} 21cm signal (solid) and thermal noise (dashed) angular power spectra for the a redshift band with $\Delta z = 0.16$ centered on $z=2$, with CHIME noise in blue and HIRAX noise in green. HIRAX can resolve smaller scales than CHIME due to its inclusion of longer baselines. At this redshift, observations of 21cm fluctuations will be signal-dominated for $j$ values that lie beyond the range of validity of the perturbative predictions in this paper, although the predictions may be carried to higher orders (or computed using simulations) to regain use of these modes.
{\em Right panel:} Lensing potential power spectrum (solid black) and combined noise per $\phi$ mode for CHIME and HIRAX, in non-bias-hardened (long-dashed) and bias-hardened (short-dashed) cases, with $j_{\rm max}$ determined by where our perturbative calculations break down (as discussed in the main text). A significant detection of the lensing potential auto spectrum will be challenging with these instruments, although detections at the few-sigma level may be possible if the information from all available redshifts is combined.
}    
\end{figure}

In Fig.~\ref{fig:chime}, we consider an example redshift band within the CHIME/HIRAX frequency range, with $\Delta z = 0.16$ (corresponding to $B\approx 25$MHz) and centered on $z=2$. The left panel shows 21cm angular power spectra for different $j$ values, along with thermal noise for CHIME and HIRAX. Due to its inclusion of longer baselines, HIRAX achieves an effective angular resolution roughly twice that of CHIME, as quantified by the scale at which the noise power spectrum exceeds the signal. Linear growth of the matter fluctuations increases the amplitude of the signal curves at lower redshifts, while the noise decreases due to the lower system temperature; both trends imply that modes with higher $\kpar$ can be resolved at lower redshift. In fact, more modes are resolvable in either case than fall within the range of validity of the tree-level calculations in this paper: for CHIME and HIRAX, our computations are valid for $j\lesssim 10$ in the band used for Fig.~\ref{fig:chime}. For comparison, modes up to $j\approx 14$ can be resolved above the noise in these surveys.  The bias-hardened estimators impose further limitations on the number of modes usable for the lensing auto spectrum: for CHIME, the effective usable $j_{\rm max}$ in this band decreases to 7, while for HIRAX, the tree-level calculation imposes a lower~$j_{\rm max}$ than bias hardening does.\footnote{
The redshift-space ``finger of God" damping primarily affects modes of the HI temperature that are beyond the range of validity of our perturbative calculations, and/or that do not contribute to the bias-hardened estimators, so the inclusion of this effect has negligible impact on the forecasts we summarize in \sec{sec:summary-and-xcorrs}.
} More details about these various restrictions for different bands can be found in App.~\ref{app:singleband21cm}.

In the right panel of Fig.~\ref{fig:chime}, we show the noise per $\phi$ mode for the $z\sim 2$ band for both CHIME and HIRAX, either with or without applying bias-hardening to the lensing estimator. For both surveys, bias-hardening increases the noise  on a measurement of $C_L^{\phi\phi}$. Even if the information from multiple redshift bands is combined, a significant detection of $C_L^{\phi\phi}$ will be challenging with these instruments, although cross-correlations with other tracers of low-redshift structure could be more promising. We will return to this point in \sec{sec:summary-and-xcorrs}.

For these forecasts, we use $j_{\rm min}=3$, which corresponds to $k_{\parallel{\rm min}}\approx 0.05\invMpc$ for 25\,MHz bands in the CHIME/HIRAX redshift range. This is a conservative choice in the sense that it is well above the expected minimum $\kpar$ associated with CHIME foreground cleaning, roughly $0.02\invMpc$~\cite{Shaw:2014khi}. If we instead take $j_{\rm min}=1$, which is right at this limit, the expected noise per $\phi$ mode from the redshift band in Fig.~\ref{fig:chime} decreases by about 30\% without bias-hardening or 60\% with bias-hardening, while if we increase $j_{\rm min}$ to 6, the noise increases by 30\% and a factor of 5, respectively.

\subsection{Single-dish survey}
\label{sec:ccat}

In the previous subsections, we have shown two examples of lensing reconstruction using 21cm maps. However, {\em any} line intensity maps will be lensed to some extent, and so we can inquire about the detectability of lensing in other intensity mapping surveys. Each emission or absorption line will be subject to different systematics or obstacles, such as different levels of shot noise, confusion with other lines that redshift into the relevant frequency range, and continuum foregrounds. These issues must be studied in detail before first detections can be achieved, let alone detections of lensing in the maps, but such studies are underway (for a recent summary, see Ref.~\cite{Kovetz:2017agg}), and the results of these studies can also likely be applied to the issues relevant for lensing. In this subsection, we will not perform a comprehensive exploration of the prospects for lensing of various forms of intensity mapping, but rather present a representative example, to motivate interest in the general prospects for such measurements.

We consider observations of the 158 $\mu$m fine-structure line in ionized carbon (CII), expected to be a good tracer of star formation at high redshifts. Ref.~\cite{Gong:2011mf} presents instrumental parameters for a proposed CII intensity mapping survey, based on a specific model for the clustered and shot-noise components of the CII emission power spectrum. However, subsequent work has explored a variety of other models that span a large amplitude range for both components of the signal (e.g.~\cite{Silva:2014ira,Lidz:2016lub,Cheng:2016yvu,Serra:2016jzs,Dumitru:2018tgh}). Therefore, we do not rely on any single signal model, but instead consider cosmic-variance-limited measurements up to some angular resolution $\ell_{\rm max}$. We incorporate uncertainty in the strength of the signal compared to the instrumental noise by considering different choices for $j_{\rm max}$, because higher S/N will imply that modes with higher $j$ will be signal-dominated.

As with 21cm radiation,  the CII signal will be dwarfed by continuum foregrounds, which in this case will be dominated by far-infrared emission from dusty galaxies and dust in the Milky Way (e.g.~\cite{Yue:2015sua}). It has been shown that the foreground-subtraction techniques developed for 21cm surveys, which rely on foregrounds being spectrally smooth, can also be successful when applied to simulated CII surveys~\cite{Yue:2015sua}. This implies that foregrounds will likely render modes with low $\kpar$ unusable for cosmology; in our forecasts in this section, we assume this is true for the first two discrete $\kpar$ values in each redshift band, so we take $j_{\rm min}=3$ (corresponding to $k_{\parallel{\rm min}}\approx 0.08$, 0.13, and $0.19\invMpc$ at $z\approx4$, 6, and 8) in each band.
 Due to uncertainty about the relative amplitudes of the clustered and shot-noise components of the power spectrum, we only consider the clustered component for simplicity. If shot noise turns out to be significant on scales relevant for lensing reconstruction, one should consider incorporating the modified estimators presented in Refs.~\cite{Pourtsidou:2013hea,Pourtsidou:2014pra}.

We will consider angular resolutions and redshift ranges based on the CII survey proposed to take place on the CCAT-prime telescope\footnote{\href{http://www.ccatobservatory.org}{http://www.ccatobservatory.org}}. Plans for this survey are broadly modeled on the recommendations of Ref.~\cite{Gong:2011mf}, observing CII emission over $3\lesssim z \lesssim 9$ with an instrument mounted on the 6m CCAT-prime dish. For simplicity, we will consider observations over three redshift bands centered on $z=4,6,8$ and each with $\Delta z=0.5$. By taking $\theta_{\rm min} \approx 1.22 \lambda_{\rm CII}(z)/D_{\rm dish}$ with $\lambda_{\rm CII}=158(1+z)$~$\mu$m, and then using $\ell_{\rm max} \approx \theta_{\rm min}/\chi(z)$, we find $\ell_{\rm max} \approx \{ 20000,14000,11000 \}$ for $z=\{ 4,6,8\}$. To take a conservative stance on instrumental uncertainties such as beam calibration, we will divide these values by 2 and take the result to be the effective angular resolution in each band. For $z=4$, this results in an $\ell_{\rm max}$ that exceeds the validity of our tree-level calculation for the lensing variance (see Fig.~\ref{fig:treelevel-validity}). Therefore, we will further reduce $\ell_{\rm max}(z=4)$ to $3700$, ensuring that our computations are valid at $j\leq 20$. (Note that this implies that our forecasts for $z=4$ may be on the pessimistic side, pending an exploration of higher-order or simulation-based calculations that extend to smaller scales.) Under these restrictions, we finally arrive at $\ell_{\rm max} \approx \{ 3700,7000,5500 \}$ at $z = \{4,6,8\}$. We also consider the same survey but for a 10m dish, which gives $\ell_{\rm max} \approx \{ 3700,9000,9000 \}$ at $z = \{4,6,8\}$ (where in this case both $z=4$ and $z=6$ are restricted by our tree-level calculation).

\begin{figure}[t]
\includegraphics[width=\textwidth, trim = 25 20 25 0 ]{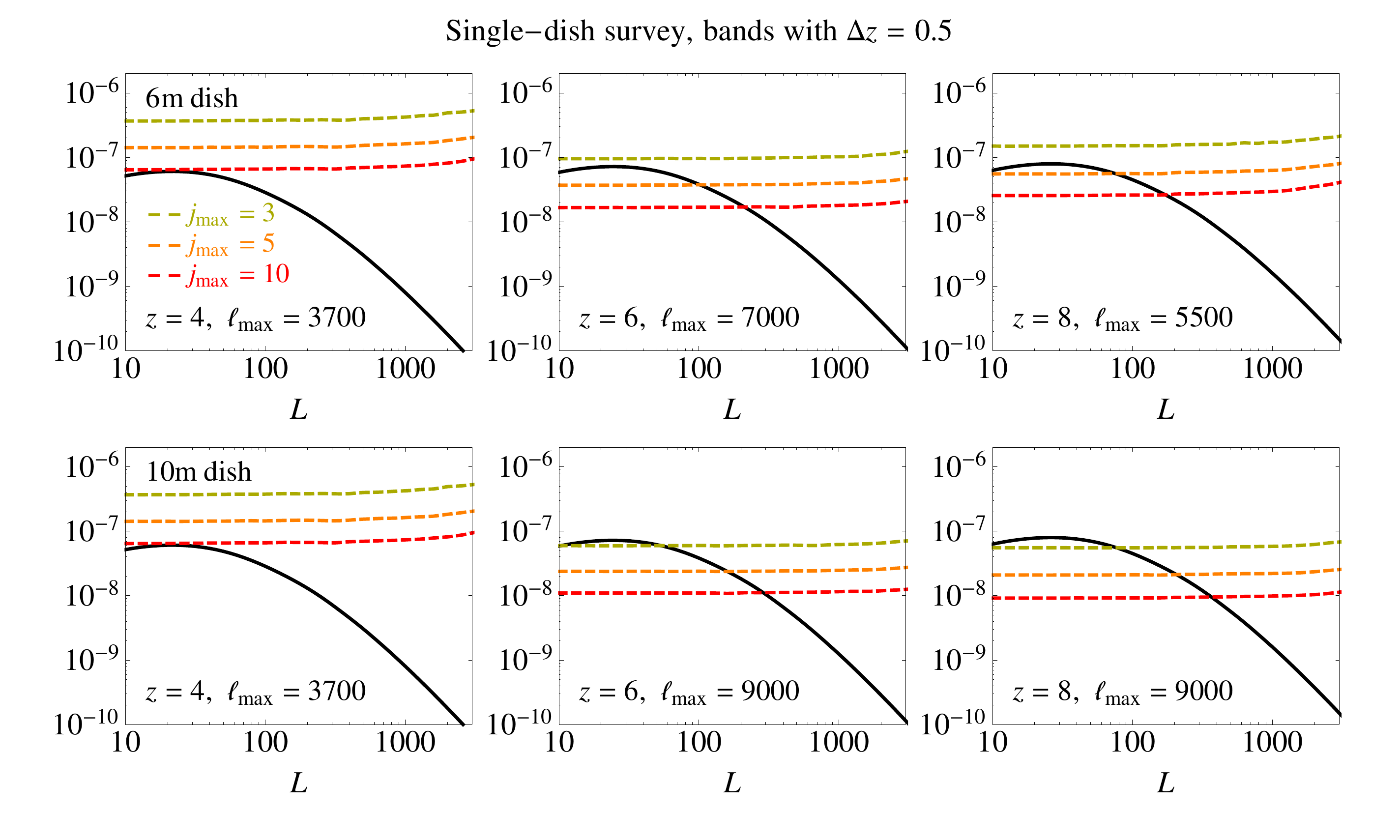}
\caption{\label{fig:ccat}
Noise per reconstructed $\phi$ mode in three different redshift bands, for the single-dish intensity mapping survey (modeled on the proposed CII survey from Ref.~\cite{Gong:2011mf}) described in the main text, with either a 6m ({\em upper panels}) or 10m ({\em lower panels}) dish. We take $j_{\rm min}=3$ to account for low-$\kpar$ modes being lost to foreground cleaning, and show a few different choices for $j_{\rm max}$ as a way of parameterizing the uncertainty in the signal-to-noise on the line intensity power spectrum. Thanks to the high angular resolution achievable with a single dish, signal-dominated lensing maps can be produced in most cases with only a modest $j_{\rm max}$. However, the limited field of view of single-dish surveys will restrict these maps to small patches of sky, and limit the detectability of the lensing power spectrum.
}    
\end{figure}

In Fig.~\ref{fig:ccat}, we show the combined noise per $\phi$ mode for different values of $j_{\rm max}$ and redshift, and for both dish sizes mentioned above. To avoid clutter, we only show the results for the bias-hardened estimator. In most cases, we find that if modes of the CII intensity with $j\leq 5$ are signal-dominated (where $j=5$ corresponds to $\kpar \approx \{ 0.13,0.22,0.32 \} \,\invMpc$ in the $z=\{ 4,6,8 \}$ bands), we can obtain signal-dominated reconstructions of the lensing potential for at least $L\lesssim 100$. This conclusion is primarily due to the small angular scales probed in these surveys, as opposed to the drift-scan 21cm interferometers like CHIME. However, this high angular resolution is offset by a small field of view (again compared to drift-scan instruments), implying that, in a survey lasting a few thousand hours, only a relatively small patch of sky (e.g.~16 deg$^2$ for the CII survey from Ref.~\cite{Gong:2011mf}) can be observed with reasonable noise levels. This implies that the total signal-to-noise on either an auto- or cross-correlation power spectrum will be correspondingly reduced, since the S/N scales like $f_{\rm sky}^{1/2}$. 

Decreasing $j_{\rm min}$ from 3 to 1 in each band improves the noise per $\phi$ mode by a factor of 2.5 for $j_{\rm max}=3$ but only a factor of 1.1 for $j_{\rm max}=10$, indicating that there is significant lensing signal available in source modes with higher $j$. These modes will be damped slightly by the ``finger of God" effect, but we expect this to have a minor impact on the detectability of lensing. We can estimate this damping using the model\footnote{
In detail, we only consider the component of this model arising from the multi-streaming regime (which we expect to be dominant over the contribution of bulk flows to $D_{\rm FoG}$ at the relevant redshifts and scales). This is given by
\beq
D_{\rm FoG}(\kpar,z) = e^{-\kpar^2\sigma_{v,{\rm multi}}^2/H^2(z)}\ ,
\eeq
where 
\beq
\sigma_{v,{\rm multi}} = a(z)^{-1} \sigma_{v,{\rm vir}} =
	102.5 \times 0.9 \times a(z)^{-1} \Delta_{\rm vir}^{1/6}(z)
	\lp \frac{H(z)}{H_0} \rp^{1/3} \lp \frac{M}{10^{13} h^{-1} M_\odot} \rp^{1/3} \text{km/s}\ ,
\eeq
we have corrected a typo in the latter formula, and $\Delta_{\rm vir}(z)$ is given below Eq.~(4.9) of Ref.~\cite{Zheng:2016xvo}.
} from Ref.~\cite{Zheng:2016xvo}, which gives the contribution to the damping factor $D_{\rm FoG}(\kpar,z)$ from the velocity dispersion inside halos of some characteristic mass. Ref.~\cite{Yue:2015sua} found that CII is mostly hosted by halos with $10^{11} \lesssim M/M_\odot \lesssim 10^{12}$ at $z\sim 5$. If we consider halos with $M\approx 10^{11.5} h^{-1} M_\odot$, we find that $D_{\rm FoG}(j=10) \approx 0.9$ and $D_{\rm FoG}(j=20) \approx 0.75$ at the relevant redshifts, which will not significantly impact the signal to noise on these higher-$j$ modes.

In the next subsection, we will present the specific S/N numbers for each survey we have considered so far.

\subsection{Summary and prospects for cross-correlations}
\label{sec:summary-and-xcorrs}

A useful way to collect the results of these forecasts is the compute the total signal to noise on a measurement of the amplitude of the lensing potential power spectrum in each case:
\beq
\lp \frac{\rm S}{\rm N} \rp^2_{\rm auto} = f_{\rm sky} \sum_L \frac{2L+1}{2}
	\lp \frac{C_L^{\phi\phi}}{C_L^{\phi\phi} + N_{\phi\phi}^{\rm (full,combined)}(L)} \rp^2\ .
\eeq
For this, we use the noise corresponding to the bias-hardened estimator, since, as we have discussed, the auto spectrum that results from the regular estimator typically has large biases from gravitational nonlinearity.

We also compute the signal to noise for a few example cross-correlations, denoted by $X$ in the following formula:
\beq
\lp \frac{\rm S}{\rm N} \rp^2_{\rm cross} = f_{\rm sky} \sum_L (2L+1)
	\frac{\lb C_L^{\kappa X} \rb^2}
	{\lb C_L^{\kappa X}  \rb^2 + \lb C_L^{XX} + N_{XX}(L) \rb
	\lb C_L^{\kappa\kappa} + N_{\kappa\kappa}^{\rm (full,combined)}(L) \rb}
\eeq
where $N_{XX}(L)$ is the noise power spectrum for $X$, and we have replaced $\phi$ with the convergence  $\kappa(\vl) = (1/2)\ell^2\phi(\vl)$, more commonly seen in galaxy lensing.   This allows us to write any cross or auto spectrum we need, in the Limber approximation, as (e.g.~\cite{Bleem:2012gm})
\beq
C_L^{XY} = \int \frac{dz}{\chi(z)^2} \frac{d\chi}{dz} W^X(\chi) W^Y(\chi) P_\delta(L/\chi(z);z)\ .
\eeq
For lensing of the CMB or a line intensity map located at a (mean) comoving radial distance $\chi_{\rm s}$, the distance kernel $W^\kappa(\chi)$ is
\beq
W^\kappa(\chi) = \frac{3}{2} \Omm H_0^2 
	\frac{\chi}{a(\chi)} \frac{\chi_{\rm s}-\chi}{\chi_{\rm s}}\ .
\eeq
We consider cross-correlations with the following tracers:
\begin{enumerate}
\item $X= g$ (galaxy clustering), with distance kernel
\beq
W^g(\chi) = \frac{dz}{d\chi} n(z) b(\chi)\ ,
\label{eq:wg}
\eeq
where $n(z)$ is the redshift distribution of galaxies, normalized such that $\int dz\, n(z)=1$, and $b(\chi)$ is the bias of the galaxy sample in question, assumed to be scale-independent. We will focus on clustering as measured by LSST, for which $n(z)$ is expected to have the rough form~\cite{Krause:2016jvl}
\beq
n(z) \propto z^{1.25} \exp\!\lb-z/0.5\rb\ .
\eeq
For the sake of simplicity, we will not consider tomography of the galaxy distribution, treating the entire galaxy sample as a single ``bin" with the redshift distribution above. Following Ref.~\cite{Schmittfull:2017ffw}, we will assume the simple linear bias model $b(\chi)=1+z(\chi)$, and a mean angular galaxy density $\bar{n} = 65$ arcmin$^{-2}$ for $z<4$. The latter determines the noise power spectrum for clustering, given by the Poisson expression $N_{gg}(L) = \bar{n}^{-1}$ with $\bar{n}$ in sr$^{-2}$.

When cross-correlating with lensing of source fluctuations at $z_{\rm min}<z<z_{\rm max}$, we only consider galaxies at $z<z_{\rm min}$, reducing $\bar{n}$ accordingly. With this restriction, the non-bias-hardened lensing estimator can be used, since the long density modes picked up by the quadratic estimator will not cross-correlate with galaxies at lower redshifts.

For cross-correlations with a single-dish IM survey, one could also consider the COSMOS photometric redshift sample~\cite{Laigle:2016jxn}, motivated by plans for the CCAT-prime CII survey patch to overlap with a deep field similar to COSMOS. The redshift distribution 
\beq
n(z) \propto z^{1.25} \exp\!\lb-(z/0.5)^{1.25}\rb\ ,
\eeq
provides a rough match to the redshift distributions shown in Ref.~\cite{Laigle:2016jxn} for different magnitude bins. From the total number of objects  ($\sim 7.7\times 10^5$) with measured photometric redshifts within the 2~deg$^2$ COSMOS field, we compute a mean galaxy density of $\bar{n} \approx 107$ arcmin$^{-2}$. We find that the S/N for cross-correlations with COSMOS galaxies would be within 10\% of that for LSST galaxies, if the COSMOS field was the same size as the IM survey, but the small size (2~deg$^2$, versus 16~deg$^2$ for the proposed CCAT-prime CII survey) will limit our ability to measure the cross power spectrum. For this reason, we will only present numbers for LSST cross-correlations.
\item $X=\gamma$ (cosmic shear; technically another measurement of the convergence $\kappa$, but we will use $X=\gamma$ to differentiate from lensing of intensity maps), with distance kernel
\beq
W^\gamma(\chi) = \frac{3}{2} \Omm H_0^2 
	\frac{\chi}{a(\chi)} \int_0^{\chi_*} d\chi' \frac{dz}{d\chi'} n(z) \frac{\chi'-\chi}{\chi'}\ ,
\eeq
where $n(z)$ is the normalized redshift distribution of source galaxies. We will again use the $n(z)$ functions from above, considering a single redshift bin spanning the entire range of each survey. The noise power spectrum is given by the shape noise term $N_{\gamma\gamma}(L) = \sigma_\epsilon^2 \bar{n}^{-1}$, where $\sigma_\epsilon=0.27$ is the intrinsic noise per ellipticity component of the observed galaxy shapes. Following Ref.~\cite{Abell:2009aa}, we use $\bar{n} = 40$ arcmin$^{-2}$ for the number density of LSST galaxies with well-measured shapes. As for the galaxy clustering cross-correlations, we will only use galaxies for cosmic shear located at lower redshifts than the source intensity field we use for lensing reconstruction, enabling the use of the non-bias-hardened lensing estimator.
\end{enumerate}

\begin{table}[t]
\centering
\begin{tabular}{lccc|ccc}
\hline \hline
\multicolumn{7}{c}{S/N on lensing power spectra for 21cm surveys}\\
\hline	
& $z$ & width of each band & $f_{\rm sky} $&  $\la\kappa\kappa\ra$ 
	& $\la\kappa g_{\rm LSST}\ra$ 
	& $\la\kappa \gamma_{\rm LSST}\ra$ \\
& & [MHz] & & & & \\
\hline
SKA1-Low	& $6<z<14$		& 5	& $6.5\times10^{-4}$		
										& 3.6		& 26		& 13 \\
CHIME		& $1.1<z<2.5$		& 25	& 0.5		& 0.25	& 34		& 27	 \\
HIRAX		& $1.35<z<2.5$	& 25	& 0.5		& 0.93	& 45		& 34	 \\
\hline \hline \\
\end{tabular}
\caption{\label{table:sn1}
Total signal to noise for a detection of either the lensing auto power spectrum ($\la\kappa\kappa\ra$) or the cross spectrum between lensing and LSST galaxy clustering ($\la\kappa g_{\rm LSST}\ra$) or cosmic shear ($\la\kappa \gamma_{\rm LSST}\ra$), for the surveys from Secs.~\ref{sec:ska} and~\ref{sec:chime} (see those sections for more details about the specifications we assume). In reality, lack of overlap between CHIME and LSST prevents cross-correlation, but we still perform forecasts in order to have an apples-to-apples comparison with HIRAX. For $\la\kappa\kappa\ra$, we use the noise per $\phi$ mode from the bias-hardened lensing estimator, while for the cross-correlations we use the non-bias-hardened noise, fixing the redshift range of the low-$z$ tracer such that gravitational effects in the source intensity field do not correlate with the tracer. A detection of the lensing auto spectrum will be weak at best in the SKA survey and impossible for CHIME and HIRAX. For all surveys, significant measurements of each cross-correlation may be possible, provided that systematics can be controlled at the appropriate level.
}
\end{table}

In Table~\ref{table:sn1}, we show our forecasts for the total signal to noise on the auto or cross spectra indicated in each column, for the 21cm surveys we have considered in Secs.~\ref{sec:ska} and~\ref{sec:chime}, assuming $f_{\rm sky}\approx 6.5\times 10^{-4}$ ($27\,{\rm deg}^2$) for the SKA1-Low reionization survey and $f_{\rm sky}=0.5$ for CHIME and HIRAX. To obtain these numbers, we first perform forecasts for a selection of redshift bands with widths listed in Table~\ref{table:sn1} (the results for these individual bands can be found in App.~\ref{app:singleband21cm}). We then interpolate these results using a cubic spline, calculate the mean redshifts of adjacent bands that completely cover the survey's redshift range, evaluate the signal to noise  for each band, and sum the results in quadrature. The lower redshift for CHIME and HIRAX indicates where the 21cm fluctuations become noise-dominated.

Even under our optimistic assumptions, the lensing auto spectrum can be detected weakly at best by the SKA survey and not at all by CHIME or HIRAX, using the estimator we have investigated. However, we find much higher signal to noise on the cross-correlations we have considered, for each of SKA, CHIME, and HIRAX (with the obvious caveat that CHIME cannot practically be cross-correlated with LSST due to lack of overlap, but our forecasts for that case would apply to an LSST-like northern survey). For all cross-correlations we consider, the lensing reconstruction noise (including non-Gaussian contributions) is the limiting factor in the overall signal to noise: for example, comparable $\langle\kappa g\rangle$ results could be achieved with a galaxy survey with a number density 20 times lower than LSST (but still covering half the sky).

We again remind the reader that these numbers represent the absolute best-case scenario for application of the lensing estimators in this paper, at the perturbative order we have computed; inevitable real-world systematics will likely degrade these numbers by a factor of a few at least. However, if these surveys are successful at detecting 21cm fluctuations at high significance, the forecasts in Table~\ref{table:sn1} motivate an investigation of lensing reconstruction using those measurements. This would further enhance the cross-correlation science possible between low-redshift 21cm and photometric surveys, adding to other existing applications such as calibration of photometric redshift distributions~\cite{Alonso:2017dgh}. 

For SKA1-Low, the S/N that we compute for $C_L^{\phi\phi}$ is a factor of $\sim$3 lower than it would be if gravitational nonlinearities in the source field were ignored, while for CHIME and HIRAX, the multiplier is at least a factor of 5. This reaffirms that these effects should be included in any lensing reconstruction forecast at these redshifts. 
Note that when we neglect nonlinearities, we find signal to noise values that are consistent with previous forecasts, e.g.~Ref.~\cite{Pourtsidou:2015mia}.

\begin{figure}
\includegraphics[width=\textwidth]{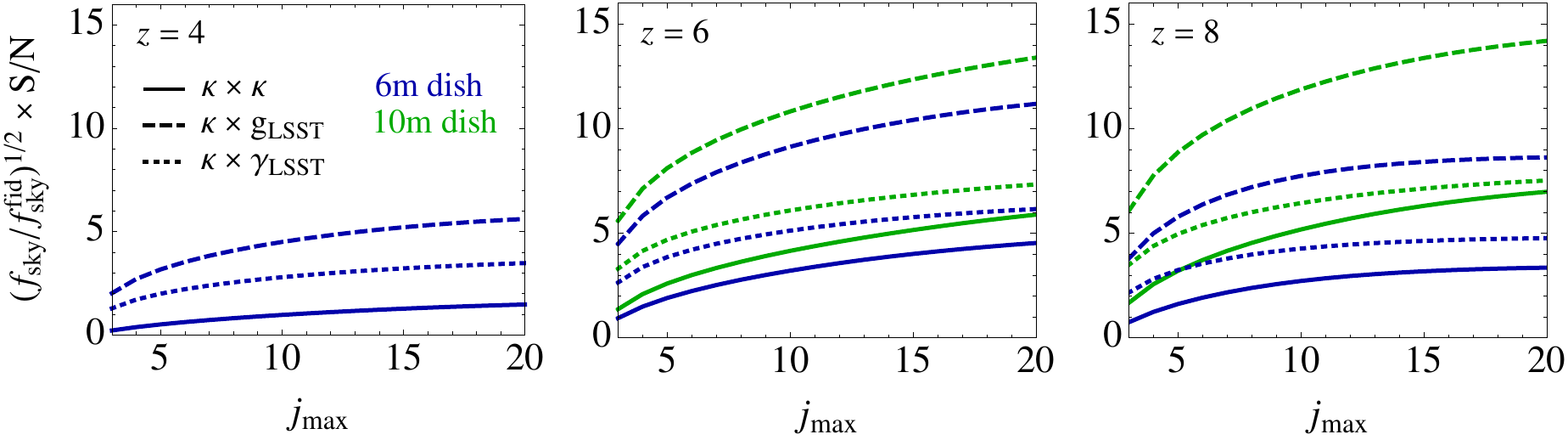}
\caption{\label{fig:ccat-sn}
As Table~\ref{table:sn1}, but computed as a function of $j_{\rm max}$ for our imagined single-dish intensity mapping survey from \sec{sec:ccat}, based on the CII survey recommended in Ref.~\cite{Gong:2011mf}. The 6m-dish version of this survey is planned for the CCAT-prime telescope, but we also forecast for a 10m dish. Higher values of $j_{\rm max}$ are achievable if the source intensity field is detected with lower noise. At the fiducial $f_{\rm sky}$ of $3.9\times10^{-4}$ (16 deg$^2$), the lensing auto spectrum will measured at low significance at best, while cross-correlations with LSST (or COSMOS, not shown) galaxy clustering or cosmic shear show better prospects for a strong detection. Increasing $f_{\rm sky}$ would boost the significance on all auto and cross spectra by $(f_{\rm sky}/f_{\rm sky}^{\rm fid})^{1/2}$, potentially bringing them within reach of strong detections.
}    
\end{figure}

As noted in \sec{sec:ccat}, the amplitudes of an intensity map's signal and noise power spectra for our imagined single-dish survey are very uncertain. In Fig.~\ref{fig:ccat-sn}, we incorporate this uncertainty by plotting the lensing signal to noise as a function of $j_{\rm max}$. We have used a fiducial $f_{\rm sky}$ of $3.9\times10^{-4}$, corresponding to 16 deg$^2$. Under this assumption, we once again find that a strong detection of the lensing auto spectrum within a single band will not be possible, despite the fact that for the highest angular resolutions, bias-hardening actually decreases the noise on $C_L^{\phi\phi}$ by as much as 70\%.

On the other hand, a cross-correlation of lensing from $z\sim 6$ with galaxy clustering from a large galaxy survey looks more promising, for both the 6m- and 10m-dish cases, with potential ${\rm S/N} \gtrsim 10$, and likewise for the 10m dish using lensing at $z\sim 8$. A cross correlation with cosmic shear from lower-redshift galaxies looks slightly less promising. However, as indicated by the figure's $y$-axis label, the signal to noise on any of these measurements scales as~$f_{\rm sky}^{1/2}$. This implies that even increasing $f_{\rm sky}$ by a factor of 4 could potentially bring a detection of the auto spectrum within reach. Furthermore, even with our fiducial $f_{\rm sky}$, we found in Fig.~\ref{fig:ccat} that it may be possible to construct lensing maps that are signal-dominated over a sizable range of multipoles, and these maps would likely be widely useful for cross-correlation science at small scales.

A first detection of lensing of a non-21cm line intensity map would represent a new and exciting science case for line intensity surveys currently being planned, and would nicely complement the more standard motivations for such surveys, which are typically related to star formation and galaxy evolution.

\section{Design considerations for future intensity mapping experiments}
\label{sec:design}

With several intensity mapping surveys underway or in various stages of planning, one can ask which characteristics of this type of survey are most important for lensing reconstruction. We will return our focus to the case of 21cm interferometers, for which a wealth of cross-correlation science will be available if a robust lensing analysis can be performed. For concreteness, we will consider various extensions of HIRAX, taking the ``base" configuration to be that described in \sec{sec:chime}, but with redshift range extended to $2<z<6$. One can imagine improving various properties of the experiment by a factor of 4: decreasing the sky area (which may or may not be a net improvement, due to a tradeoff between lower noise for fixed observing time and a smaller number of observed modes), increasing the observing time, using bigger dishes, or using more dishes.

\begin{table}[t]
\centering
\begin{tabular}{ccc|ccc}
\hline \hline
\multicolumn{6}{c}{S/N on lensing power spectra for SuperHIRAX}\\
\hline	
 $z$ & Experiment & $\ell_{\rm max}$ (derived) & $\la\kappa\kappa\ra$ & $\la\kappa g_{\rm LSST}\ra$ 
	& $\la\kappa \gamma_{\rm LSST}\ra$\\
\hline
$2<z<2.5$	& Base					& 800	& 2.3		& 36		& 27 \\
			& $0.25\times f_{\rm sky}$		& 1100	& 1.7		& 24		& 17 \\
			& $4\times t_0$				& 1100	& 3.4		& 48		& 34 \\
			& $2\times D_{\rm dish}$		& 1100	& 3.5		& 46		& 33 \\
			& $4\times N_{\rm dishes}$	& 1600	& 7.3		& 60		& 49 \\
\hline 
$3.5<z<4$	& Base					& 500	& 0.3		& 19		& 14 \\
			& $0.25\times f_{\rm sky}$		& 800	& 0.4		& 19		& 14 \\
			& $4\times t_0$				& 800	& 0.8		& 38		& 28 \\
			& $2\times D_{\rm dish}$		& 700	& 0.9		& 29		& 22 \\
			& $4\times N_{\rm dishes}$	& 1100	& 3.3		& 61		& 43 \\
\hline 
$5.5<z<6$	& Base					& 300	& $<$0.1	& 5.5		& 4.3 \\
			& $0.25\times f_{\rm sky}$		& 500	& $<$0.1	& 7.1		& 5.4 \\
			& $4\times t_0$				& 500	& 0.1		& 14		& 11 \\
			& $2\times D_{\rm dish}$		& 400	& 0.1		& 9.5		& 7.2 \\
			& $4\times N_{\rm dishes}$	& 800	& 0.6		& 25		& 18 \\
\hline
\hline \\
\end{tabular}
\caption{\label{table:superhirax}
As Table~\ref{table:sn1}, but now considering different extensions of HIRAX capable of making measurements over $2<z<6$. The ``base" configuration is described in \sec{sec:chime}. The $\ell_{\rm max}$ column states the $\ell$ value at which the 21cm signal and noise power spectra cross for each experiment. The largest improvements are to be had by increasing the number of dishes (with other survey specifications fixed), since this adds longer baselines that are more sensitive to lensing, and also decreases the noise on shorter baselines. For a fixed total collecting area, decreasing the dish size and increasing the number of dishes would also yield large gains.
}
\end{table}

In Table~\ref{table:superhirax}, we repeat our previous computations of total signal to noise on lensing auto and cross spectra, but for each version of ``SuperHIRAX" mentioned above, and for observations in three representative redshift bands. We also show an effective derived $\ell_{\rm max}$ for each experiment, determined from where the 21cm signal and noise power spectra cross in each case. We find that the largest gains in S/N are to be had by increasing $\ell_{\rm max}$, which both increases the number of small-scale modes available for use in the lensing reconstruction, and improves our ability to distinguish between lensing- and gravity-induced mode-couplings, lessening the increase in noise per mode that comes from using a bias-hardened estimator.

A few other trends are also visible in Table~\ref{table:superhirax}. At fixed $\ell_{\rm max}$, increasing the observing time and decreasing $f_{\rm sky}$ affect the lensing reconstruction noise in the same way [see Eq.~\eqref{eq:apixtpix}], but decreasing $f_{\rm sky}$ also increases sample variance in the power spectrum estimate, decreasing the overall detection significance.  (Decreasing $f_{\rm sky}$ also increases $\ell_{\rm min}$ in lensing reconstruction, but this has negligible impact on the results if $\ell_{\rm min}\ll \ell_{\rm max}$.)  Meanwhile, increasing the observation time by 4 or the dish diameter by 2 give similar results: the former lowers the thermal noise on all scales, in particular causing a few more small-scale modes to become signal-dominated, while the latter allows access to longer baselines (again translating into more small-scale modes) while increasing the noise on large-scale modes (which are negligibly important for lensing reconstruction). Ultimately, simply increasing the number of dishes (and utilizing all possible correlations between them, which will come at non-negligible computational cost) would allow for the largest improvement in lensing signal-to-noise. If the total collecting area is held fixed, {\em decreasing} the dish size while increasing the number of dishes would be the best direction to pursue, and would also benefit other science cases that come from measuring the 21cm fluctuations themselves.

\section{Conclusions}
\label{sec:conclusions}

In this paper, we have investigated several aspects of how to measure the gravitational lensing of line intensity maps. These maps will generally have angular resolution similar to maps of the CMB, but will be fully 3-dimensional, by virtue of measuring the spatial fluctuations of a redshifted line whose rest frequency is known. Unlike for the CMB, however, the statistics of these maps will generally be non-Gaussian, and these non-Gaussianities can affect the fidelity of the reconstructed lensing maps. We have particularly focused on the impact of gravitational nonlinearities on the measured intensity field, 
using perturbation theory to quantify how this impact varies with source redshift and angular resolution.

Even for measurements of the epoch of reionization, gravitational mode-couplings can significantly affect lensing reconstruction for some planned 21cm surveys, because their angular resolution will be sufficient to probe nonlinearities in the underlying density field (independently of the details of reionization itself). In some cases, our use of low-order perturbation theory imposes a restriction on the range of validity of our calculations, but allows us to retain analytical control of our predictions, and also to devise a ``bias-hardening" technique to remove the leading-order bias for reconstructed lensing maps and their power spectra.  We have identified two types of gravitational contributions that appear when performing lensing reconstruction.  The first originates from the direct response of the small-scale density field to a long-wavelength density mode; we have shown how one could use similar techniques to  reconstruct this long-wavelength mode from the same survey and then remove this contribution from a reconstructed lensing map itself.  The second contribution cannot be removed with this technique, and we show how its magnitude is in fact amplified when one performs the bias-hardening procedure.

We have performed simplistic forecasts for a selection of 21cm surveys: a 27deg$^2$ reionization survey imagined for phase one of SKA-Low, and lower-redshift observations by CHIME and HIRAX. In all cases, we find that a robust detection of the auto spectrum of the corresponding lensing potential will be out of reach.  This  conclusion is strongly affected by gravitational nonlinearities: ignoring these effects results in a forecast signal to noise that is higher by a factor $\sim$3 for SKA-Low, and by a larger factor for CHIME and HIRAX, even when including bias-hardening. On the other hand, it appears that cross-correlations between reconstructed lensing maps and galaxy clustering or galaxy lensing from a large photometric survey (such as LSST, or even something less dense but with large sky coverage) may be within reach. We have also investigated the abilities of a single-dish CII survey, such as planned for the CCAT-prime observatory, and have reached similar conclusions: the lensing auto spectrum will be challenging, but cross-correlations with other low-redshift tracers merit further study. Higher detection significance may be possible if more strongly nonlinear scales can be used, either by extending our calculation to higher order in perturbation theory, or by using other lensing estimators calibrated on simulations (e.g.~\cite{Lu:2007pk,Lu:2009je}).

In addition, we have performed forecasts for various extensions of HIRAX, as a quantitative check on our intuition about how various survey properties affect our ability to detect lensing. We have found that the single largest improvement arises from simply adding more dishes and utilizing all correlations between them: the addition of longer baselines increases $\ell_{\rm max}$ and therefore the number of small-scale modes available, while the addition of more redundant short baselines should reduce the noise on measurements of modes already accessible within the base HIRAX configuration.

In this work, we have focused exclusively on using intensity measurements for lensing reconstruction. However, it is well-known for the CMB that measurements of polarization also contain valuable information about lensing, and indeed, the precision of future CMB lensing measurements will be driven largely by this information (e.g.~\cite{Abazajian:2016yjj}). Therefore, as mentioned briefly by Ref.~\cite{Zahn:2005ap}, one could also ask about using the polarization of line intensity maps in this way. For 21cm surveys, polarization will be generated by the quadrupole seen by remote scatterers during reionization, in the same way as in the CMB~\cite{Babich:2005sb}. It has been pointed out that this polarization will undergo large amounts of Faraday rotation at the relevant frequencies, caused by galactic and extragalactic magnetic fields~\cite{De:2013wca}, and there will also be large polarized foregrounds that will further obscure the signal of interest. For other lines, it may be interesting to investigate the feasibility of measuring polarization and the lensing thereof, since, for example, Faraday rotation will generally be much less of an issue at the corresponding (much shorter) wavelengths. We leave this for future work.

We conclude by mentioning several avenues which would be useful to pursue in future studies:
\begin{itemize}
\item The gravitational mode-couplings discussed in this work will also affect attempts to reconstruct the effect of lensing on the Lyman-$\alpha$ forecast, studied recently in Refs.~\cite{Croft:2017tur,Metcalf:2017qty}. It would be interesting to quantify these effects.
\item Tracers of the density field are typically treated using a bias expansion on quasi-linear scales (e.g.~\cite{Desjacques:2016bnm}), and line intensity maps will typically be measuring the aggregate emission from these tracers. Nonlinear terms in the bias expansion will induce mode-couplings similar to those we have discussed, and these can be explored in much the same way as we have done. Weakly nonlinear effects from redshift-space distortions can likewise be explored using perturbation theory, albeit with the extra complication that effective field theory counterterms will be needed in even the tree-level four-point function~\cite{Senatore:2014vja,Lewandowski:2015ziq,delaBella:2017qjy}. At higher redshifts, the topology of reionization will need to be modeled in order to design the appropriate filters in the lensing estimator. Recent studies of simulations have indicated that a bias expansion may be valid even during reionization~\cite{Hoffmann:2018clb}, which would simplify the modeling somewhat, but in any case, further study is required.
\item At sufficiently small scales, the expansion in $\nabla\phi$ used to derive the quadratic lensing estimators breaks down, but likelihood-based methods~\cite{Hirata:2002jy} or other techniques (e.g.~\cite{Mandel:2005xh}) can circumvent this limitation. It would be useful to precisely determine the error arising from higher-order corrections in $\nabla\phi$ for different intensity mapping setups, and assess the performance of the available alternative methods for lensing reconstruction in these setups.
\item Our  investigation in \sec{sec:tidal} of reconstructing long-wavelength modes of the density field supports findings elsewhere in the literature (e.g.~\cite{Pen:2012ft,Zhu:2015zlh,Zhu:2016esh}) that this could be a very promising technique to apply to both low- and high-redshift surveys. This certainly warrants continued study. 
\item As mentioned in the introduction, curl modes  of lensing (as opposed to the familiar gradient modes sourced by scalar perturbations at linear order) have been considered as a probe of gravitational waves (e.g.~\cite{Dodelson:2003bv,Book:2011dz,Sheere:2016yqu}). However, similar to the mode-couplings studied here, tidal effects would induce mode-couplings that can mimic the curl lensing signal. Decaying gravitational waves induce a curl lensing signal that is relatively small and, similar to the gravitational wave signal itself, a steep function of scale~\cite{2012PhRvD..86h3527S}. At accessible scales, tidal effects can therefore potentially be much more important and lead to a ``fossil" imprint on the large scale structure {\it that does not decay}~\cite{Masui:2010cz,2014PhRvD..89h3507S,Masui:2017fzw}. It is possible that bias-hardening techniques can be applied to separate the tidal effects from the lensing effects for curl modes, and we hope to explore this in future work.
\end{itemize}

\acknowledgments

We wish to thank Marcelo Alvarez, Philippe Berger, Patrick Breysse, Emanuele Castorina, Anthony Challinor, Simone Ferraro, Alex Hall, Viswesh Marthi, Kavilan Moodley, Yuuki Omori, Ue-Li Pen, Emmanuel Schaan, Fabian Schmidt, Uro\v{s} Seljak, Richard Shaw, and An\v{z}e Slosar for useful discussions. We thank Emanuele Castorina for providing us with numerical results from Ref.~\cite{Castorina:2016bfm}. Computations were performed on the GPC supercomputer at the SciNet HPC Consortium~\cite{1742-6596-256-1-012026}. SciNet is funded by: the Canada Foundation for Innovation under the auspices of Compute Canada; the Government of Ontario; Ontario Research Fund - Research Excellence; and the University of Toronto. This document uses the \texttt{simpler-wick} {\LaTeX} package; we thank Joshua Ellis for making this package available at~\href{https://github.com/JP-Ellis/simpler-wick/}{https://github.com/JP-Ellis/simpler-wick/}.
J.M. and A.v.E. were supported by the Vincent and Beatrice Tremaine Fellowship. P.D.M. acknowledges support from a Senior Kavli Fellowship at the University of Cambridge and support from the Netherlands organization for scientific research (NWO) VIDI grant (dossier 639.042.730).

\appendix
\section{Perturbation theory kernels up to third order}
\label{app:kernels}

In this work, we use make use of large-scale structure perturbation theory up to third order in the linear overdensity, neglecting effective field theory corrections because they do not enter in tree-level computations. The symmetrized second- and third-order density kernels $\kerfs{2}$ and $\kerfs{3}$ can be obtained from recurrence relations found e.g.\ in Ref.~\cite{Bernardeau:2001qr}, but for convenience, we list them below.

The symmetrized second-order kernel is given directly by.
\beq
\kerfs{2}(\vk_1,\vk_2)
	= \frac{5}{7} + \frac{1}{2} \lp \frac{k_1}{k_2}+\frac{k_2}{k_1} \rp \hat{\vk}_1\cdot\hat{\vk}_2
	+ \frac{2}{7} \lp \hat{\vk}_1\cdot\hat{\vk}_2 \rp^2 \ ,
	\label{eq:f2s}
\eeq
where $\hat{\vk}_i \equiv \vk_i/k_i$. The {\em unsymmetrized} third-order kernel $F_3$ is given by
\begin{align*}
F_3(\vk_1,\vk_2,\vk_3) &= \frac{1}{18} \alpha(\vk_1,\vk_2+\vk_3) 
	\lb 5\alpha(\vk_2,\vk_3) + 2\beta(\vk_2,\vk_3) \rb \\
&\quad + \frac{1}{63} \beta(\vk_1,\vk_2+\vk_3) \lp 3\alpha(\vk_2,\vk_3) + 4\beta(\vk_2,\vk_3) \rp \\
&\quad + \frac{1}{126} \lb 3\alpha(\vk_1,\vk_2) + 4\beta(\vk_1,\vk_2) \rb
	\lb 7\alpha(\vk_1+\vk_2,\vk_3) + 2\beta(\vk_1+\vk_2,\vk_3) \rb\ ,
	\numberthis
\end{align*}
where
\beq
\alpha(\vk_1,\vk_2) \equiv \frac{\vk_1\cdot(\vk_1+\vk_2)}{k_1^2}\ ,
	\quad \beta(\vk_1,\vk_2) \equiv \frac{\vk_1\cdot\vk_2 \, |\vk_1+\vk_2|^2}{2k_1^2 k_2^2}\ ;
\eeq
the symmetrized kernel $\kerfs{3}$ is then obtained by symmetrizing $F_3(\vk_1,\vk_2,\vk_3)$ over permutations of its arguments.

\section{Contribution from ``$N^{(1)}$ bias''}
\label{app:n1bias}

\begin{figure}[t]
\includegraphics[width=\textwidth]{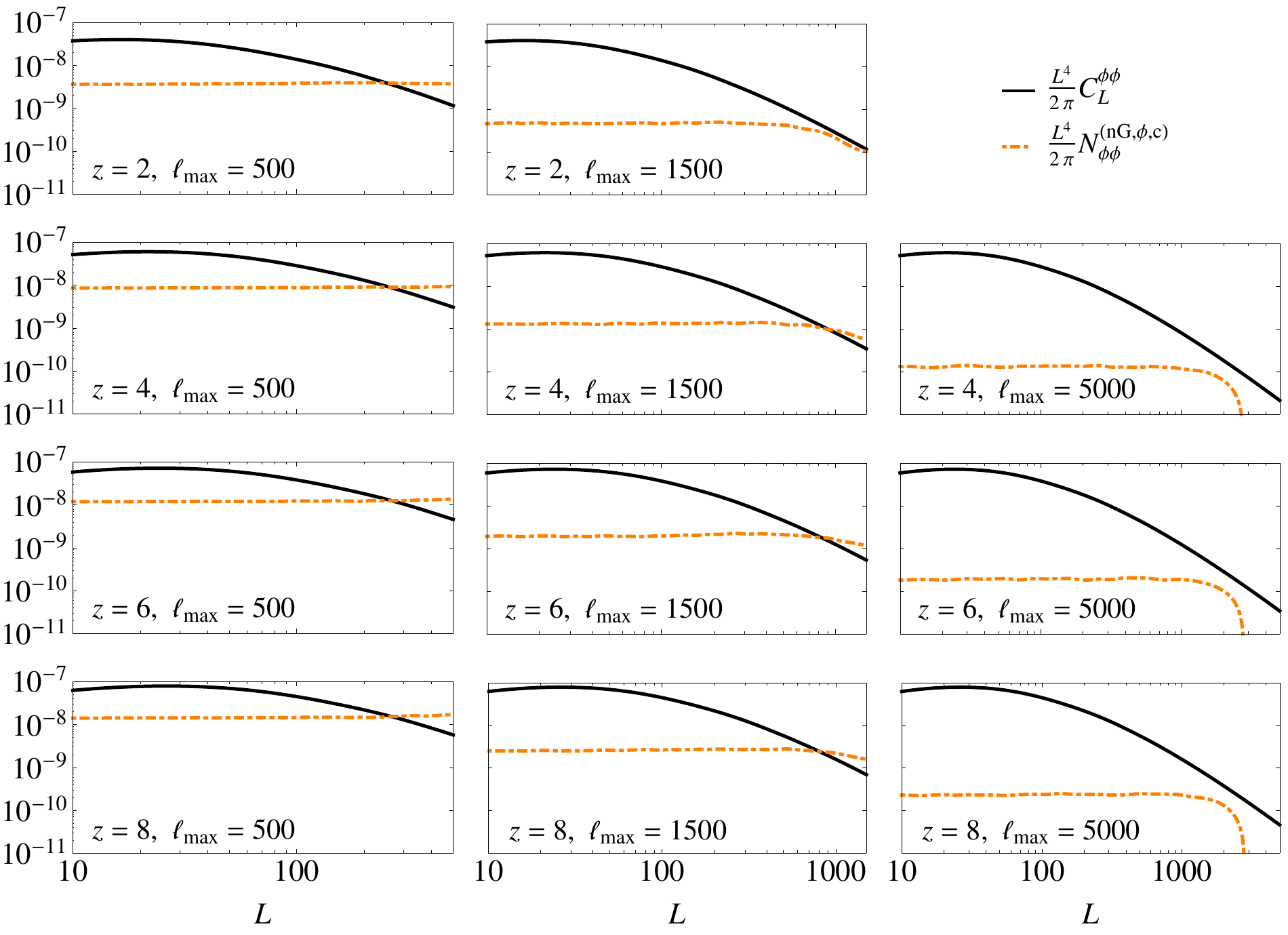}
\caption{\label{fig:singlej-n1bias}
As Fig.~\ref{fig:singlej-noiseterms}, but comparing the lensing potential power spectrum to the $N_{\phi\phi}^{({\rm nG,}\phi,{\rm c})}$ term (an integral over $C_L^{\phi\phi}$ and other factors, analogous to $N^{(1)}$ from CMB lensing). The high-$L$ downturns seen in some panels are the result of chance cancellations in the integrand at certain values of $\ell_{\rm max}$ and $\kpar$; generically, $N_{\phi\phi}^{({\rm nG,}\phi,{\rm c})}$ scales like $L^{-4}$. This term is generally much smaller than the non-Gaussian contributions from gravity investigated in the main text of this work, and does not correlate lensing estimations using modes with different $\kpar$ values. Thus, it is safe to neglect it in the forecasts we perform.
}    
\end{figure}

In \sec{sec:quadrevisited}, we find that the variance of the $\hat{\phi}$ estimator (i.e.~$\la \hat{\phi} \hat{\phi} \ra$) contains a term analogous to what is called ``$N^{(1)}$ bias" in CMB lensing~\cite{Kesden:2003cc}. This term involves an integral over $C_L^{\phi\phi}$ and several factors of the source power spectrum. In our conventions, we denote this term by $N_{XY}^{({\rm nG,}\phi,{\rm c})}$, and we find from the derivation in App.~\ref{app:grav-derivations} that it is given by
\begin{align*}
N_{\phi\phi}^{({\rm nG,}\phi,{\rm c})}(L,\kpar) &\equiv
	\int_{\vl_1} \int_{\vl_2} g_{\phi}(\vl_1,\vL-\vl_1,\kpar) g_\phi(\vl_2,\vL-\vl_2,\kpar)
	 C_{|\vl_1-\vl_2|}^{\phi\phi} \\
&\qquad\qquad \times
	f_\phi(\vl_1,-\vl_2,\kpar,\kpar) f_\phi(\vL-\vl_1,-\vL+\vl_2,\kpar,\kpar)\ .
	\numberthis 
\end{align*}
In Fig.~\ref{fig:singlej-n1bias}, we plot this term at the same redshifts and $\kpar$ values as in Fig.~\ref{fig:singlej-noiseterms}, along with $C_L^{\phi\phi}$. We find that, like $N_{\phi\phi}^{\rm (G)}$, this term scales roughly like $L^{-4}$ and $\ell_{\rm max}^{-2}$. (The high-$L$ downturns seen in some panels in Fig.~\ref{fig:singlej-n1bias} are a result of chance cancellations in the integral at specific $\ell_{\rm max}$ and $\kpar$ values -- for other $\kpar$ values than what we have plotted, we find an almost exact $L^{-4}$ scaling.)

At fixed $\kpar$, Fig.~\ref{fig:singlej-n1bias} shows that this term is generally much smaller than the other non-Gaussian contributions plotted in Fig.~\ref{fig:singlej-noiseterms}, except possibly at higher $L$. Furthermore, unlike the other non-Gaussian contributions explored in this work, $N_{\phi\phi}^{({\rm nG,}\phi,{\rm c})}$ does not correlate $\phi$ estimators evaluated at different values of $\kpar$; in other words,
\beq
\left\la \hat{\phi}(\vL,\kparone) \hat{\phi}^*(\vL,\kpartwo) \right\ra
	\supset \kro_{\kparone,\kpartwo}  N_{XY}^{({\rm nG,}\phi,{\rm c})}(L,\kparone)\ .
\eeq
This implies that when the lensing estimators for different $\kpar$ values are optimally combined, as in Eqs.~\eqref{eq:ncombng} and~\eqref{eq:noisecovmat}, the effect of this term will effectively average down like $j_{\rm max}^{-1}$, rendering it far subdominant to the contributions from gravitational mode-couplings. For high-precision observations, the bias incurred by this term will need to be modeled and subtracted away, but for the purposes of the forecasts we perform here, it is safe to neglect it.

\section{Derivation of contributions to variance of lensing estimator}
\label{app:grav-derivations}

In this appendix, we derive the various non-Gaussian contributions to the covariance of two estimators $\hat{X}$ and $\hat{Y}$ where $X,Y \in\{\phi,\delta\}$, given earlier in Eqs.~\eqref{eq:covxyfull} to~\eqref{eq:nxyngc}. This involves classifying the terms appearing in the four-point function of $\Iobs$ at different orders in $\phi$ and $\delta_1$, and then applying the double filter from Eq.~\eqref{eq:ngcov} to each term. We will make use of the following shorthand:
\beq
P[\vl,k] \equiv P_{\delta1}\!\lp \sqrt{\frac{\ell^2}{\chi^2}+k^2} \rp ,
\quad (\vl_i,k) \equiv (\ell_{i1}/\chi,\ell_{i2}/\chi,k)\ .
\label{eq:appendix-shorthands}
\eeq

\subsection{$\mathcal{O}(\phi^2 \delta_1^4)$ terms}

We begin with terms that arise from picking two factors of $\Iobs$ in the relevant four-point function and expanding each of them to $\mathcal{O}(\phi^1\delta_1^1)$, using
\beq
\Iobs(\vl,\kpar) \supset b\calL^{-1} \chi^{-2} \delta_1(\vl/\chi,\kpar)
	- b\calL^{-1}\chi^{-2} \int_{\vl'} \vl'\cdot(\vl-\vl') \phi(\vl-\vl') \delta_1(\vl'/\chi,\kpar)\ .
	\label{eq:id1phi1}
\eeq
The four $\delta_1$ factors must contract together using Wick's theorem, while the two $\phi$ factors must lie in distinct Wick contractions of the $\delta_1$ factors, or else they contribute to the disconnected four-point function, which has already been fully accounted for by the $N_{XY}^{\rm (G)}$ term. Thus, we can derive the relevant terms starting with the following grouping of factors in the four-point function,
\begin{align*}
&\left\la \Iobs(\vl_1,\kparone) \Iobs(\vL_1-\vl_1,-\kparone) 
	\Iobs(-\vl_2,-\kpartwo) \Iobs(-\vL_2+\vl_2,\kpartwo) \right\ra_{\rm c} \\
&\qquad \supset \left\la \left\la \Iobs(\vl_1,\kparone) \Iobs(\vL_1-\vl_1,-\kparone) \right\ra_\delta
	 \left\la \Iobs(-\vl_2,-\kpartwo) \Iobs(-\vL_2+\vl_2,\kpartwo) \right\ra_\delta \right\ra_{\phi} \\
&\qquad\quad + \left\la \left\la \Iobs(\vl_1,\kparone)  \Iobs(-\vl_2,-\kpartwo) \right\ra_\delta
	 \left\la \Iobs(\vL_1-\vl_1,-\kparone) \Iobs(-\vL_2+\vl_2,\kpartwo)  \right\ra_\delta \right\ra_{\phi} \\
&\qquad\quad + \left\la \left\la \Iobs(\vl_1,\kparone)  \Iobs(-\vL_2+\vl_2,\kpartwo) \right\ra_\delta
	 \left\la \Iobs(\vL_1-\vl_1,-\kparone) \Iobs(-\vl_2,-\kpartwo)  \right\ra_\delta \right\ra_{\phi} 	 
	\ , \numberthis \label{eq:i4ptclphiphi}
\end{align*}
substituting the first and second terms of Eq.~\eqref{eq:id1phi1} for either $\Iobs$ in each $\la\cdot\ra_\delta$ grouping, and evaluating the relevant contractions over $\delta_1$ and $\phi$. Since the double integral with $g_X(\vl_1,\vL_1-\vl_1,\kparone) g_Y(\vl_2,\vL_2-\vl_2,\kpartwo)$ from Eq.~\eqref{eq:ngcov} is invariant under $\vl_i\leftrightarrow\vL_i-\vl_i$, the last two lines above are equal except for the substitution $\kpartwo \leftrightarrow -\kpartwo$.

In the first line of Eq.~\eqref{eq:i4ptclphiphi}, all four choices of where to put the two $\phi$ insertions are equivalent by changes of variables in Eq.~\eqref{eq:ngcov} and by invariance of $g_X$ under $\kpar\to-\kpar$. We can evaluate each of the two $\la\cdot\ra_\delta$ groupings separately first:
\begin{align*}
&2 \left\la \Iobs(\vl_1,\kparone) \Iobs(\vL_1-\vl_1,-\kparone) \right\ra_\delta \\
&\qquad \to 2\left\la -b^2\calL^{-2} \chi^{-4} \int_{\vl'} \delta_1(\vl_1/\chi,\kparone) \vl'\cdot(\vL_1-\vl_1-\vl')
	\phi(\vL_1-\vl_1-\vl') \delta_1(\vl'/\chi,-\kparone) \right\ra_\delta \\
&\qquad = -2b^2\calL^{-2} \chi^{-4} \int_{\vl'} \vl'\cdot(\vL_1-\vl_1-\vl') \phi(\vL_1-\vl_1-\vl')
	\calL \chi^2  (2\pi)^2\dirac(\vl_1+\vl') 
	P[\vl_1,\kparone] \\
&\qquad = 2 \vl_1\cdot \vL_1 \phi(\vL_1) C_{\ell_1}(\kparone) \\
&\qquad = \vL_1\cdot \lb \vl_1 C_{\ell_1}(\kparone)
	+ \lp \vL_1-\vl_1\rp C_{|\vL_1-\vl_1|}(\kparone) \rb \phi(\vL_1) \\
&\qquad = f_\phi(\vl_1,\vL_1-\vl_1,\kparone,\kparone) \phi(\vL_1) \ ,
	\numberthis
\end{align*}
where in the fourth we used the definition of $C_\ell$ from Eq.~\eqref{eq:clidef}, also absorbing $b^2$ as mentioned in the main text, and in the fourth equality we used invariance of the $g_X g_Y$ integral under $\vl_i\leftrightarrow\vL_i-\vl_i$ to symmetrize the expression, to match the definition of $f_\phi$ in Eq.~\eqref{eq:fphidef2}. Doing the same for the second $\la\cdot\ra_\delta$ grouping, we can write
\begin{align*}
&\left\la \left\la \Iobs(\vl_1,\kparone) \Iobs(\vL_1-\vl_1,-\kparone) \right\ra_\delta
	 \left\la \Iobs(-\vl_2,-\kpartwo) \Iobs(-\vL_2+\vl_2,\kpartwo) \right\ra_\delta \right\ra_{\phi} \\
&\qquad \to f_\phi(\vl_1,\vL_1-\vl_1,\kparone,\kparone)
	f_\phi(-\vl_2,-\vL_2+\vl_2,\kpartwo,\kpartwo) \left\la \phi(\vL_1) \phi(-\vL_2)  \right\ra_{\phi} \\
&\qquad = (2\pi)^2 \dirac(\vL_1-\vL_2)  C_{L_1}^{\phi\phi} 
	\, f_\phi(\vl_1,\vL_1-\vl_1,\kparone,\kparone) f_\phi(-\vl_2,-\vL_2+\vl_2,\kpartwo,\kpartwo)\ .
	\numberthis
\end{align*}

Moving on to the second line of Eq.~\eqref{eq:i4ptclphiphi}, the first $\la\cdot\ra_\delta$ grouping can be evaluated like so:
\begin{align*}
&\left\la \Iobs(\vl_1,\kparone)  \Iobs(-\vl_2,-\kpartwo) \right\ra_\delta \\
&\qquad \to \left\la -b^2\calL^{-2} \chi^{-4} \int_{\vl'} \delta_1(\vl_1/\chi,\kparone) \vl'\cdot(-\vl_2-\vl')
	\phi(-\vl_2-\vl') \delta_1(\vl'/\chi,-\kpartwo) \right\ra_\delta \\
&\qquad\quad + \left\la -b^2\calL^{-2} \chi^{-4} \int_{\vl'} \vl'\cdot(\vl_1-\vl') \phi(\vl_1-\vl') 
	\delta_1(\vl'/\chi,\kparone) \delta_1(-\vl_2/\chi,-\kpartwo) \right\ra_\delta \\
&\qquad = -b^2\calL^{-2}\chi^{-4} \int_{\vl'} \vl'\cdot(-\vl_2-\vl')  \phi(-\vl_2-\vl') 
	\chi^2 (2\pi)^2\dirac(\vl_1+\vl') \calL \kro_{\kparone,\kpartwo} 
	P[\vl_1,\kparone] \\
&\qquad\quad -b^2\calL^{-2}\chi^{-4} \int_{\vl'} \vl'\cdot(\vl_1-\vl')  \phi(\vl_1-\vl') 
	\chi^2 (2\pi)^2\dirac(-\vl_2+\vl') \calL \kro_{\kparone,\kpartwo} 
	P[\vl_2,\kpartwo] \\ 
&\qquad = (\vl_1-\vl_2) \cdot 
	\lb \vl_1 C_{\ell_1}(\kparone) - \vl_2 C_{\ell_2}(\kpartwo) \rb
	\kro_{\kparone,\kpartwo}  \phi(\vl_1-\vl_2) \\
&\qquad = f_\phi(\vl_1,-\vl_2,\kparone,\kpartwo)  \kro_{\kparone,\kpartwo}  \phi(\vl_1-\vl_2)\ ,\	\numberthis
\end{align*}
while the second grouping can be similarly evaluated as
\begin{align*}
&\left\la \Iobs(\vL_1-\vl_1,-\kparone) \Iobs(-\vL_2+\vl_2,\kpartwo)  \right\ra_\delta \\
&\qquad =  f_\phi(\vL_1-\vl_1,-\vL_2+\vl_2,\kparone,\kpartwo)  
	\kro_{\kparone,\kpartwo}  \phi(\vL_1-\vL_2-\vl_1+\vl_2)\ ,
	\numberthis
\end{align*}
leading to
\begin{align*}
&\left\la \left\la \Iobs(\vl_1,\kparone)  \Iobs(-\vl_2,-\kpartwo) \right\ra_\delta
	 \left\la \Iobs(\vL_1-\vl_1,-\kparone) \Iobs(-\vL_2+\vl_2,\kpartwo)  \right\ra_\delta \right\ra_{\phi} \\
&\qquad \to 
	f_\phi(\vl_1,-\vl_2,\kparone,\kpartwo) f_\phi(\vL_1-\vl_1,-\vL_2+\vl_2,\kparone,\kpartwo)
	\kro_{\kparone,\kpartwo}
	 \left\la \phi(\vl_1-\vl_2) \phi(\vL_1-\vL_2-\vl_1+\vl_2) \right\ra_\phi \\
&\qquad = (2\pi)^2 \dirac(\vL_1-\vL_2) \,  C_{|\vl_1-\vl_2|}^{\phi\phi}  
	\, \kro_{\kparone,\kpartwo} 
	f_\phi(\vl_1,-\vl_2,\kparone,\kpartwo) f_\phi(\vL_1-\vl_1,-\vL_2+\vl_2,\kparone,\kpartwo)\ .
	\numberthis
\end{align*}
The third line of Eq.~\eqref{eq:i4ptclphiphi} will be the same as this, but with the $\kpartwo\to-\kpartwo$. Since we always take $\kparone,\kpartwo>0$, the Kronecker delta $\kro_{\kparone,-\kpartwo}$ will then cause this term to vanish.

Putting everything together, we end up with
\begin{align*}
&\left\la \Iobs(\vl_1,\kparone) \Iobs(\vL_1-\vl_1,-\kparone) 
	\Iobs(-\vl_2,-\kpartwo) \Iobs(-\vL_2+\vl_2,\kpartwo) \right\ra_{\rm c} \\
&\qquad \supset (2\pi)^2 \dirac(\vL_1-\vL_2)
	\lb  C_{L_1}^{\phi\phi} 
	\, f_\phi(\vl_1,\vL_1-\vl_1,\kparone,\kparone) f_\phi(-\vl_2,-\vL_2+\vl_2,\kpartwo,\kpartwo) \right. \\
&\qquad\quad +  \left.  \kro_{\kparone,\kpartwo}    C_{|\vl_1-\vl_2|}^{\phi\phi}  
	f_\phi(\vl_1,-\vl_2,\kparone,\kpartwo) f_\phi(\vL_1-\vl_1,-\vL_2+\vl_2,\kparone,\kpartwo) \rb\ .
	\numberthis
\end{align*}
Plugging this into the integral in Eq.~\eqref{eq:ngcov}, we find
\beq
{\rm Cov}_{\rm nG}\!\lb \hat{X}(\vL,\kparone),\hat{Y}^*(\vL,\kpartwo) \rb \supset
	N_{XY}^{({\rm nG,}\phi)}(L,\kparone,\kpartwo)
	+ \kro_{\kparone,\kpartwo}  N_{XY}^{({\rm nG,}\phi,{\rm c})}(L,\kparone)\ ,
\eeq
where
\beq
N_{XY}^{({\rm nG,}\phi)}(L,\kparone,\kpartwo)
	= \frac{N_{XX}^{\rm (G)}(L,\kparone) N_{YY}^{\rm (G)}(L,\kpartwo)}
	{N_{X\phi}^{\rm (G)}(L,\kparone)N_{Y\phi}^{\rm (G)}(L,\kpartwo)} C_{L}^{\phi\phi}
\eeq
and
\begin{align*}
N_{XY}^{({\rm nG,}\phi,{\rm c})}(L,\kpar) &\equiv
	\int_{\vl_1} \int_{\vl_2} g_X(\vl_1,\vL-\vl_1,\kpar) g_Y(\vl_2,\vL-\vl_2,\kpar)
	 C_{|\vl_1-\vl_2|}^{\phi\phi} \\
&\qquad\qquad \times
	f_\phi(\vl_1,-\vl_2,\kpar,\kpar) f_\phi(\vL-\vl_1,-\vL+\vl_2,\kpar,\kpar)\ .
	\numberthis \label{eq:nxyphiphi}
\end{align*}
As mentioned in the main text, Eq.~\eqref{eq:nxyphiphi}  is analogous to what is commonly called $N_L^{(1)}$ in CMB lensing~\cite{Kesden:2003cc}. It will be far subdominant to the other non-Gaussian terms we consider, and therefore we have omitted it in Eq.~\eqref{eq:covxyfull} and in our numerical results, but have included it here for completeness. As also mentioned above, there are other $\mathcal{O}(\phi^2 \delta_1^4)$ terms that will contribute to the disconnected four-point function of $\Iobs$. Such terms, and higher-order terms that also contribute to the disconnected four-point function, will automatically be included in $N_{XY}^{\rm (G)}$ provided that the $C_\ell^{\rm tot}(\kpar)$ or $P_\delta$ functions used in the filters are the nonlinear, lensed versions of the respective power spectra.

\subsection{$\mathcal{O}(\phi^0 \delta_1^6)$ terms with $\delta_2$}

This contribution arises from terms with two factors of $\kerfs{2}$ when the following expansion is inserted into the $\Iobs$ four-point function:
\beq
\Iobs(\vl,\kpar) \supset b\calL^{-1}\chi^{-2} \delta_1(\vl/\chi,\kpar)
+ b\calL^{-1} \chi^{-2} \left. \int_{\vq} \kerfs{2}(\vq,\vk-\vq) \delta_1(\vq) \delta_1(\vk-\vq)
	\right|_{\vk=(\vl/\chi,\kpar)}\ .
\eeq
These terms are easiest to identify if we start with the $\mathcal{O}(F_2^2)$ terms in the tree-level density trispectrum $T$, defined by
\beq
\left\la \delta(\vk_1) \delta(\vk_2) \delta(\vk_3) \delta(\vk_4) \right\ra
	= (2\pi)^3 \dirac(\vk_1+\vk_2+\vk_3+\vk_4) T(\vk_1,\vk_2,\vk_3,\vk_4)\ .
\eeq
If we use shorthand notation where $i$ represents a 3d wavevector $\vk_i$, $ij$ represents $\vk_i+\vk_j$, $P_i$ represents $P(\vk_i)$, and $F$ represents $\kerfs{2}$, the relevant terms in $T(\vk_1,\vk_2,\vk_3,\vk_4) \equiv T(1,2,3,4)$ are (e.g.~\cite{Bertolini:2015fya})
\begin{align*}
T(1,2,3,4) \supset\, &4 \lb P_1 P_2 \lp F(-1,13)F(-2,24)P_{13}
	+ F(-2,23)F(-1,14)P_{23} \rp \right. \\
& + P_1P_3 \lp F(-1,12)F(-3,34)P_{12} + F(-3,32)F(-1,14)P_{32} \rp \\
& + P_1P_4 \lp F(-1,13)F(-4,42)P_{13} + F(-4,43)F(-1,12)P_{43} \rp \\
& + P_2P_3 \lp F(-2,21)F(-3,34)P_{21} + F(-3,31)F(-2,24)P_{31} \rp \\
& + P_2P_4 \lp F(-2,21)F(-4,43)P_{21} + F(-4,41)F(-2,23)P_{41} \rp  \\
& \left.+ \, P_3P_4 \lp F(-3,31)F(-4,42)P_{31} + F(-4,41)F(-3,32)P_{41} \rp \rb\ .
\numberthis \label{eq:f2tri}
\end{align*}
In the trispectrum present in Eq.~\eqref{eq:ngcov}, the $\vk_i$ vectors should be substituted with
\beq
1\to(\vl_1,\kparone)\ , \quad 2\to(\vL_1-\vl_1,-\kparone)\ ,
\quad 3\to(-\vl_2,-\kpartwo)\ , \quad 4\to(-\vL_2+\vl_2,\kpartwo)\ ,
	\label{eq:ksubs}
\eeq
using the notation from Eq.~\eqref{eq:appendix-shorthands}.
We can these vector substitutions to identify three groups of terms that will each simplify together:
\begin{enumerate}
\item {\bf Terms with $P_{12}$ or $P_{34}$:} since $P_{12} = P[\vL_1,0]$ and $P_{34}=P[-\vL_2,0]$, and the trispectrum will come with an overall factor of $\dirac(\vL_1-\vL_2)$, these terms naturally group together:
\begin{align*}
&4P[\vL_1,0] \times \lb  P_1P_3  F(-1,12)F(-3,34) +  P_1P_4 F(-4,43)F(-1,12) \right. \\
&\qquad\qquad\quad + \left. P_2P_3  F(-2,21)F(-3,34) + P_2P_4  F(-2,21)F(-4,43) \rb \\
&= 4P[\vL_1,0] \times \lb F(-1,12) P_1 + F(-2,21)P_2 \rb
	\times \lb F(-3,34) P_3 +  F(-4,43) P_4 \rb \\
&= 4 P[\vL_1,0] \\
&\qquad\times 
	\lb F(-(\vl_1,\kparone),(\vL_1,0)) P[\vl_1,\kparone] 
	+ F(-(\vL_1-\vl_1,-\kparone),(\vL_1,0)) P[\vL_1-\vl_1,-\kparone] \rb \\
&\qquad\times
	\lb F(-(-\vl_2,-\kpartwo),(-\vL_2,0)) P[\vl_2,\kpartwo] \right. \\
&\qquad\qquad\left. +\, F(-(-\vL_2+\vl_2,\kpartwo),(-\vL_2,0)) P[-\vL_2+\vl_2,\kpartwo]   \rb\ .
	\numberthis
\end{align*}
Recalling the definition of $f_\delta$ from Eq.~\eqref{eq:fdeltadef}, 
we can write this expression as
\beq
\calL^4 \chi^{8} P[\vL_1,0] f_\delta(\vl_1,\vL_1-\vl_1,\kparone,-\kparone) 
	f_\delta(-\vl_2,-\vL_2+\vl_2,-\kpartwo,\kpartwo)\ .
\eeq
Finally, $f_\delta$ is invariant under parity, so we can write
\beq
\calL^4 \chi^{8} P[\vL_1,0] f_\delta(\vl_1,\vL_1-\vl_1,\kparone,-\kparone) 
	f_\delta(\vl_2,\vL_2-\vl_2,\kpartwo,-\kpartwo)\ .
\eeq
\item {\bf Terms with $P_{13}$ or $P_{24}$:} since $P_{13}=P_{24}=P[\vl_1-\vl_2,\kparone-\kpartwo]$, we can group these terms as follows:
\begin{align*}
&4P[\vl_1-\vl_2,\kparone-\kpartwo] \times \lb P_1 P_2  F(-1,13)F(-2,24) + P_1P_4  F(-1,13)F(-4,42) \right. \\
&\qquad\qquad\qquad\qquad\qquad\quad \left. +\, P_2P_3 F(-3,31)F(-2,24) + P_3P_4  F(-3,31)F(-4,42) \rb \\
&= 4P[\vl_1-\vl_2,\kparone-\kpartwo] \times \lb F(-1,13)P_1 + F(-3,31) P_3 \rb  
	\times \lb F(-2,24) P_2 + F(-4,42)P_4   \rb \\
&= 4P[\vl_1-\vl_2,\kparone-\kpartwo] \\
&\qquad\times
	\lb F(-(\vl_1,\kparone),(\vl_1-\vl_2,\kparone-\kpartwo)) P[\vl_1,\kparone] \right. \\
&\qquad\qquad\left. 
	+\, F(-(-\vl_2,-\kpartwo),(\vl_1-\vl_2,\kparone-\kpartwo)) P[-\vl_2,-\kpartwo] \rb \\
&\qquad\times
	\lb  F(-(\vL_1-\vl_1,-\kparone),(-\vl_1+\vl_2,-\kparone+\kpartwo)) P[\vL_1-\vl_1,\kparone] \right. \\
&\qquad\qquad\left.
	+\,  F(-(-\vL_2+\vl_2,\kpartwo),(-\vl_1+\vl_2,-\kparone+\kpartwo)) P[-\vL_2+\vl_2,\kpartwo] \rb \\
&= \calL^4 \chi^{8} P[\vl_1-\vl_2,\kparone-\kpartwo] 
	f_\delta(\vl_1,-\vl_2,\kparone,-\kpartwo) f_\delta(\vL_1-\vl_1,-\vL_2+\vl_2,-\kparone,\kpartwo)\ .
	\numberthis
\end{align*}
\item {\bf Terms with $P_{14}$ or $P_{13}$:} both $P_{14}$ and $P_{23}$ are equal to $P[\vL_1-\vl_1-\vl_2,\kparone+\kpartwo]$, and grouping the corresponding terms together, we get
\begin{align*}
&4P[\vL_1-\vl_1-\vl_2,\kparone+\kpartwo] \times 
	\lb P_1 P_2  F(-2,23)F(-1,14) + P_1P_3  F(-3,32)F(-1,14) \right. \\
&\qquad\qquad\qquad\qquad\qquad\qquad\quad \left. 
	+\, P_2P_4 F(-4,41)F(-2,23) + P_3P_4  F(-4,41)F(-3,32) \rb \\
&= 4P[\vL_1-\vl_1-\vl_2,\kparone+\kpartwo] \times \lb F(-1,14)P_1 + F(-4,41)  P_4\rb  
	\times \lb F(-2,23) P_2 + F(-3,32)P_3   \rb \\
&= 4P[\vL_1-\vl_1-\vl_2,\kparone+\kpartwo]  \\
&\qquad\times
	\lb F(-(\vl_1,\kparone),(-\vL_2+\vl_1+\vl_2,\kparone+\kpartwo)) P[\vl_1,\kparone] \right. \\
&\qquad\qquad \left.
	+\, F(-(-\vL_2+\vl_2,\kpartwo),(-\vL_2+\vl_1+\vl_2,\kparone+\kpartwo)) P[-\vL_2+\vl_2,\kpartwo] \rb \\
&\qquad\times
	\lb  F(-(\vL_1-\vl_1,-\kparone),(\vL_1-\vl_1-\vl_2,-\kparone-\kpartwo)) P[\vL_1-\vl_1,-\kparone] \right. \\
&\qquad\qquad\left.
	+\,  F(-(-\vl_2,-\kpartwo),(\vL_1-\vl_1-\vl_2,-\kparone-\kpartwo)) P[-\vl_2,\kpartwo] \rb \\
&= \calL^4 \chi^{8} P[\vL_1-\vl_1-\vl_2,\kparone+\kpartwo]
	f_\delta(\vl_1,-\vL_2+\vl_2,\kparone,\kpartwo) f_\delta(\vL_1-\vl_1,-\vl_2,-\kparone,-\kpartwo)\ .
	\numberthis
\end{align*}
Under the $g_X g_Y$ integral, we can change variables from $\vl_1$ to $\vL_1-\vl_1$ to get
\beq
\calL^4 \chi^{8} P[\vl_1-\vl_2,\kparone+\kpartwo]
	f_\delta(\vL_1-\vl_1,-\vL_2+\vl_2,\kparone,\kpartwo) f_\delta(\vl_1,-\vl_2,-\kparone,-\kpartwo)\ .
\eeq
\end{enumerate}

So far, we have manipulated different terms appearing in the density trispectrum $T$. We can relate these to the four-point function of $\Iobs$ via
\begin{align*}
&\left\la \Iobs(\vl_1,\kparone) \Iobs(\vL_1-\vl_1,-\kparone) 
	\Iobs(-\vl_2,-\kpartwo) \Iobs(-\vL_2+\vl_2,\kpartwo) \right\ra_{\rm c} \\
&\quad = b^4 \calL^{-4} \chi^{-8} \left\la \delta(\vl_1/\chi,\kparone) \delta([\vL_1-\vl_1]/\chi,-\kparone)
	\delta(-\vl_2/\chi,-\kpartwo) \delta([-\vL_2+\vl_2]/\chi,\kpartwo) \right\ra_{\rm c} \\
&\quad = b^4 \calL^{-4} \chi^{-8} (2\pi)^2 \dirac([\vL_1-\vL_2]/\chi) \calL \times T(\cdots) \\
&\quad \supset \chi^{-6} \calL^{-3} (2\pi)^2 \dirac(\vL_1-\vL_2)  \\
&\qquad \times \calL^{4} \chi^8 \lb  
	P[\vL_1,0] f_\delta(\vl_1,\vL_1-\vl_1,\kparone,-\kparone) f_\delta(\vl_2,\vL_2-\vl_2,\kpartwo,-\kpartwo)
	\right. \\
&\qquad\qquad\quad + \left\{ P[\vl_1-\vl_2,\kparone-\kpartwo] 
	f_\delta(\vl_1,-\vl_2,\kparone,-\kpartwo) f_\delta(\vL_1-\vl_1,-\vL_2+\vl_2,-\kparone,\kpartwo)
	\right. \\
&\qquad\qquad\qquad + \left.\left. \lb \kparone\leftrightarrow -\kparone \rb \right\} \rb\ .
	\numberthis \label{eq:i4ptdelta4pt}
\end{align*}
When inserted into the integral in Eq.~\eqref{eq:ngcov}  with $X=Y=\phi$, the first term above will yield the $P(L/\chi)$ term from Eq.~\eqref{eq:didisignal}, while the other two terms are analogous to the $N_{XY}^{({\rm nG,}\phi,{\rm c})}$ term from Eq.~\eqref{eq:nxyphiphi}, but will have much larger magnitude.

In summary, at $\mathcal{O}(\phi^0 \delta_1^6)$, the terms with two factors of $\kerfs{2}$ contribute to the estimator covariance as
\beq
{\rm Cov}_{\rm nG}\!\lb \hat{X}(\vL,\kparone),\hat{Y}^*(\vL,\kpartwo) \rb \supset
	N_{XY}^{({\rm nG,}P)}(\vL,\kparone,\kpartwo) 
	+ N_{XY}^{\rm (nG,c,2)}(\vL,\kparone,\kpartwo)\ , \numberthis
\eeq
where
\beq
N_{XY}^{({\rm nG,}P)}(L,\kparone,\kpartwo)
	= \frac{N_{XX}^{\rm (G)}(L,\kparone) N_{YY}^{\rm (G)}(L,\kpartwo)}
	{N_{X\delta}^{\rm (G)}(L,\kparone)N_{Y\delta}^{\rm (G)}(L,\kpartwo)} 
	\calL \chi^2  P_{\delta 1}(L/\chi)
\eeq
and
\begin{align*}
N_{XY}^{\rm (nG,c,2)}(\vL,\kparone,\kpartwo) &\equiv \calL \chi^2 \int_{\vl_1} \int_{\vl_2} 
	g_X(\vl_1,\vL-\vl_1,\kparone) g_Y(\vl_2,\vL-\vl_2,\kpartwo)  \\
&\quad
	 \times\lp P[\vl_1-\vl_2,\kparone-\kpartwo] 
	f_\delta(\vl_1,-\vl_2,\kparone,-\kpartwo) f_\delta(\vL-\vl_1,-\vL+\vl_2,-\kparone,\kpartwo) \right. \\
&\qquad\quad + \left.  \lb \kparone \leftrightarrow -\kparone \rb  \rp\ .
	\numberthis
	\label{eq:nxyncg2}
\end{align*}

\subsection{$\mathcal{O}(\phi^0 \delta_1^6)$ terms with $\delta_3$}

There is a contribution arising from terms with one factor of $\kerfs{3}$ when the following expansion is inserted into the $\Iobs$ four-point function:
\begin{align*}
\Iobs(\vl,\kpar) &\supset b\calL^{-1} \chi^{-2} \delta_1(\vl/\chi,\kpar) \\
&\quad + b\calL^{-1} \chi^{-2} \left. \int_{\vq} \int_{\vq}
	 \kerfs{3}(\vq,\vp,\vk-\vq-\vp) \delta_1(\vq) \delta_1(\vp) \delta_1(\vk-\vq-\vp)
	\right|_{\vk=(\vl/\chi,\kpar)} \ . \numberthis
\end{align*}
As above, the relevant terms in the tree-level density trispectrum can be written as
\begin{align*}
T(1,2,3,4) &\supset 6 \lb F_3(1,2,3)P_1 P_2 P_3 + F_3(1,2,4)P_1 P_2 P_4 \right. \\
&\qquad\left. +\, F_3(1,3,4)P_1 P_3 P_4 + F_3(2,3,4)P_2 P_3 P_4 \rb\ .
	\numberthis
\end{align*}
Using Eq.~\eqref{eq:i4ptdelta4pt} to translate between the $\Iobs$ and $\delta$ four-point functions, with appropriate changes of variables we find that
\begin{align*}
&\left\la \Iobs(\vl_1,\kparone) \Iobs(\vL_1-\vl_1,-\kparone) \Iobs(-\vl_2,-\kpartwo) 
	\Iobs(-\vL_2+\vl_2,\kpartwo) \right\ra_{\rm c} \\
&\qquad \supset b^4 \calL^{-3} \chi^{-6}  (2\pi)^2 \dirac(\vL_1-\vL_2)
	\, 6 P[\vl_1,\kparone] P[\vl_2,\kpartwo] \\
&\qquad\quad \times   \lb
	 P[\vL_1-\vl_1,\kparone] 
	\left\{ \kerfs{3}((\vl_1,\kparone),(\vL_1-\vl_1,-\kparone),(-\vl_2,-\kpartwo))
	+ \lb \kpartwo\leftrightarrow -\kpartwo \rb \right\}  \right. \\
&\qquad\qquad + \left.
	 P[-\vL_2+\vl_2,\kpartwo]
	\left\{ \kerfs{3}((\vl_1,\kparone),(-\vl_2,-\kpartwo),(-\vL_2+\vl_2,\kpartwo))
	+ \lb \kparone\leftrightarrow -\kparone \rb \right\} \rb \\
&\qquad =  \calL^{-1} \chi^{-2}  (2\pi)^2 \dirac(\vL_1-\vL_2)
	\, 6 C_{\ell_1}(\kparone) C_{\ell_2}(\kpartwo)\\
&\qquad\quad \times   \lb
	 P[\vL_1-\vl_1,\kparone] 
	\left\{ \kerfs{3}((\vl_1,\kparone),(\vL_1-\vl_1,-\kparone),(-\vl_2,-\kpartwo))
	+ \lb \kpartwo\leftrightarrow -\kpartwo \rb \right\}  \right. \\
&\qquad\qquad + \left.
	 P[-\vL_2+\vl_2,\kpartwo]
	\left\{ \kerfs{3}((\vl_1,\kparone),(-\vl_2,-\kpartwo),(-\vL_2+\vl_2,\kpartwo))
	+ \lb \kparone\leftrightarrow -\kparone \rb \right\} \rb\ .
	\numberthis
\end{align*} 
Thus, the terms with one factor of $\kerfs{3}$ contribute to the estimator variance as
\beq
{\rm Cov}_{\rm nG}\!\lb \hat{X}(\vL,\kparone),\hat{Y}^*(\vL,\kpartwo) \rb \supset
	N_{XY}^{\rm (nG,c,3)}(\vL,\kparone,\kpartwo)\ ,
\eeq
where
\begin{align*}
&N_{XY}^{\rm (nG,c,3)}(\vL,\kparone,\kpartwo) \\
&\quad\equiv \calL^{-1} \chi^{-2}
	\int_{\vl_1} \int_{\vl_2} g_X(\vl_1,\vL-\vl_1,\kparone) g_Y(\vl_2,\vL-\vl_2,\kpartwo)
	\, 6 C_{\ell_1}(\kparone) C_{\ell_2}(\kpartwo)  \\
&\qquad\times  \lb
	 P[\vL-\vl_1,\kparone]
	\left\{ \kerfs{3}((\vl_1,\kparone),(\vL-\vl_1,-\kparone),(-\vl_2,-\kpartwo))
	+ \lb \kpartwo\leftrightarrow -\kpartwo \rb \right\} \right. \\
&\qquad\qquad \left. +\, P[-\vL+\vl_2,\kpartwo] 
	\left\{ \kerfs{3}((\vl_1,\kparone),(-\vl_2,-\kpartwo),(-\vL+\vl_2,\kpartwo))
	+ \lb \kparone\leftrightarrow -\kparone \rb \right\} \rb\ .	\numberthis
	\label{eq:nxyncg3}
\end{align*}
The $N_{XY}^{\rm (nG,c)}(L,\kparone,\kpartwo)$ term in Eq.~\eqref{eq:nxyngc} is then simply the sum of Eqs.~\eqref{eq:nxyncg2} and~\eqref{eq:nxyncg3}.

\section{Additional information about mode-coupling kernels}
\label{app:fkernels}

\subsection{Large-$\ell$ limits}
\label{app:flimits}

For  reconstruction of lensing by large-scale structure, or of the tidal/density field, the modes of the source field that contribute most in typical cases will have flat-sky wavenumber $\ell$ much larger than the wavenumber $L$ of the lensing mode being reconstructed. Therefore, it is useful to derive approximate forms\footnote{For other works that make use of this approximation for various purposes, see Refs.~\cite{Zahn:2005ap,Lu:2007pk,Bucher:2010iv,Hanson:2010rp,Prince:2017sms}.}  of the filters $f_\phi$ and $f_\delta$ [see Eqs.~\eqref{eq:fphidef2} and~\eqref{eq:fdeltadef}] for the case where $\ell\gg L$.

First, note that
\beq
C_{|\vL-\vl|} \underset{\ell\gg L}{\approx}
	C_{\ell} - \vL\cdot\frac{\d\ell}{\d\vl} \frac{\d C_{\ell}}{\d\ell}
	=  C_{\ell} \lp 1- \frac{\vL\cdot\vl}{\ell^2} \alpha(\ell,\kpar)  \rp\ ,
\eeq
with
\beq
\alpha(\ell,\kpar) \equiv \frac{\d\log C_{\ell}(\kpar)}{\d\log\ell}\ .
\label{eq:alpha2d}
\eeq
From this, we can expand $f_\phi$ like so:
\begin{align*}
f_\phi(\vl,\vL-\vl,\kpar,-\kpar)
	&=  \vL\cdot \lb \vl C_{\ell}(\kpar) + [\vL-\vl] C_{|\vL-\vl|}(\kpar) \rb \\
&\!\!\underset{\ell\gg L}{\approx}
	 \vL\cdot \lb \vl C_{\ell}(\kpar) 
	+ [\vL-\vl] C_{\ell}(\kpar) \lp 1- \frac{\vL\cdot\vl}{\ell^2} \alpha(\ell,\kpar)  \rp \rb \\
&=  \vL\cdot
	\lb \vL  C_{\ell}(\kpar) - \lb\vL-\vl\rb \frac{\vL\cdot\vl}{\ell^2} \alpha(\ell,\kpar)
	C_{\ell}(\kpar) \rb \\
&\!\!\underset{\ell\gg L}{\approx}
	 L^2 \lb 1+ \lp \hat{\vL}\cdot\hat{\vl} \rp^2 \alpha(\ell,\kpar) \rb C_{\ell}(\kpar)\ .
	 \numberthis
	 \label{eq:fphisqueezed}
\end{align*}
We similarly expand $f_\delta$:
\begin{align*}
&f_{\delta}(\vl,\vL-\vl,\kpar,-\kpar) \\
&\quad= 2\calL^{-1} \chi^{-2} 
	\lb \kerfs{2}\!\lp-(\vl/\chi,\kpar),(\vL/\chi,0)\rp
	 C_{\ell}(\kpar) 
+ \, \kerfs{2}\!\lp-([\vL-\vl]/\chi,-\kpar),(\vL/\chi,0)\rp
	 C_{|\vL-\vl|}(\kpar) \rb \\
&\quad= 2\calL^{-1} \chi^{-2} 
	\lb \frac{5}{7} \lp C_{\ell}(\kpar)+C_{|\vL-\vl|}(\kpar) \rp \right. \\
&\quad\qquad\qquad\qquad - \,
	\frac{1}{2} \vl\cdot\vL \lp \frac{1}{\ell^2+\chi^2\kpar^2} + \frac{1}{L^2}  \rp C_{\ell}(\kpar) \\
&\quad\qquad\qquad\qquad - \,
	 \frac{1}{2} (\vL-\vl)\cdot\vL 
	\lp \frac{1}{|\vL-\vl|^2+\chi^2\kpar^2} + \frac{1}{L^2}  \rp C_{|\vL-\vl|}(\kpar) \\
&\quad\qquad\qquad\qquad\left. + \,
	\frac{2}{7} \frac{(\vl\cdot\vL)^2}{(\ell^2+\chi^2\kpar^2)L^2} C_{\ell}(\kpar)
	+ \frac{2}{7} \frac{([\vL-\vl]\cdot\vL)^2}{(|\vL-\vl|^2+\chi^2\kpar^2)L^2} C_{|\vL-\vl|}(\kpar) \rb \\
&\quad \!\!\underset{\ell\gg L}{\approx}
	2\calL^{-1} \chi^{-2} \lb \frac{10}{7} C_{\ell}(\kpar)
	- \frac{1}{2} C_{\ell}(\kpar)  \lp 1+ \lp \hat{\vL}\cdot\hat{\vl} \rp^2 \alpha(\ell,\kpar)  \rp
	+ \frac{4}{7} \frac{\ell^2}{\ell^2+\chi^2\kpar^2} (\hat{\vL}\cdot\hat{\vl})^2 C_{\ell}(\kpar)
	\rb \\
&\quad= 2\calL^{-1} \chi^{-2} \lb \frac{13}{14} - \frac{1}{2} \lp \hat{\vL}\cdot\hat{\vl} \rp^2 \alpha(\ell,\kpar)
	+ \frac{4}{7} \frac{\ell^2}{\ell^2+\chi^2\kpar^2} (\hat{\vL}\cdot\hat{\vl})^2 \rb C_{\ell}(\kpar)\ .
	\numberthis
	\label{eq:fdeltasqueezed}
\end{align*}
We note that $f_\phi \propto L^2$ and $f_\delta\propto L^0$ in this limit, and also that both expressions become independent of $\ell$ in the $\chi\kpar \gg \ell$ limit. These results provide helpful intuition for some of the behaviors observed in Secs.~\ref{sec:ideal-im} and~\ref{sec:bhlensing}.

\subsection{Relationship to convergence and shear}
\label{app:fconshear}

We can further use the limits derived above to isolate the dependence of each mode-coupling kernel on the angle between $\vl$ and $\vL$, and therefore the anisotropy of the two-point function of small-scale intensity modes in the presence of long lensing or density modes [recall Eq.~\eqref{eq:iiavgbothfixed}]. We can rewrite Eq.~\eqref{eq:alpha2d} for $\alpha$  in terms of the tilt $\alpha_P$ of the 3d power spectrum $P_I$ as
\beq
\alpha(\ell,\kpar) = \frac{\ell^2}{\ell^2+\chi^2\kpar^2}  \alpha_P(\ell,\kpar)\ ,
\quad \alpha_P(\ell,\kpar) \equiv
	\left. \frac{\d\log P_I(k)}{\d\log k} \right|_{k = \sqrt{\ell^2/\chi^2+\kpar^2}} \ .
\eeq
If we also write $\cos\theta \equiv \hat{\vL}\cdot\hat{\vl}$, then Eq.~\eqref{eq:fphisqueezed} for $f_\phi$ can be rewritten as
\begin{align*}
f_\phi(\vl,\vL-\vl,\kpar,-\kpar) &\underset{\ell\gg L}{\approx}
	\frac{1}{2} L^2 \left\{  2 + \frac{\ell^2}{\ell^2+\chi^2\kpar^2} \alpha_P(\ell,\kpar)  \right. \\
&\qquad\qquad\quad\left. 
	+\,   \frac{\ell^2}{\ell^2+\chi^2\kpar^2} \alpha_P(\ell,\kpar)  \cos 2\theta \right\} C_\ell(\kpar)\ .
	\numberthis
	\label{eq:fphisep}
\end{align*}
Eq.~\eqref{eq:fdeltasqueezed} can similarly be rewritten as
\begin{align*}
f_\delta(\vl,\vL-\vl,\kpar,-\kpar) &\underset{\ell\gg L}{\approx}
	2\calL^{-1} \chi^{-2} \left\{  \frac{13}{14} 
	+ \frac{\ell^2}{\ell^2+\chi^2\kpar^2} \lb 
	-\frac{1}{4}\alpha_P(\ell,\kpar) + \frac{2}{7}   \rb \right. \\
&\qquad\qquad\qquad\quad \left. 
	+\, \frac{\ell^2}{\ell^2+\chi^2\kpar^2} \lb  -\frac{1}{4}\alpha_P(\ell,\kpar) + \frac{2}{7}  \rb 
	\cos 2\theta \right\} C_\ell(\kpar)\ .
	\numberthis
	\label{eq:fdeltasep}
\end{align*}
The first lines of Eqs.~\eqref{eq:fphisep} and~\eqref{eq:fdeltasep} are monopole-type distortions of the small-scale statistics of the $I$ field, while the second lines are quadrupole-type distortions, otherwise respectively known as {\em convergence} and {\em shear} in the context of lensing. Recall that the $f_\delta$ involves the $\kerfs{2}$ perturbation theory kernel in Eq.~\eqref{eq:f2s}. The three terms in this kernel correspond to an isotropic ``growth" effect, a ``shift" that has the same form as a local coordinate transformation at leading order, and an anisotropic distortion term~\cite{Sherwin:2012nh}. The shift term has exactly the same form as lensing, while the growth term adds an extra contribution to the $f_\delta$ monopole, and the anisotropic term adds to both the monopole and quadrupole.

Our ability to distinguish lensing from second-order gravitational nonlinearities arises from two differences between $f_\phi$ and $f_\delta$:
\begin{enumerate}
\item the different relative contributions of the monopole and quadrupole to each mode-coupling kernel, and
\item the different scale-dependences of each monopole and quadrupole.
\end{enumerate}
It is not necessary to have both of these differences in order to distinguish lensing from gravitational nonlinearity at this order, however.  For particular values of $\alpha_P$ and ranges of $\ell/\chi$ and $\kpar$, these two differences will provide varying degrees of discriminating power between lensing and gravitational nonlinearity. 

Separate convergence and shear estimators (e.g.~\cite{Bucher:2010iv,Lu:2007pk,Lu:2009je}) would separate the monopole and quadrupole distortions in a particularly transparent manner, and may also aid in distinguishing other sources of mode-coupling. We leave an investigation of this to future work.

\section{Results of single-band 21cm lensing forecasts}
\label{app:singleband21cm}

In this appendix, we present our signal-to-noise forecasts for the lensing auto spectrum and the cross spectra described in \sec{sec:summary-and-xcorrs}, for individual redshift bands within the redshift ranges of the 21cm surveys we consider in the main text. Tables~\ref{table:singleband-ska},~\ref{table:singleband-chime}, and~\ref{table:singleband-hirax} contain the results for our fiducial SKA1-Low reionization survey, with 5MHz bands; CHIME, with 25MHz bands; and HIRAX, also with 25MHz bands, respectively.

We have not explicitly performed forecasts for a set of bands that completely cover each survey's redshift range; instead, we have chosen a representative sample of bands. To compute the total signal to noise for an entire survey, we 1) construct a cubic spline that interpolates the S/N per band as a function of the band's central redshift, 2) compute the central redshifts of bands that would have the desired width and cover the entire redshift range, 3) evaluate the spline at each central redshift, and 4) sum the resulting S/N values in quadrature. We report these results in Table~\ref{table:sn1}.

In Tables~\ref{table:singleband-ska},~\ref{table:singleband-chime}, and~\ref{table:singleband-hirax}, we have listed several derived values for each band, in addition to the S/N values we have computed. The effective angular resolution $\ell_{\rm max}$ is the maximum $\ell$ value at which the 21cm angular power spectrum is less than the instrument's thermal noise (i.e.\ the typical value where the solid and dashed curves cross in the analogs of the left panels of Figs.~\ref{fig:ska} or~\ref{fig:chime}). We also quote several values of $j_{\rm max}$, based on different criteria:
\begin{itemize}
\item ``$j_{\max}$ from noise": the $j$ value above which the 21cm power spectrum falls completely below the instrument's thermal noise.
\item ``$j_{\max}$ from PT": at larger $j$ values than this, the 3d wavenumber $k = \lb (\ell/\chi)^2+(2\pi j/\calL)^2 \rb^{1/2}$ surpasses the maximum wavenumber at which the tree-level approximation in perturbation theory is valid (see Table~\ref{fig:treelevel-validity}), if $\ell$ is set to $\ell_{\rm max}$ as listed in Tables~\ref{table:singleband-ska}-\ref{table:singleband-hirax}. For many CHIME/HIRAX bands, this $j_{\rm max}$ can be lower than the value determined by thermal noise, meaning that it is our tree-level computation rather than the instrument's sensitivity that is limiting the range of the lensing estimator. This could be improved by going to higher order in perturbation theory, or investigating estimators calibrated with simulations, as in Refs.~\cite{Lu:2007pk,Lu:2009je}.
\item ``$j_{\rm max}$ from BH": the maximum $j$ value which will contribute any signal to the bias-hardened lensing estimator. As discussed in \sec{sec:bh}, we use $j_{\rm max} \sim 2\ell_{\rm max}/\chi$ for this, but the lensing signal begins to drop off already around $j\sim \ell_{\rm max}/\chi$. For cross-correlations with low-redshift tracers, this is typically not a limitation, since the range of the tracers can be truncated such that bias-hardening is not needed.
\end{itemize}

\begin{table}[t]
\centering
\begin{tabular}{ccc|ccc|ccc}
\hline \hline
\multicolumn{9}{c}{Single-band forecasts for SKA1-Low reionization survey}\\
\hline	
$z$ & $\Delta z$ &  $\ell_{\rm max}$ & \multicolumn{3}{c}{$j_{\rm max}$} &  \multicolumn{3}{c}{S/N}  \\
& & & from noise & from PT & from BH &  $\la\kappa\kappa\ra$ 
	& $\la\kappa g_{\rm LSST}\ra$ 
	& $\la\kappa \gamma_{\rm LSST}\ra$\\
\hline
6	& 0.17	& 5000	& 17	& 12	& 14	& 1.4	 	& 6.5 	 & 3.8 \\
7	& 0.23	& 4000	& 15	& -	& 12	& 1.1	 	& 7.0		 & 3.3 \\
8	& 0.3		& 3200	& 13	& -	& 10	& 0.76	 & 6.5	 & 3.2 \\
10	& 0.45	& 2500	& 10	& -	& 8	& 0.35	 & 5.2	 & 2.8 \\
12	& 0.6		& 2000	& 7	& -	& 6	& 0.14	 & 3.8	 & 2.1 \\
14	& 0.8		& 1100	& 4	& -	& 4	& 0.017	 & 1.6	 & 1.0 \\
\hline \hline \\
\end{tabular}
\caption{\label{table:singleband-ska}
Forecasts for the signal to noise on detections of $\la\kappa\kappa\ra$, $\la\kappa g_{\rm LSST}\ra$, or $\la\kappa \gamma_{\rm LSST}\ra$ (calculated as described in \sec{sec:summary-and-xcorrs}), for our fiducial SKA1-Low reionization survey from \sec{sec:ska}, and for redshift bands indicated by the first two columns. We also include several derived values: an effective angular resolution $\ell_{\rm max}$, and the $j$ (line-of-sight wavenumber) values at which the 21cm angular power spectrum drops below the thermal noise, exceeds the range of validity of our perturbative calculation, or fails to add signal to the bias-hardened lensing estimator.
}
\end{table}

We have also examined the effect of using wider bands for CHIME and HIRAX. Using two 100MHz bands, corresponding to $1.4<z<1.85$ and $1.85<z<2.5$, we find that the combined S/N for $\la\kappa g_{\rm LSST}\ra$ or $\la\kappa \gamma_{\rm LSST}\ra$ are within $\sim$20\% of the combined results from 25MHz bands. However, we find that the S/N on $\la\kappa\kappa\ra$ improves by about a factor of 3 compared to the 25MHz-band case. This is because wider bands contain more low-$\kpar$ modes, and these modes have a large impact on the performance of the bias-hardened lensing estimator. However, both the thermal noise in the receivers and the 21cm power spectrum will vary more strongly over these wider bands, decreasing the validity of our assumption that both can be evaluated at the mean redshift of each band; in this sense, the forecasts with narrower bands are more realistic.

\begin{table}[t]
\centering
\begin{tabular}{ccc|ccc|ccc}
\hline \hline
\multicolumn{9}{c}{Single-band forecasts for CHIME}\\
\hline	
$z$ & $\Delta z$ &  $\ell_{\rm max}$ & \multicolumn{3}{c}{$j_{\rm max}$} &  \multicolumn{3}{c}{S/N}  \\
& & & from noise & from PT & from BH &  $\la\kappa\kappa\ra$ 
	& $\la\kappa g_{\rm LSST}\ra$ 
	& $\la\kappa \gamma_{\rm LSST}\ra$\\
\hline
1.2	& 0.085	& 700	& 17	& 5	& 11	& 0.077	 & 6.0	 & 4.1 \\
1.4	& 0.1		& 650	& 15	& 7	& 10	& 0.091	 & 8.1	 & 6.2 \\
1.6	& 0.12	& 600	& 15	& 8	& 9	& 0.090	 & 10	 	& 7.9 \\
1.8	& 0.14	& 550	& 15	& 10	& 8	& 0.081	 & 12	 	& 9.7 \\
2.0	& 0.16	& 500	& 14	& 11	& 7	& 0.067	 & 13	 	& 11 \\
2.2	& 0.18	& 480	& 13	& 13	& 7	& 0.053	 & 14		 & 11 \\
2.4	& 0.2		& 450	& 12	& 14	& 6	& 0.043	 & 14		 & 11 \\
\hline \hline \\
\end{tabular}
\caption{\label{table:singleband-chime}
As Table~\ref{table:singleband-ska}, but for individual 25MHz bands in CHIME. (See \sec{sec:chime} for the assumptions used for these forecasts.)
}
\end{table}

\begin{table}[t]
\centering
\begin{tabular}{ccc|ccc|ccc}
\hline \hline
\multicolumn{9}{c}{Single-band forecasts for HIRAX}\\
\hline	
$z$ & $\Delta z$ &  $\ell_{\rm max}$ & \multicolumn{3}{c}{$j_{\rm max}$} &  \multicolumn{3}{c}{S/N}  \\
& & & from noise & from PT & from BH &  $\la\kappa\kappa\ra$ 
	& $\la\kappa g_{\rm LSST}\ra$ 
	& $\la\kappa \gamma_{\rm LSST}\ra$\\
\hline
1.4	& 0.1		& 1100	& 17	& 3	& 16	& 0.10	 & 6.3	 & 7.1 \\
1.6	& 0.12	& 1000	& 16	& 6	&15	& 0.32	 & 12	 	& 9.4 \\
1.8	& 0.14	& 950	& 15	& 8	& 14	& 0.39	 & 15 	 & 12 \\
2.0	& 0.16	& 900	& 15	& 10	& 13	& 0.40	 & 18 	 & 14 \\
2.2	& 0.18	& 850	& 14	& 12	& 12	& 0.37	 & 20		 & 15 \\
2.4	& 0.2		& 800	& 13	& 13	& 11	& 0.32	 & 21		 & 16 \\
\hline \hline \\
\end{tabular}
\caption{\label{table:singleband-hirax}
As Table~\ref{table:singleband-ska}, but for individual 25MHz bands in HIRAX.}
\end{table}

\bibliographystyle{JHEP_sjf}
\bibliography{references}

\providecommand{\href}[2]{#2}\begingroup\raggedright\begin{thebibliography}{100}

\bibitem{Smith:2007rg}
K.~M. Smith, O.~Zahn, and O.~Dore, {\it {Detection of Gravitational Lensing in
  the Cosmic Microwave Background}},  {\em Phys. Rev.} {\bf D76} (2007) 043510,
  [\href{http://arxiv.org/abs/0705.3980}{{\tt arXiv:0705.3980}}].

\bibitem{Hirata:2008cb}
C.~M. Hirata, S.~Ho, N.~Padmanabhan, U.~Seljak, and N.~A. Bahcall, {\it
  {Correlation of CMB with large-scale structure: II. Weak lensing}},  {\em
  Phys. Rev.} {\bf D78} (2008) 043520,
  [\href{http://arxiv.org/abs/0801.0644}{{\tt arXiv:0801.0644}}].

\bibitem{Ade:2015zua}
{\bf Planck} Collaboration, P.~A.~R. Ade et~al., {\it {Planck 2015 results. XV.
  Gravitational lensing}},  {\em Astron. Astrophys.} {\bf 594} (2016) A15,
  [\href{http://arxiv.org/abs/1502.01591}{{\tt arXiv:1502.01591}}].

\bibitem{Abazajian:2016yjj}
{\bf CMB-S4} Collaboration, K.~N. Abazajian et~al., {\it {CMB-S4 Science Book,
  First Edition}},  \href{http://arxiv.org/abs/1610.02743}{{\tt
  arXiv:1610.02743}}.

\bibitem{Kaplinghat:2003bh}
M.~Kaplinghat, L.~Knox, and Y.-S. Song, {\it {Determining neutrino mass from
  the CMB alone}},  {\em Phys. Rev. Lett.} {\bf 91} (2003) 241301,
  [\href{http://arxiv.org/abs/astro-ph/0303344}{{\tt astro-ph/0303344}}].

\bibitem{Troxel:2017xyo}
{\bf DES} Collaboration, M.~A. Troxel et~al., {\it {Dark Energy Survey Year 1
  Results: Cosmological Constraints from Cosmic Shear}},
  \href{http://arxiv.org/abs/1708.01538}{{\tt arXiv:1708.01538}}.

\bibitem{Mandelbaum:2017jpr}
R.~Mandelbaum, {\it {Weak lensing for precision cosmology}},
  \href{http://arxiv.org/abs/1710.03235}{{\tt arXiv:1710.03235}}.

\bibitem{Croft:2017tur}
R.~A.~C. Croft, A.~Romeo, and R.~B. Metcalf, {\it {Weak lensing of the
  Lyman-alpha forest}},  \href{http://arxiv.org/abs/1706.07870}{{\tt
  arXiv:1706.07870}}.

\bibitem{Metcalf:2017qty}
R.~B. Metcalf, R.~A.~C. Croft, and A.~Romeo, {\it {Noise Estimates for
  Measurements of Weak Lensing from the Lyman-alpha Forest}},
  \href{http://arxiv.org/abs/1706.08939}{{\tt arXiv:1706.08939}}.

\bibitem{Schaan:2018yeh}
E.~Schaan, S.~Ferraro, and D.~N. Spergel, {\it {Weak Lensing of Intensity
  Mapping: the Cosmic Infrared Background}},
  \href{http://arxiv.org/abs/1802.05706}{{\tt arXiv:1802.05706}}.

\bibitem{Furlanetto:2006jb}
S.~Furlanetto, S.~P. Oh, and F.~Briggs, {\it {Cosmology at Low Frequencies: The
  21 cm Transition and the High-Redshift Universe}},  {\em Phys. Rept.} {\bf
  433} (2006) 181--301, [\href{http://arxiv.org/abs/astro-ph/0608032}{{\tt
  astro-ph/0608032}}].

\bibitem{2012RPPh...75h6901P}
J.~R. {Pritchard} and A.~{Loeb}, {\it {21 cm cosmology in the 21st century}},
  {\em Reports on Progress in Physics} {\bf 75} (Aug., 2012) 086901,
  [\href{http://arxiv.org/abs/1109.6012}{{\tt arXiv:1109.6012}}].

\bibitem{Chang:2007xk}
T.-C. Chang, U.-L. Pen, J.~B. Peterson, and P.~McDonald, {\it {Baryon Acoustic
  Oscillation Intensity Mapping as a Test of Dark Energy}},  {\em Phys. Rev.
  Lett.} {\bf 100} (2008) 091303, [\href{http://arxiv.org/abs/0709.3672}{{\tt
  arXiv:0709.3672}}].

\bibitem{Cooray:2003ar}
A.~R. Cooray, {\it {Lensing studies with diffuse backgrounds}},  {\em New
  Astron.} {\bf 9} (2004) 173--187,
  [\href{http://arxiv.org/abs/astro-ph/0309301}{{\tt astro-ph/0309301}}].

\bibitem{Pen:2003yv}
U.-L. Pen, {\it {Gravitational lensing of pre-reionization gas}},  {\em New
  Astron.} {\bf 9} (2004) 417--424,
  [\href{http://arxiv.org/abs/astro-ph/0305387}{{\tt astro-ph/0305387}}].

\bibitem{Zhang:2005pu}
P.~Zhang and U.-L. Pen, {\it {Mapping dark matter with cosmic magnification}},
  {\em Phys. Rev. Lett.} {\bf 95} (2005) 241302,
  [\href{http://arxiv.org/abs/astro-ph/0506740}{{\tt astro-ph/0506740}}].

\bibitem{Zhang:2005eb}
P.~Zhang and U.-L. Pen, {\it {Precision measurement of cosmic magnification
  from 21 cm emitting galaxies}},  {\em Mon. Not. Roy. Astron. Soc.} {\bf 367}
  (2006) 169--178, [\href{http://arxiv.org/abs/astro-ph/0504551}{{\tt
  astro-ph/0504551}}].

\bibitem{Sigurdson:2005cp}
K.~Sigurdson and A.~Cooray, {\it {Cosmic 21-cm delensing of microwave
  background polarization and the minimum detectable energy scale of
  inflation}},  {\em Phys. Rev. Lett.} {\bf 95} (2005) 211303,
  [\href{http://arxiv.org/abs/astro-ph/0502549}{{\tt astro-ph/0502549}}].

\bibitem{Zahn:2005ap}
O.~Zahn and M.~Zaldarriaga, {\it {Lensing reconstruction using redshifted 21cm
  fluctuations}},  {\em Astrophys. J.} {\bf 653} (2006) 922--935,
  [\href{http://arxiv.org/abs/astro-ph/0511547}{{\tt astro-ph/0511547}}].

\bibitem{Hu:2001tn}
W.~Hu, {\it {Mapping the dark matter through the CMB damping tail}},  {\em
  Astrophys. J.} {\bf 557} (2001) L79--L83,
  [\href{http://arxiv.org/abs/astro-ph/0105424}{{\tt astro-ph/0105424}}].

\bibitem{Hu:2001kj}
W.~Hu and T.~Okamoto, {\it {Mass reconstruction with cmb polarization}},  {\em
  Astrophys. J.} {\bf 574} (2002) 566--574,
  [\href{http://arxiv.org/abs/astro-ph/0111606}{{\tt astro-ph/0111606}}].

\bibitem{Metcalf:2006ji}
R.~B. Metcalf and S.~D.~M. White, {\it {High-resolution imaging of the cosmic
  mass distribution from gravitational lensing of pregalactic HI}},  {\em Mon.
  Not. Roy. Astron. Soc.} (2006)
  [\href{http://arxiv.org/abs/astro-ph/0611862}{{\tt astro-ph/0611862}}]. [Mon.
  Not. Roy. Astron. Soc.381,447(2007)].

\bibitem{Metcalf:2008gq}
R.~B. Metcalf and S.~D.~M. White, {\it {Cosmological Information in the
  Gravitational Lensing of Pregalactic HI}},  {\em Mon. Not. Roy. Astron. Soc.}
  {\bf 394} (2009) 704--714, [\href{http://arxiv.org/abs/0801.2571}{{\tt
  arXiv:0801.2571}}].

\bibitem{Seljak:1998aq}
U.~Seljak and M.~Zaldarriaga, {\it {Measuring dark matter power spectrum from
  cosmic microwave background}},  {\em Phys. Rev. Lett.} {\bf 82} (1999)
  2636--2639, [\href{http://arxiv.org/abs/astro-ph/9810092}{{\tt
  astro-ph/9810092}}].

\bibitem{Zaldarriaga:1998te}
M.~Zaldarriaga and U.~Seljak, {\it {Reconstructing projected matter density
  from cosmic microwave background}},  {\em Phys. Rev.} {\bf D59} (1999)
  123507, [\href{http://arxiv.org/abs/astro-ph/9810257}{{\tt
  astro-ph/9810257}}].

\bibitem{Lu:2007pk}
T.~Lu and U.-L. Pen, {\it {Precision of diffuse 21-cm lensing}},  {\em Mon.
  Not. Roy. Astron. Soc.} {\bf 388} (2008) 1819,
  [\href{http://arxiv.org/abs/0710.1108}{{\tt arXiv:0710.1108}}].

\bibitem{Lu:2009je}
T.~Lu, U.-L. Pen, and O.~Dore, {\it {Dark Energy from Large-Scale Structure
  Lensing Information}},  {\em Phys. Rev.} {\bf D81} (2010) 123015,
  [\href{http://arxiv.org/abs/0905.0499}{{\tt arXiv:0905.0499}}].

\bibitem{Pourtsidou:2013hea}
A.~Pourtsidou and R.~B. Metcalf, {\it {Weak lensing with 21cm intensity mapping
  at $z \sim 2-3$}},  {\em Mon. Not. Roy. Astron. Soc.} {\bf 439} (2014)
  L36--L40, [\href{http://arxiv.org/abs/1311.4484}{{\tt arXiv:1311.4484}}].

\bibitem{Pourtsidou:2014pra}
A.~Pourtsidou and R.~B. Metcalf, {\it {Gravitational lensing of cosmological 21
  cm emission}},  {\em Mon. Not. Roy. Astron. Soc.} {\bf 448} (2015)
  2368--2383, [\href{http://arxiv.org/abs/1410.2533}{{\tt arXiv:1410.2533}}].

\bibitem{Romeo:2017zwt}
A.~Romeo, R.~B. Metcalf, and A.~Pourtsidou, {\it {Simulations for 21 cm
  radiation lensing at EoR redshifts}},
  \href{http://arxiv.org/abs/1708.01235}{{\tt arXiv:1708.01235}}.

\bibitem{Mandel:2005xh}
K.~S. Mandel and M.~Zaldarriaga, {\it {Weak gravitational lensing of
  high-redshift 21 cm power spectra}},  {\em Astrophys. J.} {\bf 647} (2006)
  719--736, [\href{http://arxiv.org/abs/astro-ph/0512218}{{\tt
  astro-ph/0512218}}].

\bibitem{Hilbert:2007jda}
S.~Hilbert, R.~B. Metcalf, and S.~D.~M. White, {\it {Imaging the Cosmic Matter
  Distribution using Gravitational Lensing of Pregalactic HI}},  {\em Mon. Not.
  Roy. Astron. Soc.} {\bf 382} (2007) 1494,
  [\href{http://arxiv.org/abs/0706.0849}{{\tt arXiv:0706.0849}}].

\bibitem{Kovetz:2012pt}
E.~D. Kovetz and M.~Kamionkowski, {\it {Galaxy-Cluster Masses via 21st-Century
  Measurements of Lensing of 21-cm Fluctuations}},  {\em Phys. Rev.} {\bf D87}
  (2013), no.~6 063516, [\href{http://arxiv.org/abs/1210.3041}{{\tt
  arXiv:1210.3041}}].

\bibitem{Kovetz:2012jq}
E.~D. Kovetz and M.~Kamionkowski, {\it {21-cm Lensing and the Cold Spot in the
  Cosmic Microwave Background}},  {\em Phys. Rev. Lett.} {\bf 110} (2013),
  no.~17 171301, [\href{http://arxiv.org/abs/1211.4610}{{\tt
  arXiv:1211.4610}}].

\bibitem{Sheere:2016yqu}
C.~Sheere, A.~van Engelen, P.~D. Meerburg, and J.~Meyers, {\it {Establishing
  the origin of CMB B-mode polarization}},  {\em Phys. Rev.} {\bf D96} (2017),
  no.~6 063508, [\href{http://arxiv.org/abs/1610.09365}{{\tt
  arXiv:1610.09365}}].

\bibitem{Book:2011dz}
L.~Book, M.~Kamionkowski, and F.~Schmidt, {\it {Lensing of 21-cm Fluctuations
  by Primordial Gravitational Waves}},  {\em Phys. Rev. Lett.} {\bf 108} (2012)
  211301, [\href{http://arxiv.org/abs/1112.0567}{{\tt arXiv:1112.0567}}].

\bibitem{Kovetz:2017agg}
E.~D. Kovetz et~al., {\it {Line-Intensity Mapping: 2017 Status Report}},
  \href{http://arxiv.org/abs/1709.09066}{{\tt arXiv:1709.09066}}.

\bibitem{Namikawa:2012pe}
T.~Namikawa, D.~Hanson, and R.~Takahashi, {\it {Bias-Hardened CMB Lensing}},
  {\em Mon. Not. Roy. Astron. Soc.} {\bf 431} (2013) 609--620,
  [\href{http://arxiv.org/abs/1209.0091}{{\tt arXiv:1209.0091}}].

\bibitem{Pen:2012ft}
U.-L. Pen, R.~Sheth, J.~Harnois-Deraps, X.~Chen, and Z.~Li, {\it {Cosmic
  Tides}},  \href{http://arxiv.org/abs/1202.5804}{{\tt arXiv:1202.5804}}.

\bibitem{Zhu:2015zlh}
H.-M. Zhu, U.-L. Pen, Y.~Yu, X.~Er, and X.~Chen, {\it {Cosmic tidal
  reconstruction}},  {\em Phys. Rev.} {\bf D93} (2016), no.~10 103504,
  [\href{http://arxiv.org/abs/1511.04680}{{\tt arXiv:1511.04680}}].

\bibitem{Zhu:2016esh}
H.-M. Zhu, U.-L. Pen, Y.~Yu, and X.~Chen, {\it {Recovering lost 21 cm radial
  modes via cosmic tidal reconstruction}},
  \href{http://arxiv.org/abs/1610.07062}{{\tt arXiv:1610.07062}}.

\bibitem{Limber:1953}
D.~N. Limber, {\it {The Analysis of Counts of the Extragalactic Nebulae in
  Terms of a Fluctuating Density Field}},  {\em Astrophys. J.} {\bf 117} (1953)
  134.

\bibitem{Kaiser:1991qi}
N.~Kaiser, {\it {Weak gravitational lensing of distant galaxies}},  {\em
  Astrophys. J.} {\bf 388} (1992) 272.

\bibitem{Bucher:2010iv}
M.~Bucher, C.~S. Carvalho, K.~Moodley, and M.~Remazeilles, {\it {CMB Lensing
  Reconstruction in Real Space}},  {\em Phys. Rev.} {\bf D85} (2012) 043016,
  [\href{http://arxiv.org/abs/1004.3285}{{\tt arXiv:1004.3285}}].

\bibitem{vanEngelen:2013rla}
A.~van Engelen, S.~Bhattacharya, N.~Sehgal, G.~P. Holder, O.~Zahn, and
  D.~Nagai, {\it {CMB Lensing Power Spectrum Biases from Galaxies and Clusters
  using High-angular Resolution Temperature Maps}},  {\em Astrophys. J.} {\bf
  786} (2014) 13, [\href{http://arxiv.org/abs/1310.7023}{{\tt
  arXiv:1310.7023}}].

\bibitem{Osborne:2013nna}
S.~J. Osborne, D.~Hanson, and O.~Dor\'{e}, {\it {Extragalactic Foreground
  Contamination in Temperature-based CMB Lens Reconstruction}},  {\em JCAP}
  {\bf 1403} (2014) 024, [\href{http://arxiv.org/abs/1310.7547}{{\tt
  arXiv:1310.7547}}].

\bibitem{Bernardeau:2001qr}
F.~Bernardeau, S.~Colombi, E.~Gaztanaga, and R.~Scoccimarro, {\it {Large scale
  structure of the universe and cosmological perturbation theory}},  {\em Phys.
  Rept.} {\bf 367} (2002) 1--248,
  [\href{http://arxiv.org/abs/astro-ph/0112551}{{\tt astro-ph/0112551}}].

\bibitem{Baumann:2010tm}
D.~Baumann, A.~Nicolis, L.~Senatore, and M.~Zaldarriaga, {\it {Cosmological
  Non-Linearities as an Effective Fluid}},  {\em JCAP} {\bf 1207} (2012) 051,
  [\href{http://arxiv.org/abs/1004.2488}{{\tt arXiv:1004.2488}}].

\bibitem{Carrasco:2012cv}
J.~J.~M. Carrasco, M.~P. Hertzberg, and L.~Senatore, {\it {The Effective Field
  Theory of Cosmological Large Scale Structures}},  {\em JHEP} {\bf 09} (2012)
  082, [\href{http://arxiv.org/abs/1206.2926}{{\tt arXiv:1206.2926}}].

\bibitem{Pajer:2013jj}
E.~Pajer and M.~Zaldarriaga, {\it {On the Renormalization of the Effective
  Field Theory of Large Scale Structures}},  {\em JCAP} {\bf 1308} (2013) 037,
  [\href{http://arxiv.org/abs/1301.7182}{{\tt arXiv:1301.7182}}].

\bibitem{Carrasco:2013sva}
J.~J.~M. Carrasco, S.~Foreman, D.~Green, and L.~Senatore, {\it {The 2-loop
  matter power spectrum and the IR-safe integrand}},  {\em JCAP} {\bf 1407}
  (2014) 056, [\href{http://arxiv.org/abs/1304.4946}{{\tt arXiv:1304.4946}}].

\bibitem{Baldauf:2014qfa}
T.~Baldauf, L.~Mercolli, M.~Mirbabayi, and E.~Pajer, {\it {The Bispectrum in
  the Effective Field Theory of Large Scale Structure}},  {\em JCAP} {\bf 1505}
  (2015), no.~05 007, [\href{http://arxiv.org/abs/1406.4135}{{\tt
  arXiv:1406.4135}}].

\bibitem{Angulo:2014tfa}
R.~E. Angulo, S.~Foreman, M.~Schmittfull, and L.~Senatore, {\it {The One-Loop
  Matter Bispectrum in the Effective Field Theory of Large Scale Structures}},
  {\em JCAP} {\bf 1510} (2015) 039, [\href{http://arxiv.org/abs/1406.4143}{{\tt
  arXiv:1406.4143}}].

\bibitem{Baldauf:2015zga}
T.~Baldauf, E.~Schaan, and M.~Zaldarriaga, {\it {On the reach of perturbative
  methods for dark matter density fields}},  {\em JCAP} {\bf 1603} (2016),
  no.~03 007, [\href{http://arxiv.org/abs/1507.02255}{{\tt arXiv:1507.02255}}].

\bibitem{McQuinn:2015tva}
M.~McQuinn and M.~White, {\it {Cosmological perturbation theory in 1+1
  dimensions}},  {\em JCAP} {\bf 1601} (2016), no.~01 043,
  [\href{http://arxiv.org/abs/1502.07389}{{\tt arXiv:1502.07389}}].

\bibitem{Foreman:2015lca}
S.~{Foreman}, H.~{Perrier}, and L.~{Senatore}, {\it {Precision Comparison of
  the Power Spectrum in the EFTofLSS with Simulations}},  {\em JCAP} {\bf 1605}
  (2016) 027, [\href{http://arxiv.org/abs/1507.05326}{{\tt arXiv:1507.05326}}].

\bibitem{Bertolini:2015fya}
D.~Bertolini, K.~Schutz, M.~P. Solon, J.~R. Walsh, and K.~M. Zurek, {\it
  {Non-Gaussian Covariance of the Matter Power Spectrum in the Effective Field
  Theory of Large Scale Structure}},  {\em Phys. Rev.} {\bf D93} (2016), no.~12
  123505, [\href{http://arxiv.org/abs/1512.07630}{{\tt arXiv:1512.07630}}].

\bibitem{Perko:2016puo}
A.~Perko, L.~Senatore, E.~Jennings, and R.~H. Wechsler, {\it {Biased Tracers in
  Redshift Space in the EFT of Large-Scale Structure}},
  \href{http://arxiv.org/abs/1610.09321}{{\tt arXiv:1610.09321}}.

\bibitem{Takahashi:2012em}
R.~Takahashi, M.~Sato, T.~Nishimichi, A.~Taruya, and M.~Oguri, {\it {Revising
  the Halofit Model for the Nonlinear Matter Power Spectrum}},  {\em Astrophys.
  J.} {\bf 761} (2012) 152, [\href{http://arxiv.org/abs/1208.2701}{{\tt
  arXiv:1208.2701}}].

\bibitem{Kesden:2003cc}
M.~H. Kesden, A.~Cooray, and M.~Kamionkowski, {\it {Lensing reconstruction with
  CMB temperature and polarization}},  {\em Phys. Rev.} {\bf D67} (2003)
  123507, [\href{http://arxiv.org/abs/astro-ph/0302536}{{\tt
  astro-ph/0302536}}].

\bibitem{Hanson:2010rp}
D.~Hanson, A.~Challinor, G.~Efstathiou, and P.~Bielewicz, {\it {CMB temperature
  lensing power reconstruction}},  {\em Phys. Rev.} {\bf D83} (2011) 043005,
  [\href{http://arxiv.org/abs/1008.4403}{{\tt arXiv:1008.4403}}].

\bibitem{Lewis:1999bs}
A.~Lewis, A.~Challinor, and A.~Lasenby, {\it {Efficient computation of CMB
  anisotropies in closed FRW models}},  {\em Astrophys. J.} {\bf 538} (2000)
  473--476, [\href{http://arxiv.org/abs/astro-ph/9911177}{{\tt
  astro-ph/9911177}}].

\bibitem{Carlson:2009it}
J.~Carlson, M.~White, and N.~Padmanabhan, {\it {A critical look at cosmological
  perturbation theory techniques}},  {\em Phys. Rev.} {\bf D80} (2009) 043531,
  [\href{http://arxiv.org/abs/0905.0479}{{\tt arXiv:0905.0479}}].

\bibitem{Hahn:2004fe}
T.~Hahn, {\it {CUBA: A Library for multidimensional numerical integration}},
  {\em Comput. Phys. Commun.} {\bf 168} (2005) 78--95,
  [\href{http://arxiv.org/abs/hep-ph/0404043}{{\tt hep-ph/0404043}}].

\bibitem{Dvorkin:2008tf}
C.~Dvorkin and K.~M. Smith, {\it {Reconstructing Patchy Reionization from the
  Cosmic Microwave Background}},  {\em Phys. Rev.} {\bf D79} (2009) 043003,
  [\href{http://arxiv.org/abs/0812.1566}{{\tt arXiv:0812.1566}}].

\bibitem{Takada:2013bfn}
M.~Takada and W.~Hu, {\it {Power Spectrum Super-Sample Covariance}},  {\em
  Phys. Rev.} {\bf D87} (2013), no.~12 123504,
  [\href{http://arxiv.org/abs/1302.6994}{{\tt arXiv:1302.6994}}].

\bibitem{Li:2014jra}
Y.~Li, W.~Hu, and M.~Takada, {\it {Super-Sample Signal}},  {\em Phys. Rev.}
  {\bf D90} (2014), no.~10 103530, [\href{http://arxiv.org/abs/1408.1081}{{\tt
  arXiv:1408.1081}}].

\bibitem{Moodley:inprep}
K.~Moodley et~al., {\it {in preparation}}, .

\bibitem{Prince:2017sms}
H.~Prince, K.~Moodley, J.~Ridl, and M.~Bucher, {\it {Real space lensing
  reconstruction using cosmic microwave background polarization}},  {\em JCAP}
  {\bf 1801} (2018), no.~01 034, [\href{http://arxiv.org/abs/1709.02227}{{\tt
  arXiv:1709.02227}}].

\bibitem{Barreira:2017sqa}
A.~Barreira and F.~Schmidt, {\it {Responses in Large-Scale Structure}},  {\em
  JCAP} {\bf 1706} (2017), no.~06 053,
  [\href{http://arxiv.org/abs/1703.09212}{{\tt arXiv:1703.09212}}].

\bibitem{Barreira:2017kxd}
A.~Barreira and F.~Schmidt, {\it {Response Approach to the Matter Power
  Spectrum Covariance}},  {\em JCAP} {\bf 1711} (2017), no.~11 051,
  [\href{http://arxiv.org/abs/1705.01092}{{\tt arXiv:1705.01092}}].

\bibitem{Pritchard:2015fia}
{\bf EoR/CD-SWG, Cosmology-SWG} Collaboration, J.~Pritchard et~al., {\it
  {Cosmology from EoR/Cosmic Dawn with the SKA}},  {\em PoS} {\bf AASKA14}
  (2015) 012, [\href{http://arxiv.org/abs/1501.04291}{{\tt arXiv:1501.04291}}].

\bibitem{Sarkar:2018gcb}
D.~Sarkar and S.~Bharadwaj, {\it {Modelling redshift-space distortion in the
  post-reionization ${\rm HI}$ 21-cm power spectrum}},  {\em Mon. Not. Roy.
  Astron. Soc.} {\bf 476} (2018) 96,
  [\href{http://arxiv.org/abs/1801.07868}{{\tt arXiv:1801.07868}}].

\bibitem{Pober:2014lva}
J.~C. Pober, {\it {The Impact of Foregrounds on Redshift Space Distortion
  Measurements With the Highly-Redshifted 21 cm Line}},  {\em Mon. Not. Roy.
  Astron. Soc.} {\bf 447} (2015), no.~2 1705--1712,
  [\href{http://arxiv.org/abs/1411.2050}{{\tt arXiv:1411.2050}}].

\bibitem{Koopmans2017}
A.~Ghosh, F.~Mertens, and L.~V.~E. Koopmans, {\it {Deconvolving the Wedge:
  Maximum-Likelihood Power Spectra via Spherical-Wave Visibility Modeling}},
  {\em Mon. Not. Roy. Astron. Soc.} {\bf 474} (2018) 4552,
  [\href{http://arxiv.org/abs/1709.06752}{{\tt arXiv:1709.06752}}].

\bibitem{2014SPIE.9145E..22B}
K.~{Bandura} et~al., {\it {Canadian Hydrogen Intensity Mapping Experiment
  (CHIME) pathfinder}},  in {\em Ground-based and Airborne Telescopes V},
  vol.~9145 of {\em Proc.\ SPIE}, p.~914522, July, 2014.
\newblock \href{http://arxiv.org/abs/1406.2288}{{\tt arXiv:1406.2288}}.

\bibitem{Newburgh:2016mwi}
L.~B. Newburgh et~al., {\it {HIRAX: A Probe of Dark Energy and Radio
  Transients}},  {\em Proc. SPIE Int. Soc. Opt. Eng.} {\bf 9906} (2016) 99065X,
  [\href{http://arxiv.org/abs/1607.02059}{{\tt arXiv:1607.02059}}].

\bibitem{Shaw:2013wza}
J.~R. Shaw, K.~Sigurdson, U.-L. Pen, A.~Stebbins, and M.~Sitwell, {\it {All-Sky
  Interferometry with Spherical Harmonic Transit Telescopes}},  {\em Astrophys.
  J.} {\bf 781} (2014) 57, [\href{http://arxiv.org/abs/1302.0327}{{\tt
  arXiv:1302.0327}}].

\bibitem{Masui:2012zc}
K.~W. Masui et~al., {\it {Measurement of 21 cm brightness fluctuations at z
  $\sim$ 0.8 in cross-correlation}},  {\em Astrophys. J.} {\bf 763} (2013) L20,
  [\href{http://arxiv.org/abs/1208.0331}{{\tt arXiv:1208.0331}}].

\bibitem{Castorina:2016bfm}
E.~Castorina and F.~Villaescusa-Navarro, {\it {On the spatial distribution of
  neutral hydrogen in the Universe: bias and shot-noise of the HI Power
  Spectrum}},  {\em Mon. Not. Roy. Astron. Soc.} {\bf 471} (2017) 1788,
  [\href{http://arxiv.org/abs/1609.05157}{{\tt arXiv:1609.05157}}].

\bibitem{Seo:2009fq}
H.-J. Seo, S.~Dodelson, J.~Marriner, D.~Mcginnis, A.~Stebbins, C.~Stoughton,
  and A.~Vallinotto, {\it {A ground-based 21cm Baryon acoustic oscillation
  survey}},  {\em Astrophys. J.} {\bf 721} (2010) 164--173,
  [\href{http://arxiv.org/abs/0910.5007}{{\tt arXiv:0910.5007}}].

\bibitem{2017arXiv171008591M}
K.~W. {Masui}, J.~R. {Shaw}, C.~{Ng}, K.~M. {Smith}, K.~{Vanderlinde}, and
  A.~{Paradise}, {\it {Algorithms for FFT Beamforming Radio Interferometers}},
  {\em ArXiv e-prints} (Oct., 2017)
  [\href{http://arxiv.org/abs/1710.08591}{{\tt arXiv:1710.08591}}].

\bibitem{Shaw-comm}
J.~R. Shaw. personal communication.

\bibitem{Shaw:2014khi}
J.~R. Shaw, K.~Sigurdson, M.~Sitwell, A.~Stebbins, and U.-L. Pen, {\it {Coaxing
  cosmic 21 cm fluctuations from the polarized sky using m-mode analysis}},
  {\em Phys. Rev.} {\bf D91} (2015), no.~8 083514,
  [\href{http://arxiv.org/abs/1401.2095}{{\tt arXiv:1401.2095}}].

\bibitem{Gong:2011mf}
Y.~Gong, A.~Cooray, M.~Silva, M.~G. Santos, J.~Bock, M.~Bradford, and
  M.~Zemcov, {\it {Intensity Mapping of the [CII] Fine Structure Line during
  the Epoch of Reionization}},  {\em Astrophys. J.} {\bf 745} (2012) 49,
  [\href{http://arxiv.org/abs/1107.3553}{{\tt arXiv:1107.3553}}].

\bibitem{Silva:2014ira}
M.~B. Silva, M.~G. Santos, A.~Cooray, and Y.~Gong, {\it {Prospects for
  Detecting C$\scriptsize{II}$ Emission During the Epoch of Reionization}},
  {\em Astrophys. J.} {\bf 806} (2015), no.~2 209,
  [\href{http://arxiv.org/abs/1410.4808}{{\tt arXiv:1410.4808}}].

\bibitem{Lidz:2016lub}
A.~Lidz and J.~Taylor, {\it {On Removing Interloper Contamination from
  Intensity Mapping Power Spectrum Measurements}},  {\em Astrophys. J.} {\bf
  825} (2016) 143, [\href{http://arxiv.org/abs/1604.05737}{{\tt
  arXiv:1604.05737}}].

\bibitem{Cheng:2016yvu}
Y.-T. Cheng, T.-C. Chang, J.~Bock, C.~M. Bradford, and A.~Cooray, {\it
  {Spectral Line De-confusion in an Intensity Mapping Survey}},  {\em
  Astrophys. J.} {\bf 832} (2016), no.~2 165,
  [\href{http://arxiv.org/abs/1604.07833}{{\tt arXiv:1604.07833}}].

\bibitem{Serra:2016jzs}
P.~Serra, O.~Dor\'{e}, and G.~Lagache, {\it {Dissecting the high-z interstellar
  medium through intensity mapping cross-correlations}},  {\em Astrophys. J.}
  {\bf 833} (2016), no.~2 153, [\href{http://arxiv.org/abs/1608.00585}{{\tt
  arXiv:1608.00585}}].

\bibitem{Dumitru:2018tgh}
S.~Dumitru, G.~Kulkarni, G.~Lagache, and M.~G. Haehnelt, {\it {Predictions and
  sensitivity forecasts for reionization-era [C II] line intensity mapping}},
  \href{http://arxiv.org/abs/1802.04804}{{\tt arXiv:1802.04804}}.

\bibitem{Yue:2015sua}
B.~Yue, A.~Ferrara, A.~Pallottini, S.~Gallerani, and L.~Vallini, {\it
  {Intensity mapping of [CII] emission from early galaxies}},  {\em Mon. Not.
  Roy. Astron. Soc.} {\bf 450} (2015), no.~4 3829--3839,
  [\href{http://arxiv.org/abs/1504.06530}{{\tt arXiv:1504.06530}}].

\bibitem{Zheng:2016xvo}
Y.~Zheng, P.~Zhang, and M.~Oh, {\it {Quantification of the multi-streaming
  effect in Redshift Space Distortion}},  {\em JCAP} {\bf 1705} (2017), no.~05
  030, [\href{http://arxiv.org/abs/1611.09075}{{\tt arXiv:1611.09075}}].

\bibitem{Bleem:2012gm}
L.~E. Bleem et~al., {\it {A Measurement of the Correlation of Galaxy Surveys
  with CMB Lensing Convergence Maps from the South Pole Telescope}},  {\em
  Astrophys. J.} {\bf 753} (2012) L9,
  [\href{http://arxiv.org/abs/1203.4808}{{\tt arXiv:1203.4808}}].

\bibitem{Krause:2016jvl}
E.~Krause and T.~Eifler, {\it {CosmoLike - Cosmological likelihood analyses for
  photometric galaxy surveys}},  {\em Mon. Not. Roy. Astron. Soc.} {\bf 470}
  (2017), no.~2 2100--2112, [\href{http://arxiv.org/abs/1601.05779}{{\tt
  arXiv:1601.05779}}].

\bibitem{Schmittfull:2017ffw}
M.~Schmittfull and U.~Seljak, {\it {Parameter constraints from
  cross-correlation of CMB lensing with galaxy clustering}},
  \href{http://arxiv.org/abs/1710.09465}{{\tt arXiv:1710.09465}}.

\bibitem{Laigle:2016jxn}
C.~Laigle et~al., {\it {The COSMOS2015 Catalog: Exploring the $1 < z < 6$
  Universe with half a million galaxies}},  {\em Astrophys. J. Suppl.} {\bf
  224} (2016), no.~2 24, [\href{http://arxiv.org/abs/1604.02350}{{\tt
  arXiv:1604.02350}}].

\bibitem{Abell:2009aa}
{\bf LSST Science, LSST Project} Collaboration, P.~A. Abell et~al., {\it {LSST
  Science Book, Version 2.0}},  \href{http://arxiv.org/abs/0912.0201}{{\tt
  arXiv:0912.0201}}.

\bibitem{Alonso:2017dgh}
D.~Alonso, P.~G. Ferreira, M.~J. Jarvis, and K.~Moodley, {\it {Calibrating
  photometric redshifts with intensity mapping observations}},  {\em Phys.
  Rev.} {\bf D96} (2017), no.~4 043515,
  [\href{http://arxiv.org/abs/1704.01941}{{\tt arXiv:1704.01941}}].

\bibitem{Pourtsidou:2015mia}
A.~Pourtsidou, D.~Bacon, R.~Crittenden, and R.~B. Metcalf, {\it {Prospects for
  clustering and lensing measurements with forthcoming intensity mapping and
  optical surveys}},  {\em Mon. Not. Roy. Astron. Soc.} {\bf 459} (2016), no.~1
  863--870, [\href{http://arxiv.org/abs/1509.03286}{{\tt arXiv:1509.03286}}].

\bibitem{Babich:2005sb}
D.~Babich and A.~Loeb, {\it {Polarization of 21cm radiation from the epoch of
  reionization}},  {\em Astrophys. J.} {\bf 635} (2005) 1--10,
  [\href{http://arxiv.org/abs/astro-ph/0505358}{{\tt astro-ph/0505358}}].

\bibitem{De:2013wca}
S.~De and H.~Tashiro, {\it {Galactic Faraday rotation effect on polarization of
  21 cm lines from the epoch of reionization}},  {\em Phys. Rev.} {\bf D89}
  (2014), no.~12 123002, [\href{http://arxiv.org/abs/1307.3584}{{\tt
  arXiv:1307.3584}}].

\bibitem{Desjacques:2016bnm}
V.~Desjacques, D.~Jeong, and F.~Schmidt, {\it {Large-Scale Galaxy Bias}},
  \href{http://arxiv.org/abs/1611.09787}{{\tt arXiv:1611.09787}}.

\bibitem{Senatore:2014vja}
L.~Senatore and M.~Zaldarriaga, {\it {Redshift Space Distortions in the
  Effective Field Theory of Large Scale Structures}},
  \href{http://arxiv.org/abs/1409.1225}{{\tt arXiv:1409.1225}}.

\bibitem{Lewandowski:2015ziq}
M.~Lewandowski, L.~Senatore, F.~Prada, C.~Zhao, and C.-H. Chuang, {\it {On the
  EFT of Large Scale Structures in Redshift Space}},
  \href{http://arxiv.org/abs/1512.06831}{{\tt arXiv:1512.06831}}.

\bibitem{delaBella:2017qjy}
L.~F. de~la Bella, D.~Regan, D.~Seery, and S.~Hotchkiss, {\it {The matter power
  spectrum in redshift space using effective field theory}},  {\em JCAP} {\bf
  1711} (2017), no.~11 039, [\href{http://arxiv.org/abs/1704.05309}{{\tt
  arXiv:1704.05309}}].

\bibitem{Hoffmann:2018clb}
K.~Hoffmann, Y.~Mao, H.~Mo, and B.~D. Wandelt, {\it {Signatures of Cosmic
  Reionization on the 21cm 2- and 3-point Correlation Function I: Quadratic
  Bias Modeling}},  \href{http://arxiv.org/abs/1802.02578}{{\tt
  arXiv:1802.02578}}.

\bibitem{Hirata:2002jy}
C.~M. Hirata and U.~Seljak, {\it {Analyzing weak lensing of the cosmic
  microwave background using the likelihood function}},  {\em Phys. Rev.} {\bf
  D67} (2003) 043001, [\href{http://arxiv.org/abs/astro-ph/0209489}{{\tt
  astro-ph/0209489}}].

\bibitem{Dodelson:2003bv}
S.~Dodelson, E.~Rozo, and A.~Stebbins, {\it {Primordial gravity waves and weak
  lensing}},  {\em Phys. Rev. Lett.} {\bf 91} (2003) 021301,
  [\href{http://arxiv.org/abs/astro-ph/0301177}{{\tt astro-ph/0301177}}].

\bibitem{2012PhRvD..86h3527S}
F.~Schmidt and D.~Jeong, {\it {Cosmic Rulers}},  {\em Phys. Rev.} {\bf D86}
  (2012) 083527, [\href{http://arxiv.org/abs/1204.3625}{{\tt
  arXiv:1204.3625}}].

\bibitem{Masui:2010cz}
K.~W. Masui and U.-L. Pen, {\it {Primordial gravity wave fossils and their use
  in testing inflation}},  {\em Phys. Rev. Lett.} {\bf 105} (2010) 161302,
  [\href{http://arxiv.org/abs/1006.4181}{{\tt arXiv:1006.4181}}].

\bibitem{2014PhRvD..89h3507S}
F.~Schmidt, E.~Pajer, and M.~Zaldarriaga, {\it {Large-Scale Structure and
  Gravitational Waves III: Tidal Effects}},  {\em Phys. Rev.} {\bf D89} (2014),
  no.~8 083507, [\href{http://arxiv.org/abs/1312.5616}{{\tt arXiv:1312.5616}}].

\bibitem{Masui:2017fzw}
K.~W. Masui, U.-L. Pen, and N.~Turok, {\it {Two- and Three-Dimensional Probes
  of Parity in Primordial Gravity Waves}},  {\em Phys. Rev. Lett.} {\bf 118}
  (2017), no.~22 221301, [\href{http://arxiv.org/abs/1702.06552}{{\tt
  arXiv:1702.06552}}].

\bibitem{1742-6596-256-1-012026}
C.~Loken et~al., {\it Scinet: Lessons learned from building a power-efficient
  top-20 system and data centre},  {\em Journal of Physics: Conference Series}
  {\bf 256} (2010), no.~1 012026.

\bibitem{Sherwin:2012nh}
B.~D. Sherwin and M.~Zaldarriaga, {\it {The Shift of the Baryon Acoustic
  Oscillation Scale: A Simple Physical Picture}},  {\em Phys. Rev.} {\bf D85}
  (2012) 103523, [\href{http://arxiv.org/abs/1202.3998}{{\tt
  arXiv:1202.3998}}].

\end{thebibliography}\endgroup

\end{document}